\documentclass[twocolumn,aps,prl,superscriptaddress,showpacs,secnumroman,showkeys]{revtex4}
\usepackage{amsmath,bm}%
\usepackage{graphicx}%

\begin{document}
\title{Reaction Dynamics and Multifragmentation in Fermi Energy Heavy Ion Reactions}  

\author{R. Wada}
\email[E-mail at:]{wada@comp.tamu.edu}
\author{T. Keutgen}
\thanks{Now at FNRS and IPN, Universit\'e Catholique de Louvain, B-1348 Louvain-Neuve, Belgium}
\author{K. Hagel}
\author{Y. G. Ma}
\thanks{on leave from Shanghai Institute of Nuclear Research,
Chinese Academy of Sciences, Shanghai 201800, China}
\author{J. Wang}
\author{M. Murray}
\thanks{Now at University of Kansas, Lawrence, Kansas 66045-7582}
\author{L. Qin}
\author{P. Smith}
\author{J. B. Natowitz}
\affiliation{Cyclotron Institute, Texas A\&M University, College Station, Texas 77843}

\author{R. Alfarro}
\affiliation{Instituto de Fisica, Universidad National Autonoma de Mexico, Apactado Postal 20-364 01000, Mexico City, Mexico}
\author{J. Cibor}
\affiliation{Institute of Nuclear Physics, ul. Radzikowskiego 152, PL-31-342 Krakow, Poland}
\author{M. Cinausero}
\affiliation{INFN, Laboratori Nazionali di Legnaro, I-35020 Legnaro, Italy}
\author{Y. El Masri}
\affiliation{FNRS and IPN, Universit\'e Catholique de Louvain, B-1348 Louvain-Neuve, Belgium}
\author{D. Fabris}
\affiliation{INFN and Dipartimento di Fisica dell' Universit\'a di Padova, I-35131 Padova, Italy}
\author{E. Fioretto}
\affiliation{INFN and Dipartimento di Fisica dell' Universit\'a di Padova, I-35131 Padova, Italy}
\author{A. Keksis}
\affiliation{Cyclotron Institute, Texas A\&M University, College Station, Texas 77843}
\author{ M. Lunardon}
\affiliation{INFN and Dipartimento di Fisica dell' Universit\'a di Padova, I-35131 Padova, Italy}
\author{A. Makeev}
\author{N. Marie}
\thanks{Now at LCP Caen, ISMRA, IN2P3-CNRS, F-14050 Caen, France}
\author{ E. Martin}
\affiliation{Cyclotron Institute, Texas A\&M University, College Station, Texas 77843}
\author{A. Martinez-Davalos}
\author{A. Menchaca-Rocha}
\affiliation{Instituto de Fisica, Universidad National Autonoma de Mexico, Apactado Postal 20-364 01000, Mexico City, Mexico}
\author{G. Nebbia}
\affiliation{INFN and Dipartimento di Fisica dell' Universit\'a di Padova, I-35131 Padova, Italy}
\author{G. Prete}
\affiliation{INFN, Laboratori Nazionali di Legnaro, I-35020 Legnaro, Italy}
\author{V. Rizzi}
\affiliation{INFN and Dipartimento di Fisica dell' Universit\'a di Padova, I-35131 Padova, Italy}
\author{A. Ruangma}
\author{D. V. Shetty}
\author{G. Souliotis}
\affiliation{Cyclotron Institute, Texas A\&M University, College Station, Texas 77843}
\author{P. Staszel}
\affiliation{Jagellonian University, M Smoluchowski Institute of Physics, PL-30059, Krakow, Poland}
\author{M. Veselsky}
\affiliation{Cyclotron Institute, Texas A\&M University, College Station, Texas 77843}
\author{G. Viesti}
\affiliation{INFN and Dipartimento di Fisica dell' Universit\'a di Padova, I-35131 Padova, Italy}
\author{E. M. Winchester}
\author{S. J. Yennello}
\affiliation{Cyclotron Institute, Texas A\&M University, College Station, Texas 77843}
\author{Z. Majka}
\affiliation{Jagellonian University, M Smoluchowski Institute of Physics, PL-30059, Krakow, Poland}
\collaboration {\bf The NIMROD collaboration}
\noaffiliation
\author{~and~A.~Ono}
\affiliation{ Department of Physics, Tohoku University, Sendai 980-8578, Japan}

\date{\today}

\begin{abstract}

The reaction systems, $^{64}$Zn + $^{58}$Ni, $^{64}$Zn + $^{92}$Mo, 
$^{64}$Zn + $^{197}$Au, at 26A, 35A and 47A MeV, have been studied 
both in experiments with a 4$\pi$ detector array, NIMROD, 
and with Antisymmetrized Molecular Dynamics model calculations employing  
effective interactions corresponding to soft and stiff equations of state 
(EOS). Direct experimental observables, such as multiplicity 
distributions, charge distributions, energy spectra and velocity spectra, have 
been compared in detail with those of the calculations and a reasonable 
agreement is obtained. The velocity distributions of $\alpha$ particles and 
fragments with Z $\geq$ 3 show distinct differences in calculations with 
the soft EOS and the stiff EOS. The velocity distributions of $\alpha$ 
particle and Intermediate Mass Fragments (IMF's) are best described by 
the stiff EOS. Neither of the above direct observables nor the strength of the
elliptic flow are sensitive to changes in the in-medium nucleon-nucleon (NN) 
cross sections. A detailed analysis of the central collision events calculated
with the stiff EOS revealed that multifragmentation with cold fragment 
emission is a common feature predicted for all reactions studied here. 
A possible multifragmentation scenario is presented; 
after the preequilibrium emission ceases in the composite system, 
cold light fragments are formed 
in a hotter gas of nucleons and stay cold until the composite system
underdoes multifragmentation. For reaction 
with $^{197}$Au at 47A MeV a significant radial expansion takes place. 
For reactions with $^{58}$Ni and $^{92}$Mo at 47A MeV semi-transparency 
becomes prominent. The differing reaction dynamics drastically change 
the kinematic characteristics of emitted fragments. This scenario gives 
consistent 
explanations for many existing experimental results in the Fermi energy domain.
\end{abstract}
 
\pacs{25.70Pq}

\keywords{Intermediate Heavy ion reactions, 4$\pi$ detector array, 
antisymmetrized molecular dynamics model calculations, nuclear equation 
of state, in-medium nucleon-nucleon cross section, cold fragment emission, 
semi-transparency, radial expansion, elliptic flow}

\maketitle
 
\section*{I. INTRODUCTION}

One of the main aims of the study of heavy ion reactions is to explore the 
properties of nuclear matter at various densities and temperatures. 
In intermediate heavy ion reactions (a few tens of MeV/nucleon to a few 
hundreds of MeV/nucleon), it is generally expected that the composite system 
of projectile and target nuclei is compressed and excited 
in the early stage of the reaction, and  
that the hot-dense nuclear system expands and 
breaks up by a multifragmentation process. Recently many studies have been
undertaken to elucidate possible critical behaviors for such  
matter~\cite{Chomaz01}. In many of these studies, thermal and/or chemical
equilibrium is assumed~\cite{Bonasera00}. 
However in order to reach high enough 
in excitation energy and/or in density, the collisions become very violent 
and the collision processes become very complicated. Therefore it is 
indispensable to establish reliable microscopic dynamical models for the 
study of the properties of the highly excited matter produced in heavy ion 
reactions. 

In order to elucidate the reaction mechanism many microscopic dynamical 
models for nuclear collisions have been proposed~\cite{Bonasera00,Moretto93}. 
Among such models, the molecular dynamics models are well suited to
deal with the multifragmentation process. In the Classical Molecular Dynamics 
model, particles are treated as 
point particles and their transport is governed by a classical equation of 
motion in a given mean field~\cite{Bondorf76,Wilets77,Bodmer77}. 
Nucleon-Nucleon collisions are taken into 
account and treated as hard-sphere scatterings without Pauli-blocking. 
In the Quantum Molecular Dynamics (QMD) model, each particle is 
described by a Gaussian wave packet~\cite{QMD,QMD1}. 
Initial nuclei are constructed 
by ensuring that there is less than one nucleon in each phase space cell of 
1/h$^{3}$. During the time evolution of the wave packets the Pauli 
principle is respected only by the Liouville theorem of classical 
mechanics. In the model nucleon-nucleon (NN) collisions are allowed
and the Pauli blocking is treated
in an approximate manner. During the propagation of 
the wave packets, however, the time evolution based on the classical 
equation of motion eventually leads the initial state 
into a Pauli-forbidden zone and the occupation number of nucleons in phase 
space often significantly exceeds 1/h$^{3}$~\cite{CoMD}.

There have been several attempts to respect the Pauli principle more strictly 
during the propagation of the wave packets within the classical mechanics. 
Introduction of a Pauli potential is one such 
approach~\cite{Boal,EQMD,CHIMERA}. 
The Pauli potential is a non-physical repulsive potential introduced to 
avoid the overlap of the wave packets in the phase space during the time 
evolution. However, when it is applied in heavy ion collisions, the Pauli 
potential operates as a spurious repulsive force to increase the nuclear 
stopping, especially during the early stages of the collisions~\cite{wada00}. 
Recently, Papa {\it et al.} introduced a new procedure to 
remove such overlaps of the wave packet in the phase space at each 
time step, without introducing the Pauli potential. The model is called the
Constraint Molecular Dynamics model~\cite{CoMD}. 
Although the idea is interesting and the computation time is short, 
no extensive application has yet been made for comparisons with experimental 
results. 

In order to resolve the problem from the quantum mechanical side, 
the Fermionic Molecular Dynamics model (FMD) and the Antisymmetrized 
Molecular Dynamics (AMD) model have been proposed~\cite{FMD,ono92}. 
In both models the total wave function of the system is antisymmetrized and 
described by a Slater determinant of Gaussian wave packets. The time evolution
of the centroid of the wave packets is
treated in a classical manner. In FMD the width of the wave 
packets is treated as variable in time and NN collisions are treated 
as potential scatterings. Until now the calculations have only been made 
with a harmonic oscillator potential and no application to 
heavy ion collisions has yet been made~\cite{FMD1}.
  
In AMD-V, an improved version of AMD used in this study, 
the quantum nature of nucleons in the wave packet propagation 
is incorporated as follows~\cite{ono96,ono99}: 

\begin{itemize}{}
\item {The total wave function of the system is 
antisymmetrized and therefore the Pauli principle is respected 
at all times.}

\item {The Pauli blocking in stochastic NN collisions is taken 
into account in an unambiguous manner. }

\item {The probabilistic nature of the wave packet is taken into account 
as a diffusion process during the wave packet propagation. The diffusion 
process is formulated in a manner to take into account 
the quantum branching
to many final states of multifragmentation channels.}

\end{itemize}

AMD-V has been applied
for intermediate heavy ion collisions and found to reproduce 
reasonably well the experimental results~\cite{wada00,wada98,ono02}. 
For example in the previous study of the $^{64}$Zn + 
$^{58}$Ni reactions at 35A-79A MeV~\cite{wada00}, the calculated multiplicity, 
charge distribution and energy spectra of the reaction products were in
good agreement with the experimental results. In that study it is  
pointed out that nuclear semi-transparency plays an important role 
in the multifragmentation process.
In the present paper we extend the earlier study with both experiments and 
model calculations on additional systems. 
Experimental reaction measurements have been extended both to
heavier systems and to lower incident energies. 
For the AMD-V calculations, the effective interaction and in-medium 
nucleon-nucleon (NN) cross section are the two important ingredients. 
In this work two different effective interactions with different stiffnesses 
of the equation of state (EOS) 
and two different formulations of the in-medium NN cross 
sections have been employed.

A goal of the present work is to elucidate the reaction mechanisms,
especially focusing on the equilibration and multifragmentation processes,
using a reliable dynamical model for intermediate heavy ion 
reactions. For this purpose, the best parameter set
for AMD-V calculations is chosen by comparing different calculations 
to the experimental results. Then employing the best parameter set in AMD-V,
the reaction mechanisms are studied in detail for the calculated events.   
Along these lines this paper is organized as follows;
In Sec.II the experiment is described. 
In Sec.III a brief description of the AMD approach, effective interactions
and in-medium NN cross sections is presented.
In Sec.IV some remarks on the data analysis both for experimental and 
calculated results are given.
In Sec.V detailed comparisons between the experimental results and calculated 
results are presented. In Sec.VI, a detailed analysis of the underline 
reaction mechanisms is presented and 
possible multifragmentation scenarios are proposed. 
In Sec.VII, the proposed 
multifragmentation mechanisms and existing data are discussed. 
In Sec.VIII a summary is given.

In this paper nine  reactions, $^{64}$Zn + $^{58}$Ni, $^{64}$Zn + $^{92}$Mo, 
$^{64}$Zn + $^{197}$Au, at 26A, 35A and 47A MeV, have been studied. In order 
to refer to each reaction system in the text, the target name and the incident
energy are used throughout the paper for simplicity, thus the 
$^{64}$Zn + $^{58}$Ni reaction at 47A MeV becomes $^{58}$Ni at 47A MeV.  
 
\section*{II. EXPERIMENT}

The experiment was performed at the K-500 super-conducting cyclotron facility
at Texas A\&M University, using the 4$\pi$ detector array, NIMROD, 
(Neutron Ion Multi-detector for Reaction Oriented Dynamics). $^{64}$Zn 
projectiles were incident on $^{58}$Ni, $^{92}$Mo and $^{197}$Au targets at 
energies of 26A, 35A and 47A MeV. NIMROD consists of 
a charged particle array set inside a neutron ball. The charged particle 
array is made of 166 segments in 12 concentric rings around the beam axis. 
Eight forward rings have the same geometrical design as the INDRA 
detector, but have less granularity~\cite{INDRA}. The angle, 
number of segments in 
each ring and solid angle of each CsI segment are given in Table I.  

The eight forward rings are covered by ionization chambers (IC). Furthermore
in each of these rings two of the segments have two Si detectors between 
the IC and CsI detectors (super telescopes) and three have one Si detector.
Each super telescope is further subdivided into two parts.
The CsI detector is a Tl doped crystal read by a photo-multiplier tube. 
A pulse shape discrimination method is employed to identify particles, 
using different responses of fast and slow components of the light output of 
the CsI crystals for different charged particles~\cite{CsI-detector}. 
The ionization chambers were made of fiber-glass 
(G10) and were filled with 30 Torr of CF$_4$ gas. Front and back windows were 
made of 2.0 $\mu$m aluminized Mylar foil. The signals were read by 5 to 10 fine 
wires, arranged perpendicular to the particle direction in the active 
volume. In the CsI detector Hydrogen and Helium isotopes are clearly 
identified and Li fragments are also isolated from the heavier fragments.
In the super telescopes, all isotopes with atomic number Z $\leq$ 8 are 
clearly identified and in all telescopes, particles are identified in atomic 
number.

\begin{table}
\caption{NIMROD Charge Particle Array}
\begin{tabular}{cccc}
\hline
Ring\ \ \ \ \ & Angle \ \ \ & No. of segments\ \ \  & Solid Angle\\
  & (deg.) &  & (msr)\\
\hline
1  & 4.3 & 12  & 0.96\\
2   & 6.4 & 12 & 2.67\\
3   & 9.4 & 12 & 4.26\\
4   & 12.9 & 12 & 7.99\\
5   & 18.2 & 12 & 16.1\\
6   & 24.5 & 24 & 12.7\\
7   & 32.1 & 12 & 33.6\\
8   & 40.4 & 24 & 27.6\\
9   & 61.2 & 16 & 154.0\\
10  & 90.0 & 14 & 207.0\\
11  & 120.0 & 8 & 378.0\\
12  & 152.5 & 8 & 241.0\\
\hline
\end{tabular}
\end{table}

The energy calibration of the Si detectors was made with a $^{228}$Th source 
and the observed punch through energies of identified particles. The 
punch through energies are calculated using a Range-Energy 
table~\cite{ORSAY-table}. Since the energy losses of light particles, 
especially high energy Hydrogen isotopes, are rather small in the Si 
detectors, evaluation of the energy deposited in the CsI crystal requires 
special care. 
Therefore an additional energy calibration was performed as a separate run 
using a few telescopes in an 80 cm diameter scattering chamber. 
In the calibration run, the reaction of $^{64}$Zn + $^{92}$Mo at 47A MeV 
was chosen as the standard. 
Si-Si telescopes backed by CsI detectors of three different lengths 
(1 cm, 3 cm and 5 cm) were used to measure the inclusive energy spectra of 
light charged particles. The energy spectra were measured at all angles 
corresponding to those of the 12 rings and solid angles were adjusted to 
be similar to those in Table I at each angle. 
The energy calibrations for high energy particles were made using the 
punch-through energies of different lengths of the CsI crystals. 
The energy calibration for heavier fragments were made, using the Si 
detector calibration. The extracted energy spectra of  
fragments with atomic charge between 4 to 10 have been compared with those in 
reference~\cite{wada00} for the $^{64}$Zn + $^{58}$Ni reaction at 35A MeV, 
in which Si detectors were used as $\Delta$E-E telescopes, 
and a good agreement is obtained.    
      
Neutron multiplicity was measured with the 4$\pi$ neutron ball 
surrounding the charged particle array. The neutron ball consists of two 
hemispherical end caps and a central cylindrical section. The hemispheres 
are 150 cm in diameter with beam pipe holes in the center and they are 
upstream and downstream of the charged particle array. The central 
cylindrical section is 1.25m long with an inner hole of 60 cm diameter 
and 150 cm outer diameter. It is divided into 4 segments in the azimuthal 
angle direction. Between the hemispheres and the central section, 
there are 20 cm air gaps for cables and a duct for a pumping station.
The neutron ball is filled with a pseudocumene based liquid scintillator
mixed with 0.3 weight percent of Gd salt (Gd 2-ethyl hexanoate)~\cite{Schmitt}.
Scintillation from a thermal neutron captured by Gd is detected by 
five 5-in phototubes in each hemisphere and three phototubes in each segment 
of the central section.      

In the experiment, data have been taken in two different trigger modes. One is
the minimum bias trigger in which at least one of the CsI detectors fired. 
The other is the high multiplicity trigger which required that at least 3-5 
CsI detectors fired. The minimum bias trigger was scaled down, typically 
by a factor of 10, to reduce the rate of peripheral events. In order to reduce 
the neutron background, the beam was swept away in the injection line between 
the ECR source and the K-500 cyclotron for 1 msec following detection of an 
event. 
 
\section*{III. MODEL CALCULATIONS}

A brief description of the AMD model, including recent 
improvements related to the present work, is given in this section.
Two important ingredients, effective interactions and in-medium 
nucleon-nucleon (NN) cross sections, are also described. In order to show 
possible effects on the experimental observables resulting from different 
ingredients, changes in nuclear semi-transparency, a characteristic 
feature predicted by the model for intermediate heavy ion collisions, 
are explored. 

\subsection*{A. AMD-V model}

In AMD a reaction system with N nucleons is described by a wave function 
which is a single Slater determinant of N Gaussian wave packets~\cite{ono92},
 
\begin{equation}
\Phi(Z)=
\det\Bigl[
  \exp\Bigl\{
    -\nu\Bigl({\bf r}_j - \frac{{\bf Z}_i}{\sqrt\nu}\Bigr)^2
    +\frac{1}{2}{\bf Z}_i^2
  \Bigr\}\chi_{\alpha_i}(j)
\Bigr],
\label{eq:AMDWaveFunction}
\end{equation}
where the complex variables $Z\equiv\{{\bf Z}_i;\
i=1,\ldots,N\}=\{Z_{i\sigma};\ i=1,\ldots,N,\ \sigma=x,y,z\}$
represent the centroids of the wave packets.  
{\bf Z$_i$} can be described by 
\begin{eqnarray}
{\bf Z}_i = \sqrt{\nu} {\bf D}_i + {i \over{2\hbar\sqrt{\nu}}} {\bf K}_i
\end{eqnarray}
  
The width parameter $\nu$ is 
taken as $\nu=0.16$ $\rm{fm}^{-2}$ and 
$\chi_{\alpha_i}$ represents the spin and isospin states of $p\uparrow$, 
$p\downarrow$, $n\uparrow$, or $n\downarrow$. For a dilute nuclear gas system, 
{\bf D$_i$} and {\bf K$_i$} correspond to the position and momentum of
each nucleon. Inside the nucleus, however, these quantities do not have
physical meanings because of the antisymmetrization.
The time evolution of $Z$ is 
determined by the time-dependent variational principle and the two nucleon 
collision process. The equation of motion for $Z$ derived from the 
time-dependent variational principle is 

\begin{equation}
  i\hbar\sum_{j\tau}C_{i\sigma,j\tau}{dZ_{j\tau}\over dt}=
  {\partial{\cal H}\over\partial Z_{i\sigma}^*}.
  \label{eq:AMDEqOfMotion}
\end{equation}
$C_{i\sigma,j\tau}$ is a hermitian matrix defined by
\begin{equation}
C_{i\sigma,j\tau}=
\frac{\partial^2}{\partial Z_{i\sigma}^*\partial Z_{j\tau}}
\log\langle\Phi(Z)|\Phi(Z)\rangle,
\end{equation}
and $\cal H$ is the expectation value of the Hamiltonian after the
subtraction of the spurious kinetic energy of the zero-point
oscillation of the center-of-masses of fragments.
Two nucleon collisions are introduced by the use of the physical coordinates
W$\equiv\Bigl\{W_i\Bigr\}$ which are defined as
\begin{eqnarray}
{\bf W}_i = \sum_{j=1}^{A} ( \sqrt{Q} )_{ij} {\bf Z}_i,
  \label{eq:AMDphysicalK}
\end{eqnarray}
and Q$_{ij}$ is defined as
\begin{eqnarray}
Q_{ij}=\frac{\partial}{\partial({\bf Z}_i^*\cdot{\bf Z}_j)}
       \log\langle\Phi(Z)|\Phi(Z)\rangle.
\end{eqnarray}
In molecular dynamics models with Gaussian wave packets, the i-th nucleon 
at time t=t$_0$ is represented in phase space by
\begin{eqnarray}
f_i({\bf r},{\bf p},t_0)
=8exp\Bigl\{
  -2\nu({\bf r}-{\bf R}_i(t_0))^2 
  - {({\bf p}-{\bf P}_i(t_0))^2\over{2\hbar^2\nu}}
\Bigr\}
\end{eqnarray}
with the centroid ${\bf R}_i$ and ${\bf P}_i$. The total one-body distribution
function is the sum of $f_i$. In AMD, this representation of nucleon as a
simple Gaussian wave packet is only approximately valid when the physical
coordinate 
\begin{eqnarray}
{\bf W}_i = \sqrt{\nu} {\bf R}_i + {i \over{2\hbar\sqrt{\nu}}} {\bf P}_i
\end{eqnarray}
is used for the centroids. 

In order to properly treat reactions with many branching channels, such as 
multifragmentation processes, 
AMD has been extended by introducing the wave
packet diffusion effect as a quantum branching process. This extended AMD is
called AMD-V, since the wave packet diffusion effect is calculated with the
Vlasov equation~\cite{ono96}. 
The AMD-V code has been further improved in order to save 
CPU time in the numerical calculations and to be applicable to heavier 
reaction systems. In the newly developed code, used for all
calculations in this paper, the calculation of the wave packet 
diffusion effect has been reformulated and a triple-loop 
approximation has been incorporated~\cite{ono99}.

The calculations 
were performed in the VPP700E supercomputer facility in RIKEN, Japan. 
For the $^{64}$Zn + $^{58}$Ni and $^{64}$Zn + $^{92}$Mo reactions, 
about 5000 and 3000 events, respectively, were generated  
at each energy in the impact parameter range of 0-12 fm for a given parameter 
set. For the $^{64}$Zn + $^{197}$Au case, about 1000 events were generated 
in the impact parameter range of 0-14 fm. 
The calculations were started with a distance of 15 fm along the beam direction 
between centers of the projectile and the target. 
The calculation for each event was carried
out typically up to t=300 fm/c for the reactions at 35A and 47A MeV
and up to 500 fm/c for those at 26A MeV. In the following text, t=0 is 
set as the time at which the projectile and the target touch each other.
 
\subsection*{B. Effective interaction and in-medium NN cross section}

\begin{table*}
\caption{Parameters for the stiff effective interaction}
\begin{tabular}{ccccccccc}
\hline
 & $a_k$ & $W_k$ & $B_k$ & $H_k$ & $M_k$ & $\sigma$ & $t_\rho$ & $t_\mathrm{surf}$ \\ 
 \ \ \ \ \ \ &\ \ [fm]\ \ &\ \ [MeV]\ \ &\ \ [MeV]\ \ &\ \ [MeV]\ \ &\ \ [MeV]\ \ &  & [MeV fm$^{3(1+\sigma)}$]& [MeV fm$^5$] \\
\hline
$\ k=1\ $&0.7&$-402.4$&$-100.0$&$-496.2$&$-23.56$&     &      &\\
$k=2$& 1.2&    $0.96$& $51.575$& $46.535$& $-33.41$&     &      &\\
     &    &          &         &         &         &\ \ 1.24\ \ &1896&75\\
\hline
\label {StiffPara}
\end{tabular}

\end{table*}

The Gogny interaction~\cite{Decharge} has been used successfully in the 
previous analyses~\cite{wada00,ono96,wada98,ono02}.
This interaction gives a soft equation of state (EOS) with 
an incompressibility value K of 228 MeV for infinite nuclear matter 
and has a momentum dependent mean field.
In the present work, a modified Gogny interaction with a stiff EOS 
with K=360 MeV is also applied~\cite{ono00}. The parameter set adopted for 
K=360 MeV was that used by Haddad et al.~\cite{Haddad} and labeled as D1-G3.
However this new interaction gives neither the proper charge radius nor the
correct binding energy for the nuclear ground state in AMD.  In order to get
proper ground state properties of nuclei, it is necessary to add a
surface term in the Hamiltonian.  Furthermore we correct the two body
interaction term, so that, the force produces a reasonable equation of
state for asymmetric nuclear matter.  The expectation value of our
stiff Gogny force $V$ is given by

\begin{equation}
\langle V\rangle
=\langle \sum_{i<j}v_2(\mathbf{r}_i,\mathbf{r}_j)\rangle
+\langle \sum_{i<j}v_\rho(\mathbf{r}_i,\mathbf{r}_j)\rangle
+ \mathcal{V}_\mathrm{surf},
\label{eq:E_internal}
\end{equation}
where
\begin{widetext}
\begin{eqnarray}
v_2(\mathbf{r}_i,\mathbf{r}_j)
&=&\sum_{k=1,2} (W_k+B_kP_\sigma-H_kP_\tau-M_kP_\sigma P_\tau)
  e^{-(\mathbf{r}_i-\mathbf{r}_j)^2/a_k^2},\\
v_\rho(\mathbf{r}_i,\mathbf{r}_j)
&=& t_\rho(1+P_\sigma)\rho(\mathbf{r}_i)^\sigma
    \delta(\mathbf{r}_i-\mathbf{r}_j),\\
\mathcal{V}_\mathrm{surf}
&=&
t_\mathrm{surf}
      \int d^3\mathbf{r}
      \sum_{\alpha,\beta}\langle\alpha\beta|P_\sigma|\alpha\beta\rangle
    [\nabla\rho_\alpha(\mathbf{r})]\cdot[\nabla\rho_\beta(\mathbf{r})].
\end{eqnarray}
\end{widetext}

The indices $\alpha$ and $\beta$ take four states of spin and isospin,
$p\uparrow$, $p\downarrow$, $n\uparrow$, and $n\downarrow$.  The
parameters are shown in Table~\ref{StiffPara}.  
This new effective interaction is
used in all calculations with the stiff EOS in the present study.  The
incompressibility $K$ is still 360 MeV, because the surface term does
not contribute to the incompressibility of infinite nuclear matter and
the correction to the two-body part does not affect the symmetric
nuclear matter.

Calculated binding energies for the soft EOS deviate from the experimental 
values, which were taken from \cite{MassTable}, by about 0.5 MeV/nucleon for 
the mass number A $\leq$ 30 and about 
0.2 MeV/nucleon for the heavier fragments with A $\leq$ 100. 
The agreement to the experimental values become better for the heavier 
fragments. The calculated binding energies for the stiff EOS, on the other 
hand, show larger deviations and they are up to 1 MeV/nucleon less bound 
for all fragments, comparing to the experimental values.
 
\begin{figure}
\includegraphics[scale=0.55]{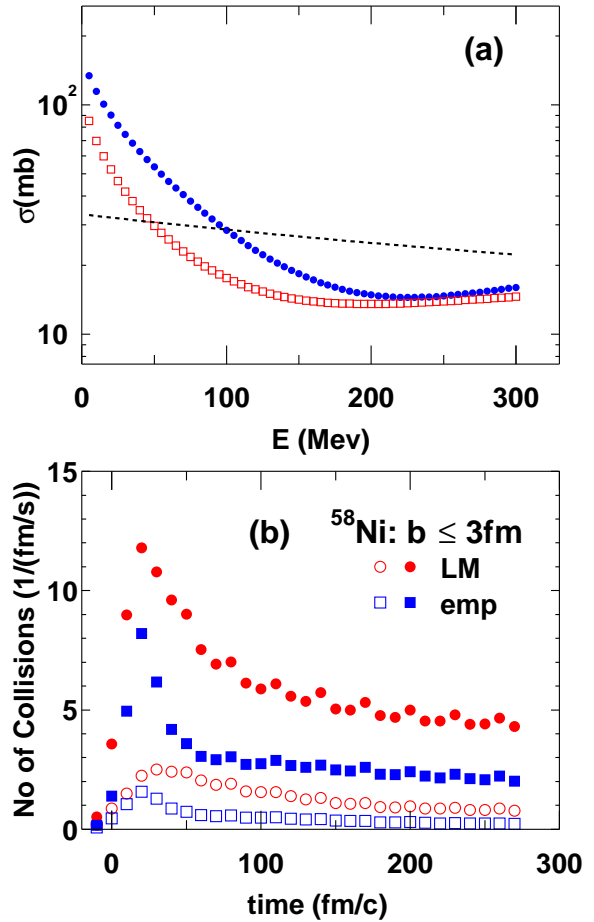}
\caption{\footnotesize (a) Calculated in-medium NN cross sections for  
normal density nuclear matter. The cross sections calculated by the 
empirical formula Eq.\ (\ref{eq:NNempirical}) are depicted by a dashed line 
and those of the Li-Machleidt formulae 
Eqs.\ (\ref{eq:NN-LMpn}),(\ref{eq:NN-LMpp}) 
are given by dots(np) and squares (nn,pp). 
(b) Number of collisions as a function of reaction 
time for central collisions of $^{64}$Zn + $^{58}$Ni at 47A MeV. Events 
with b$\leq$3 fm are analyzed. Solid symbols indicate the number of 
attempted collisions and open symbols indicate the number of Pauli-allowed 
collisions. Circles show the results of the Li-Machleidt formulae and squares 
show those of the empirical formula. 
}
\label{Ncollision}
\end{figure}

In the previous analyses, an empirical in-medium nucleon-nucleon cross 
section was used~\cite{wada00,ono96,wada98}. This cross 
section is given by~\cite{ono93} 
\begin{eqnarray}
\sigma_{pn} = \sigma_{pp} \ \ \ \ \ \ \ \ \ \ \ \ \ \ \ \ \ \ \ \ \ \ \ \ \ \ \ \ \ \ \ \ \ \ \ \ \ \ \ \ \ \ \ \ \ \ \ \nonumber\\
= {100 \over{1+E/(200MeV) + 2min((\rho/\rho_o)^{1/ 2},1)}}
  \label{eq:NNempirical}
\end{eqnarray}
where $\rho$ is the nuclear density and $\rho_o$ is its normal value. 
The cross section is given in mbarn. 
The calculated cross section for the normal nuclear density is shown 
in Fig.~\ref{Ncollision}(a) by a dashed line. 
In the cross sections no distinction was made
between p-p(n-n) and n-p collisions. The density 
dependence is taken into account, but the dependence is rather small
in the range of the density variation expected for the reactions studied 
in this work. For $\rho/\rho_0=1.5$, the cross section decreases by 
about 10\%.

Li and Machleidt calculated the in-medium NN cross sections 
based on the Dirac-Brueckner approach~\cite{Li93,Li94}. 
For np collisions, the in-medium cross section is given by 
\begin{eqnarray}
\sigma_{pn} = [31.5 +{0.092 \times |20.2-E^{0.53}|^{2.9}}]\nonumber\\
\times {1.0+0.0034E^{1.51}\rho^2\over{1.0+21.55\rho^{1.34}}}
  \label{eq:NN-LMpn}
\end{eqnarray}
and for pp, the cross section is given by
\begin{eqnarray}
\sigma_{pp} = [23.5 +{0.00256 \times |18.2-E^{0.5}|^{4.0}}]\nonumber\\
\times {1.0+0.1667E^{1.05}\rho^3\over{1.0+9.704\rho^{1.2}}}
  \label{eq:NN-LMpp}
\end{eqnarray}
A normal nuclear density of 0.18 fm$^{-3}$ is assumed in the paper. 
The calculated 
cross sections for normal nuclear matter are represented by symbols 
in Fig.~\ref{Ncollision}(a). At low energies these cross sections are about 
two to four times greater than those resulting from the empirical formula 
previously employed.
Above 100 MeV, both pp and np cross sections become smaller than that of the 
empirical prescription. For $\rho$=1.5$\rho_o$, the cross 
section decreases by about 35\% at E=0 MeV and 25\% at E=100 MeV.  

In Fig.~\ref{Ncollision}(b) the number of attempted and 
Pauli-allowed collisions are shown for actual calculations with the empirical 
cross section and with
the Li-Machleidt cross section. The calculations were made for the central 
collision events ( b $\leq$ 3 fm) of $^{64}$Zn + 
$^{58}$Ni at 47A MeV. The soft Gogny interaction was used.   
For the empirical formula, the number of attempted collisions reaches $\sim$8 
collisions/(fm/c) at a time of 30 fm/c when the two nuclei totally overlap 
and decreases quickly to 3-4 collisions/( fm/c). Only about 20\% of the 
attempted collisions are Pauli-allowed during this process. For the 
Li-Machleidt cross sections, the number of attempted collisions
increases by almost a factor of two. About 30\% of the attempted
collisions are Pauli-allowed in this case.

\subsection*{C. Nuclear semi-transparency}

It has been reported that nuclear transparency plays an important 
role for the multifragmentation process in intermediate heavy ion 
collisions~\cite{wada00,Nebauer}. The stiffness of the effective interaction 
and the in-medium NN cross section are both important ingredients for 
determining the degree of the transparency in the calculations. 
In the present study three 
parameter sets have been investigated for the calculations. They are 
\begin{itemize}{}
\item{soft EOS + NN$_{emp}$ (empirical NN cross section)} 
\item{stiff EOS + NN$_{emp}$ } 
\item{stiff EOS + NN$_{LM}$ (Li-Machleidt cross section).} 
\end{itemize}

\begin{figure}
\includegraphics[scale=0.45]{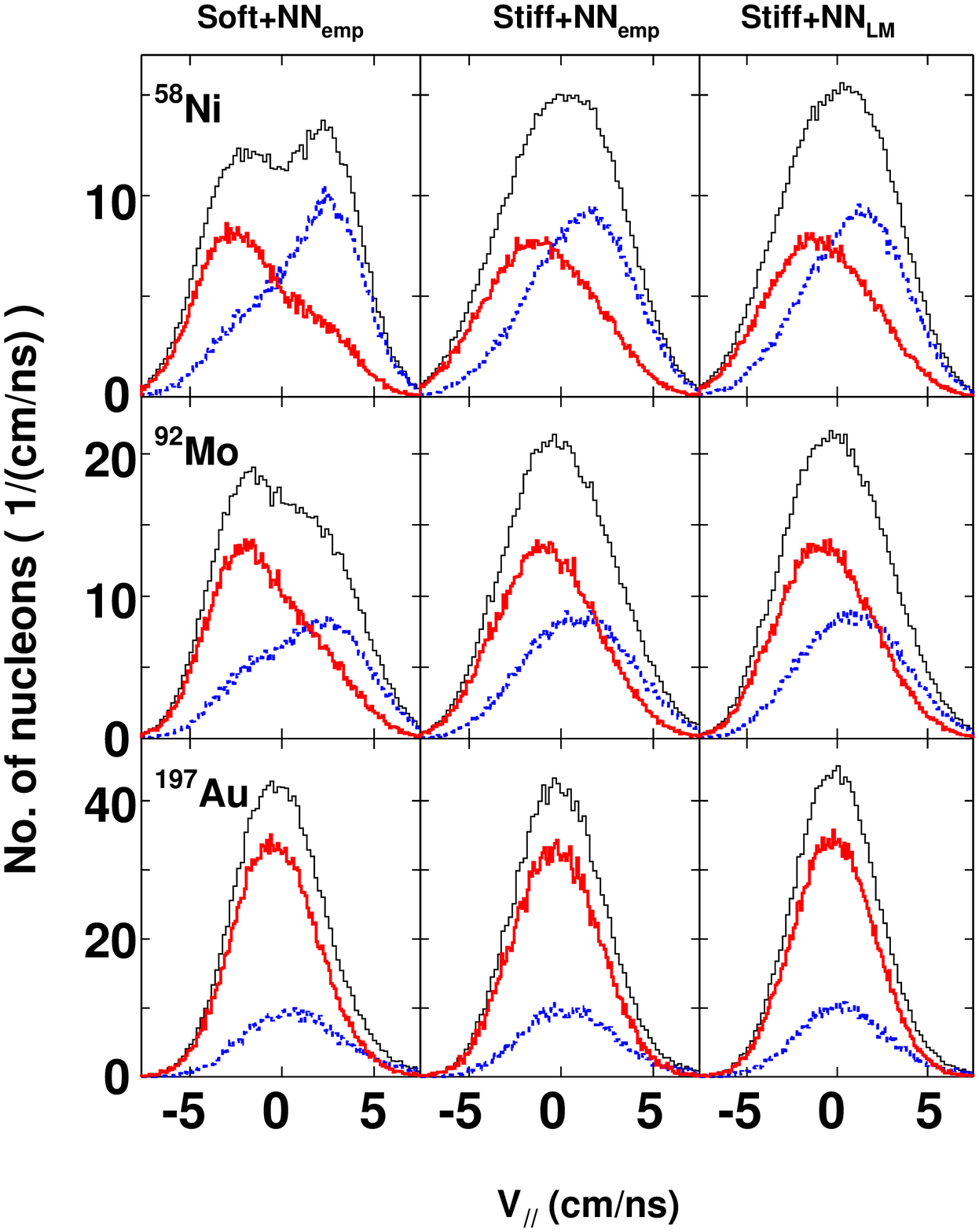}
\caption{\footnotesize Calculated parallel velocity distributions of
nucleons at 280 fm/c are plotted
in the center of mass system for $^{58}$Ni, $^{92}$Mo 
and $^{197}$Au at 47A MeV from top to bottom, respectively. 
The results for soft EOS + NN$_{emp}$, stiff + NN$_{emp}$ 
and stiff + NN$_{LM}$ are plotted from left to right, respectively.
Thick solid and dashed lines indicate the contributions 
of nucleons from the target and the projectile, respectively. 
Thin lines show the sum of these two contributions.  
}
\label{Vpara}

\includegraphics[scale=0.45]{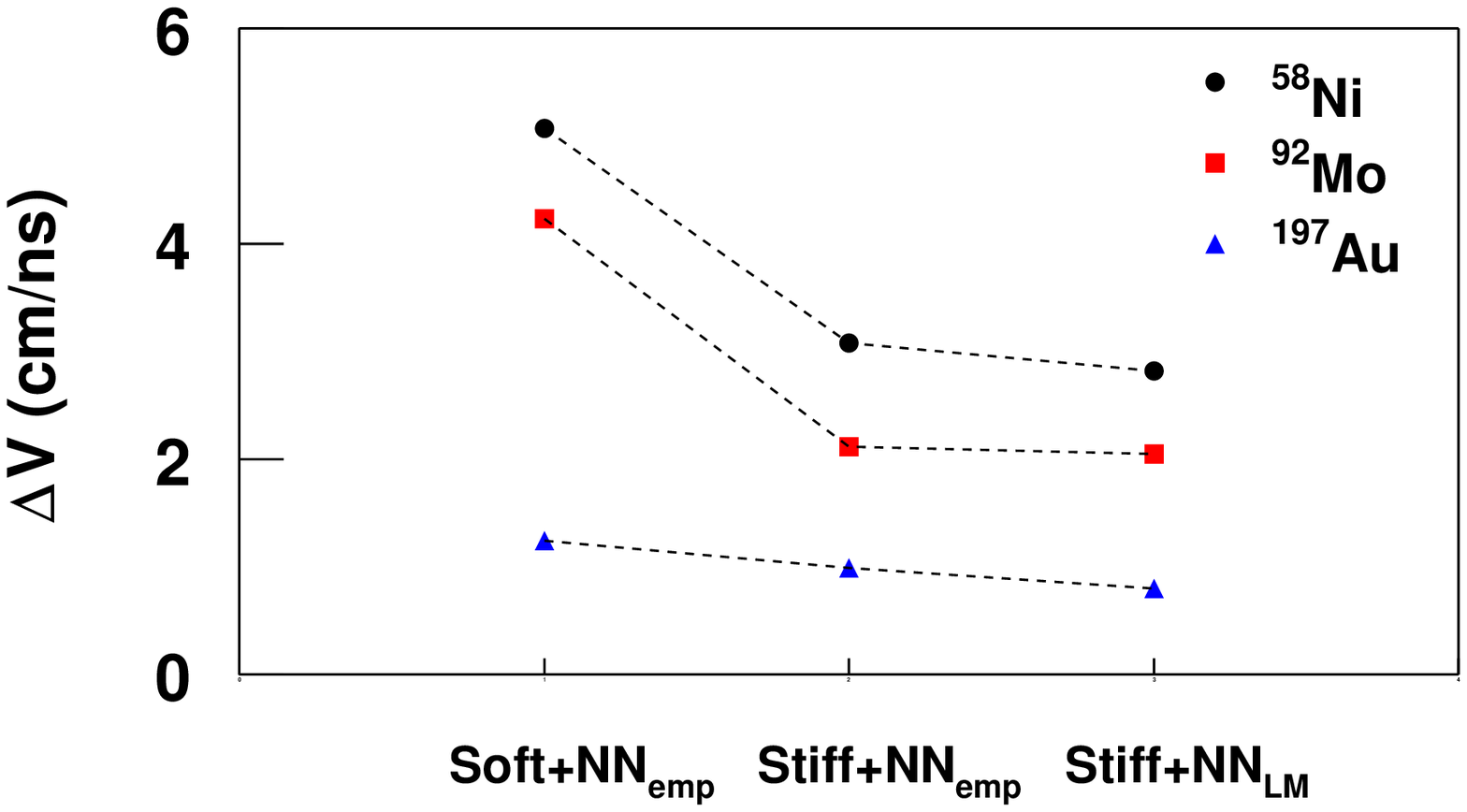}
\caption{\footnotesize Difference of the peak values of the parallel velocity 
distribution of nucleons from the projectile and from the target. Peak 
values are obtained by a Gaussian fit around each peak. Dashed lines are to 
guide the eye.
}
\label{Vp_mean}

\end{figure}

In order to show how much nuclear semi-transparency changes with 
the different parameter sets, parallel velocity distributions for all 
nucleons (free nucleons and nucleons bound in fragments) are shown in 
Fig.~\ref{Vpara}, at a time t = 280 fm/c, for central collisions in reactions 
at 47A MeV. In the case of $^{58}$Ni (top row), the majority of 
the projectile and target nucleons move in the same direction after the 
collisions. For the soft EOS (left column) the projectile nucleons  
exhibit a broad distribution, centered at about half of the incident 
velocity after penetrating through the target nucleus, indicating significant 
transparency. The transparency is reduced for the stiff EOS 
as seen in the right column of the figure. It is 
interesting that only a small difference is observed between the results 
calculated with NN$_{emp}$ (middle) and 
those with NN$_{LM}$ (right). Therefore, in the framework of the AMD-V, 
the transparency depends significantly on the stiffness 
of EOS, but depends only weakly on the in-medium NN cross sections. 
As the result the 
parallel velocity distributions of all nucleons show a two-peak structure 
for the soft EOS and these two peaks tend to merge into one peak for the 
stiff EOS. When the target becomes heavier, the two-peak structure for the 
soft EOS is also less prominent and becomes one peak 
for the $^{197}$Au target. 
In that case no notable difference is observed between the soft 
and stiff equation of state. The differences of the peak velocities between 
nucleons from the projectile and those from the target are also plotted 
in Fig.~\ref{Vp_mean}. Enhancement of the semi-transparency for the soft EOS 
is clearly seen as the sharp increases of the differences for $^{58}$Ni 
and $^{92}$Mo.    
           
\subsection*{D. Afterburner and switching time}

The fragments generated in AMD-V calculations are generally in an 
excited state at a time of 280 fm/c. AMD-V should properly treat 
the cooling of these fragments in a quantum statistical manner~\cite{ono96_1}.
However in order to cool the fragments down to the ground state, 
a great deal of CPU time is needed. Instead of continuing the AMD-V calculation
for such a long time, we stopped the calculation at t = 280 fm/c 
(which corresponds to a realistic CPU time to get a few thousand events 
in the VPP700E) and the fragment cooling was followed, using a statistical decay code 
as an afterburner. A modified version of GEMINI~\cite{Charity} was used as 
the afterburner. In this modified version, discrete levels of the excited 
states of light fragments with Z $\leq$ 15 are taken into account and the 
Hauser-Feshbach formalism is extended to the particle decay of these
fragments when the excitation energy is below 50 MeV. 
Each AMD-V event was used 100 times in the afterburner in order to 
sample all possible decay paths of the excited fragments. This also provides
enough statistics for detailed comparisons to the experimental results.
The switching time of t = 280 fm/c was chosen only for 
practical reason of the computation time in the VPP700E. In the AMD-V 
calculations, later switching times are preferable, because 
the particle evaporation occurs 
in the quantum statistical manner~\cite{ono96_1}. As discussed in 
reference~\cite{wada00} the switching time of t = 280 fm/c is late enough 
so that the final results do not depend significantly on this choice. 

\section*{IV. DATA Analysis}

In order to perform direct comparisons between the experimental results
and the calculations, the efficiency of the experimental acceptance, such as 
the neutron ball efficiency and multi hit effects, have to be evaluated.
Event classification is also crucial for
the comparisons, because many observables change drastically 
depending on the impact parameter. In this section these 
experimental conditions and event classification are described in detail.

\subsection*{A. Neutron ball efficiency}

Neutron balls have been applied to measure neutron multiplicity 
in heavy ion reactions for the last two decades~\cite{Neutron-Ball}.
In order to simulate the neutron ball efficiency the program DENIS has
been widely used~\cite{DENIS}. This code is designed for low energy
neutrons and no secondary neutron generation is taken into account.
Recently Trzcinski {\it et al.} developed a more sophisticated program MSX 
which takes into account secondary neutron generation~\cite{Neutron-Ball}.
These codes have been applied to neutron energies up to a few tens of MeV
for neutron balls containing a minimum amount of material inside the 
scattering chamber. However NIMROD has a large amount of material inside the 
chamber. This can result in scattering, absorption and generation of neutrons.
To address these possibilities we have used  
the GCALOR code coupled to the GEANT-3 simulation package  
to simulate the neutron ball in NIMROD~\cite{GCalor,Geant}.  
The GCALOR code is designed to simulate low
energy neutrons. It makes use of the latest compilation of cross section data.
In the program the geometry and material of the Neutron Ball 
and the charged particle array have been taken into account in detail. 
Only cables, phototubes and Si detectors are neglected. 
The calculations were made for neutrons emitted isotropically 
from the target with energies from 2 to 100 MeV in 2 MeV steps. For each
event, one neutron is emitted at the target and one million events were 
generated for each case. Each neutron
was followed until either it escaped from the neutron ball
or became thermalized ( E $\leq$ 0.03 eV). If it thermalized 
inside the liquid scintillator, we assumed the neutron was captured by the 
Gadolinium. When additional neutrons are generated, all neutrons were followed 
until they were thermalized or escaped. 
The detection efficiency for a neutron of a given initial energy is determined 
by the ratio of all detected neutrons to primary neutrons. The light detection 
efficiency inside the neutron ball was also simulated 
separately by generating $\gamma$ rays emitting from a point source 
scattered inside the neutron ball. For a threshold of 300 KeV 
the detection efficiency of a captured neutron was nearly 100\%. 
The program was run in two modes, i.e., with the 
charged particle array and without the array 
to specify different contributions to the efficiency.

\begin{figure}
\includegraphics[scale=0.45]{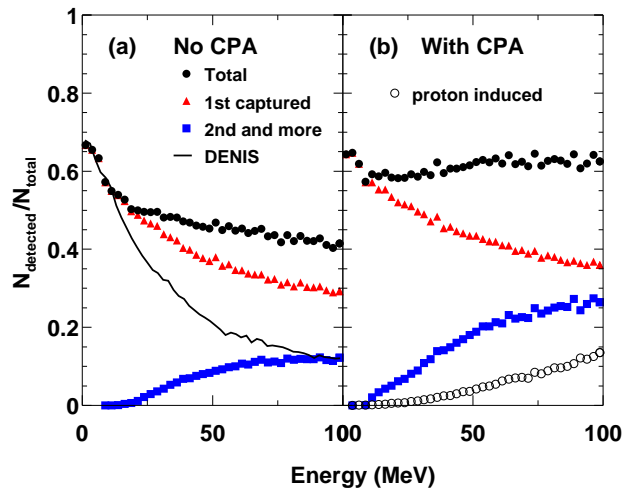}
\caption{Calculated neutron ball efficiency without the charged particle 
array (a) and with the charged particle array (b). Intrinsic neutron efficiency 
at a given energy is shown by dots. The contribution 
of the first detected neutrons is shown by triangles and that of
the second and higher order detected neutrons is shown by squares. 
The solid line in the left figure shows results from the DENIS code.
The open circles in the right figure display the detection efficiency of 
generated neutrons when a proton of an initial energy, given on x-axis, is 
emitted at the target.
}
\label{NBLEff}
\end{figure}

In Fig.~\ref{NBLEff} the calculated results for the neutron ball 
efficiencies are shown. For comparison the DENIS code predictions are also 
shown by a solid line for the case without the charged 
particle array~\cite{Schmitt}. 
Below 10 MeV both calculations agree reasonably well. 
For the higher energy neutrons, the efficiency calculated by DENIS drops below 
the results for the first detected neutrons in the GEANT simulation 
(triangles), because there is no neutron generation in DENIS. 
In other words, even 
for the first captured neutrons in the GEANT simulation, generated neutrons 
start to contribute to the detection efficiency above 10 MeV, as one can see 
in the figure. This indicates that a significant number of neutrons are 
generated in the liquid scintillator for these high energy neutrons. 
The contribution from the generation of more than two neutrons becomes 
significant above 20 MeV. Since the contribution of the third detected neutron 
is only 10\% of that of the second detected neutron (not shown), the main
contribution comes from one or two neutron generation in this energy range.
With the charged particle array, the contribution of generated neutrons
almost doubles. The total efficiency of the neutron ball slightly increases 
above 20 MeV as the neutron energy increases and becomes even approximately 
constant with respect to energy.
 
Protons also generate secondary neutrons at high energy.
However the generation of neutrons by protons is much less efficient
than that by neutrons. This is because protons lose
their energy very quickly by ionization processes and the
cross section for neutron generation decreases rapidly as the proton
energy decreases. Only a 5\% contribution to the neutron 
efficiency is observed at a proton energy of 50 MeV and 14\% at 100 MeV.

The efficiencies shown above are those averaged over all angles. However in 
the intermediate heavy ion reactions the angular distribution of the neutrons 
has a significant forward peak, especially for higher energy neutrons.
Since the charged particle array has more material at forward angles and two 
air gaps between the central part and the two hemispheres are not symmetric 
relative to the target position, the angular dependence of the 
neutron efficiency has to be taken into account for the actual application. 
Therefore the neutron efficiency was calculated as a function of both neutron 
energy and polar angle. The azimuthal angular dependence was neglected.
Since the proton multiplicity is much smaller than that of neutrons, 
the proton contribution to the neutron ball efficiency was not taken into 
account in the present analysis. 
           
\subsection*{B. Particle identification and multi-hit events}

\begin{figure}
\includegraphics[scale=0.575]{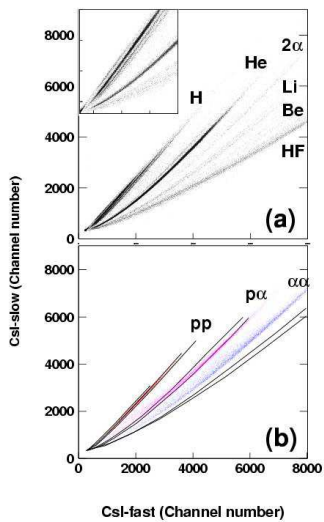}
\caption{\footnotesize (a) Typical experimental two dimensional plot of 
fast vs slow components of charge integrated light output from a CsI detector. 
Insert shows the expanded spectrum for Hydrogen isotopes. Corresponding elements
are indicated in the figure. HF stands for heavy fragments with Z $\geq$ 5. 
(b) Distributions of artificially generated double-hit events for pp, 
p$\alpha$ and $\alpha\alpha$. Each particle combination is indicated in 
the figure. Solid lines correspond to the locus of the ridges of p, d, t, i$^3$He,
$\alpha$, $^6$Li and $^7$Li from (a). 
}
\label{CsIFS}
\end{figure}

In NIMROD, light charged particles with atomic number Z $\leq$ 3 are 
identified by a pulse shape discrimination method in the CsI detector.
A typical two dimensional plot of fast versus slow charge integrated signals is shown 
in Fig.~\ref{CsIFS}(a). Each type of particle lies along a specific curve and 
one can clearly identify different particles as indicated in the figure. 
In the insert the spectrum is expanded for Hydrogen isotopes. The events
along the far left line correspond to $\gamma$ rays (and accidental cosmic muons). 
Hydrogen isotopes are clearly identified down to a few MeV/nucleon.

Special care has to be taken in the pulse discrimination method when more than one 
charged particle hits a CsI detector.
In order to simulate double hit events in an actual detector, 
two experimental events were artificially mixed and 
the two outputs for a given combination of light charged particles were 
added both for the fast and slow components. The loci of double-hit events 
for pp, p$\alpha$ and $\alpha\alpha$ are plotted in Fig.~\ref{CsIFS}(b). 
The pp events 
are scattered along the deuteron and triton lines. Therefore 
these double hits are identified as a single deuteron or a triton 
in the present experimental data analysis. 
The main part of p$\alpha$ events lie near the alpha line. 
Most of these events, therefore, are identified as a single $\alpha$ particle. 
When two $\alpha$ particles hit a detector, these events lie between the 
$\alpha$ and $^6$Li lines and are easily identified as two $\alpha$ 
particles, though only their summed energy is given experimentally. 
This is clearly seen in the experimental spectrum in Fig.~\ref{CsIFS}(a) 
and indicated by 2$\alpha$.
   
\begin{figure}
\includegraphics[scale=0.475]{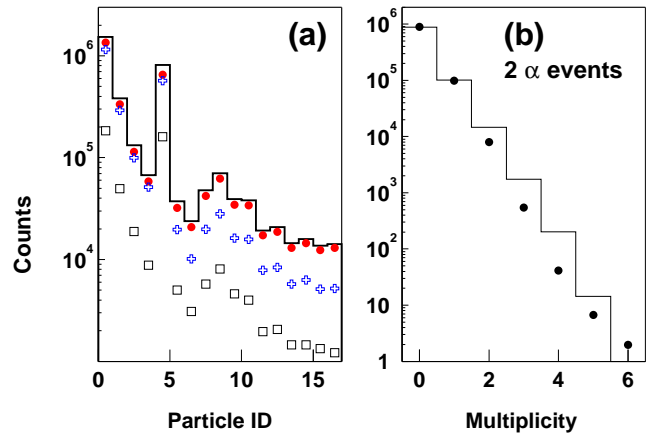}
\caption{\footnotesize (Left) Number of particles generated 
by the calculation for $^{92}$Mo at 47A MeV. Total number of charged particles
is given by a histogram as a function of particle ID. The particle ID is given
by ID = 0, 1, 2, 3, 4 for p, d, t, $^3$He, $\alpha$ and ID=Z+2 for Z$\ge$3, 
respectively. The number of single hit particles is given by dots and that of 
double hit particles is shown by squares. The number of the detected single 
hit particle is shown by crosses.
(Right) Number of double $\alpha$ hit events is shown 
as a function of multiplicity for the experiment (dots)
and the simulation (histogram). In both cases one million events were analyzed.
The calculation is done for the stiff EOS + NN$_{LM}$ case. 
The calculated events have been treated by the experimental filter.   
}
\label{DHits}
\end{figure}

The rate of multiple hits has been simulated using AMD-V. 
Results are shown in Fig.~\ref{DHits}(a) 
for the case of $^{92}$Mo at 47A MeV. 
The total number of each generated particle, the number of 
single hits and that of double hits are given by different 
histograms. For the double hits both of the particles are counted separately. 
According to this simulation,
about 12\% of protons and 20\% of $\alpha$ particles hit a CsI detector which 
is hit by other charged particles. The rate of two $\alpha$ particles 
is larger, mainly because of the two $\alpha$ decay of $^8$Be. 
The rate increases as the detector angle increases, 
because the solid angle of each segment increases rapidly as the angle 
increases as seen in Table I. The actual detected number of each particle 
as a single hit, after filtering through the 
experimental filter, is also given in the figure.
About 85\% of single-hit light charged particles were detected and identified.

In order to verify the double-hit rate in the above simulation, a comparison 
was made between two-$\alpha$ events in the experiment and in
the simulation. The results are shown in Fig.~\ref{DHits}(b). 
A reasonable agreement between the experimental and calculated results is 
obtained. About 10\% of the events have one double 
$\alpha$ hit, both in the experiment and in the simulation. 

\begin{figure}
\includegraphics[scale=0.5]{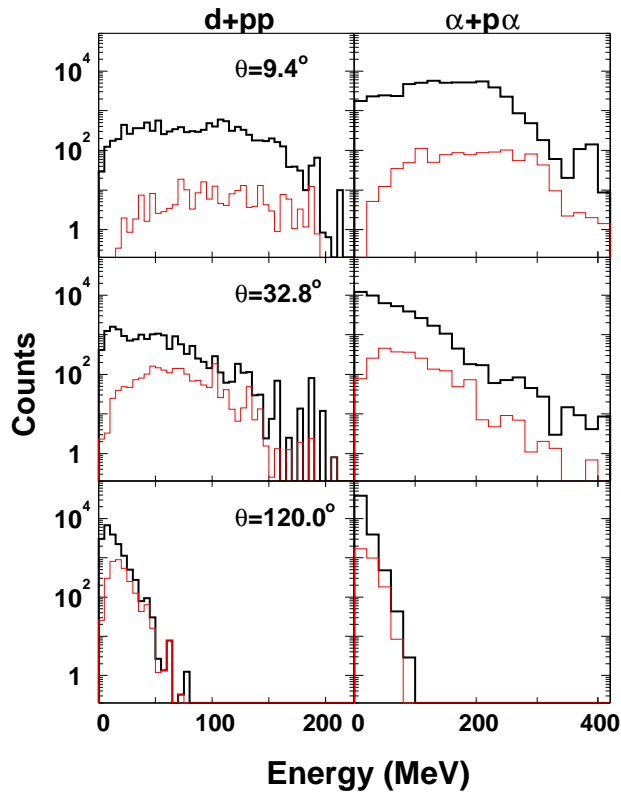}
\caption{\footnotesize Calculated effects of double hits on the energy 
spectra of deuterons (left) and $\alpha$ (right) particles at three different 
detection angles, as indicated in each figure. Reaction for $^{92}$Mo 
at 47A MeV is used. Thick line histograms show the single hit spectra 
and thin line histograms indicate the spectra of pp in the deuteron spectra 
and p$\alpha$ in the $\alpha$ spectra. The energy of the double hit events 
is given by the sum of the energy of the two particles.    
}
\label{DHitsEspectra}
\end{figure}

The effect of multi-hit events on the energy spectra of light charged 
particles has also been studied. In Fig.~\ref{DHitsEspectra}, 
deuteron and alpha spectra at three different detection angles 
are shown. In the calculation all pp-double hits are assigned as a 
deuteron hit with the summed energy and all p$\alpha$ hits are assigned 
as a single $\alpha$ hit. The contribution
of the double hits at $\theta=9.4^o$ is a few percent in both cases, and gradually 
increases as angle increases.
At backward angles, especially for deuterons, the contribution 
becomes of the same order as that of the single particle hits. 
The contributions for the $\alpha$ spectra are slightly less.

For heavier fragments with Z $\ge$ 4, particle identification was made 
by the $\Delta$E-E method and no multi-hit problem in particle identification 
occurs. (The energy loss of a proton or an $\alpha$ particle is much smaller 
than for a heavier fragment.) However heavier fragments from the target-like 
source have rather small kinetic energies and may be stopped in the $\Delta$E  
detector. In this experiment heavy particles 
emitted at angles larger than $\theta=45^o$ were not identified by charge. 
As seen in Fig.~\ref{DHits}(a) heavy fragments
with Z $\ge$ 3 were detected with an efficiency of 40-50\%.    

In the experimental filter applied to the calculated events, 
all pp hits are identified as a deuteron, all p$\alpha$ are 
identified as an $\alpha$ and $\alpha\alpha$ events are treated as 
two-$\alpha$ hits. A heavy particle ( Z $\ge$ 3) with light charged 
particles in one detector is identified as a single heavy particle hit and 
the light charged particles are not counted in the charged particle
multiplicity. Events with more than two hits
in a single detector are treated as not identified for the largest fragment with 
Z $\leq$ 2 and as a single hit for the largest with 
Z $\geq$ 3, though the rate of these events is very small in the reactions 
studied in this paper.

\subsection*{C. Event classification}

\begin{figure}
\includegraphics[scale=0.5]{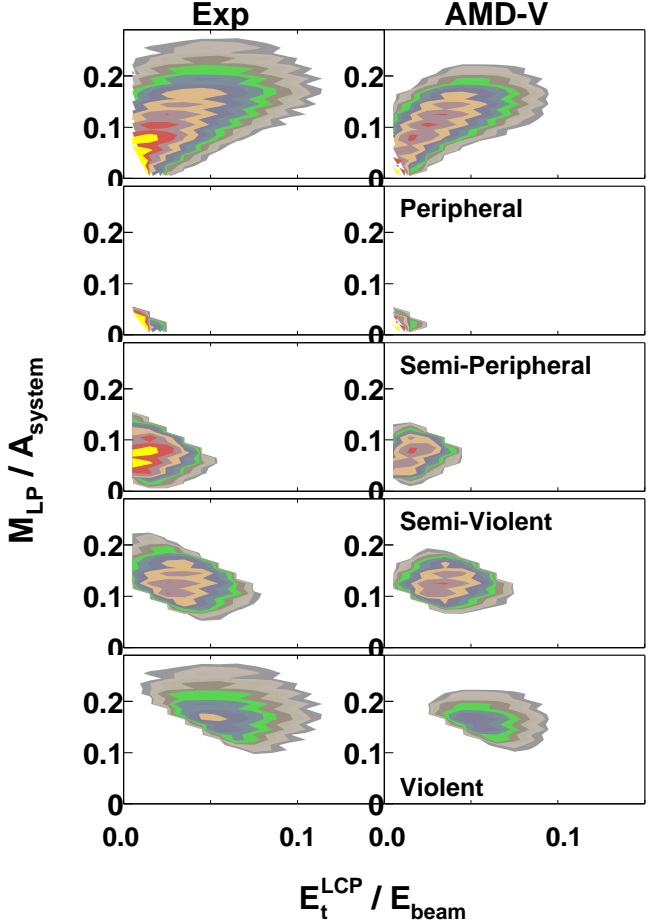}
\caption{\footnotesize 2D plots of the experimental (right) 
and the calculated (left) M$_{LP}$/A$_{system}$ vs E$_{t}^{LCP}$/E$_{beam}$
for $^{92}$Mo at 47A MeV. M$_{LP}$ is the observed 
multiplicity of light particles, including neutrons, and A$_{system}$ is the 
total nucleon number of the reaction system. 
E$_{t}^{LCP}$ is the sum of transverse 
energy of the light charged particles with Z $\le$ 2 and E$_{beam}$ is 
the projectile incident energy. 
Generated events by AMD-V has been 
filtered through the experimental acceptance. Contours are in logarithmic scale
and the scale is set arbitrarily.
}
\label{EventClass}
\end{figure}

Detected events have been classified in four groups (Violent, Semi-Violent, 
Semi-Peripheral and Peripheral), depending on the violence of 
collisions. This assignment is based upon the neutron and light charged 
particle multiplicities and the total transverse energy of the light 
charged particles. 
In Fig.~\ref{EventClass} typical two dimensional plots of the normalized total
multiplicity M$_{LP}/$A$_{system}$ versus the normalized total transverse 
energy E$_{t}^{LPC}$/E$_{beam}$ are shown 
for $^{92}$Mo at 47A MeV, both for the experimental and the
calculated results. In both cases the experimental inclusive distribution 
(top) shows a slightly broader distribution than the distribution calculated 
with AMD-V. For the multiplicity axis, 
this is mainly induced by the neutron multiplicity
distribution as seen in the next section. 
The top 5\% (3\%) of the minimum bias events in the experiments were 
assigned to ``Violent" collisions for $^{58}$Ni and $^{92}$Mo
($^{197}$Au). The next 20\% (10\%) events 
were assigned as ``Semi-Violent" collisions and the following 20\% (10\%) of 
events were assigned as ``Semi-Peripheral". The rest were assigned as 
"Peripheral" collisions. The same boundaries were applied to the 
calculated events. The resulting distributions corresponding to these cuts 
are compared in Fig.~\ref{EventClass}.

\begin{figure}
\includegraphics[scale=0.5]{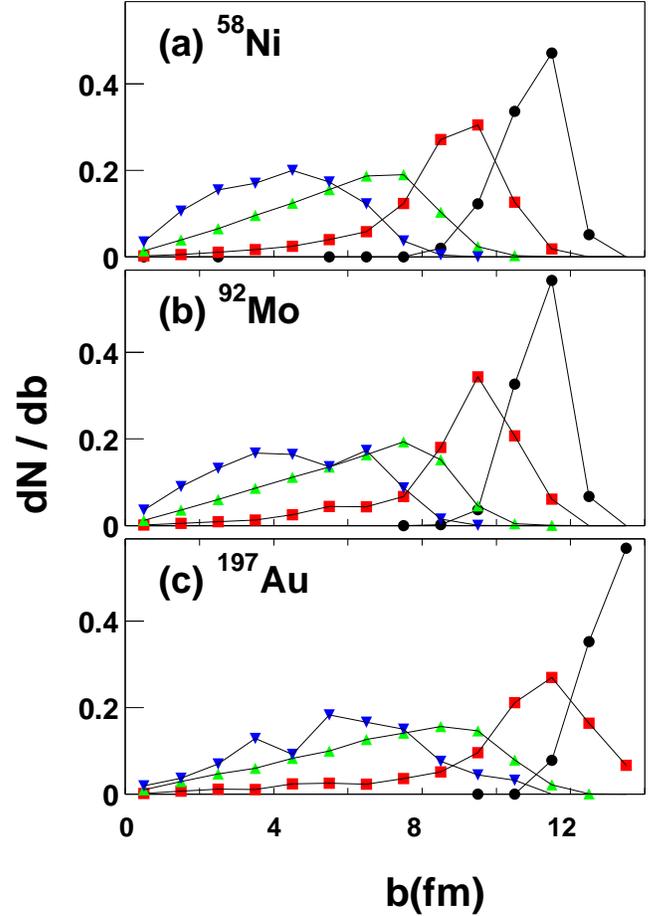}
\caption{\footnotesize Impact parameter distributions for different 
classes of events for (a) $^{58}$Ni (b) $^{92}$Mo and (c) $^{197}$Au at 
47A MeV. Triangles (down), triangles (up), squares and dots indicate the 
results for ``Violent",``Semi-Violent",``Semi-Peripheral" and "Peripheral"
classes, respectively. The area of each distribution is normalized to 1.    
}
\label{BimpDist}
\end{figure}

The distributions of impact parameter for the four different class of events 
were studied using AMD-V and are shown in Fig.~\ref{BimpDist}. 
For the ``Violent" and ``Semi-Violent" classes the calculated distributions are
very broad and about 50\% of events of each class overlap with the 
distribution of another class. 
In the ``Violent" class for $^{58}$Ni, 
the distribution reaches up to 8 fm and about 70\% of events are distributed 
in the impact parameter range of b $\leq$ 5 fm. The distributions become  
broader for the heavier targets. 
In the ``Violent" class for $^{197}$Au, 
the distribution reaches up to 10 fm and only about 35\% of events originate 
from collisions with b $\leq$ 5 fm. On the other hand
the events in the ``Semi-Peripheral" and Peripheral classes show 
a rather localized distribution with a full width at a half maximum (FWHM) 
of 2-3 fm, indicating 
that most of these events actually originated from collisions with a large 
impact parameter. Therefore  
throughout this paper, the word ``Violent" is used 
instead of "Central" for the class of events with the highest multiplicity 
and the largest transverse energy. The word "Peripheral" 
is used as is customary.

\section*{V. COMPARISON BETWEEN EXPERIMENT AND MODEL}

Detailed comparisons of the experimental and calculated results
are presented in this section for the direct observables, 
such as multiplicity distributions, 
charge distributions, energy and velocity spectra. Comparisons
for the elliptic flow are also shown. In order to make these comparisons,
all calculated results have been filtered through 
the experimental conditions, unless otherwise specified.

\subsection*{A. Multiplicity distributions}

\begin{figure}
\includegraphics[scale=0.475]{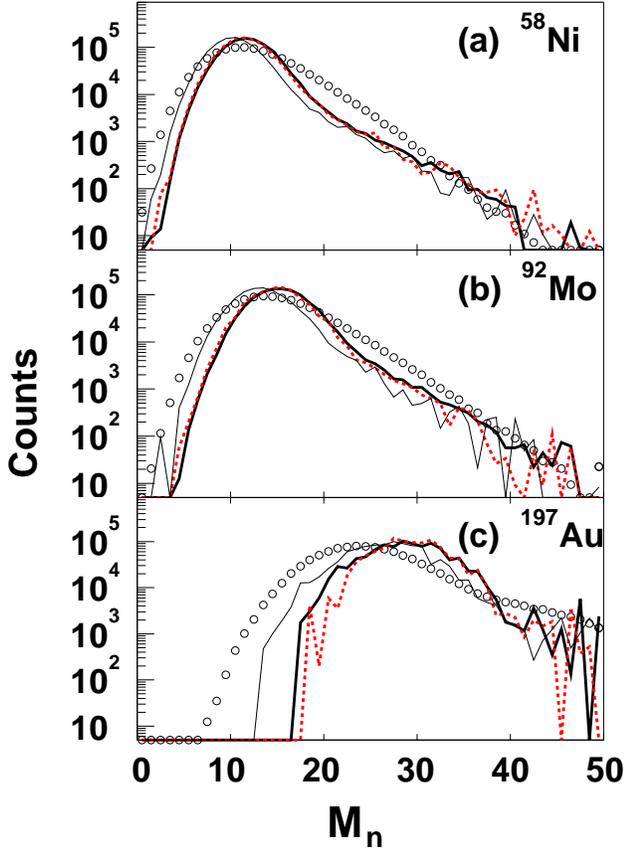}
\caption{\footnotesize Neutron multiplicity distributions for events 
in the ``Violent" class of events of the reactions at 47A MeV 
for (a) $^{58}$Ni, (b) $^{92}$Mo and (c) $^{197}$Au. 
Experimental results are shown by circles and calculated 
results are shown by different lines.
Thin solid, dashed and thick solid lines indicate 
the results of Soft+NN$_{emp}$, 
Stiff+NN$_{emp}$ and Stiff+NN$_{LM}$, respectively. 
No efficiency and background corrections were applied 
for the experimental distributions, whereas the calculated results 
have been treated with the experimental filter. All distributions are 
normalized to one million events in total.          
}
\label{NeutronMul}
\end{figure} 

\begin{figure}
\includegraphics[scale=0.475]{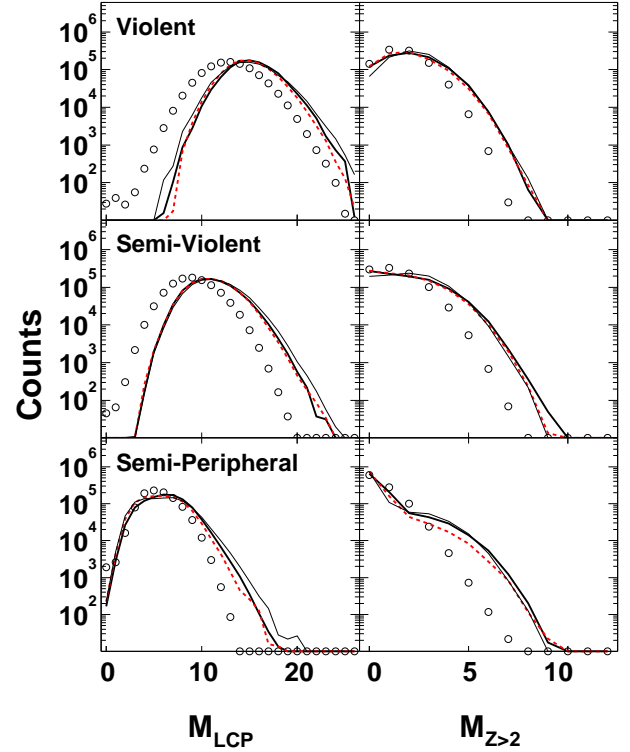}
\caption{\footnotesize Multiplicity distributions of light charged particles 
(Z $\le$2) (left) and heavier charged particles (Z $\ge$3) (right) for events
in the ``Violent", ``Semi-Violent" and ``Semi-Peripheral" classes, 
from top to bottom, respectively, for $^{92}$Mo at 47A MeV. Selected 
classes are indicated in each figure. 
Experimental results are shown by circles and calculated 
results are shown by different lines.
Thin solid, dashed and thick solid lines indicate 
the results of Soft+NN$_{emp}$, 
Stiff+NN$_{emp}$ and Stiff+NN$_{LM}$, respectively. 
All calculated results have been treated with the experimental filter.
All distributions are normalized to one million events in total. 
}
\label{CPMul}
\end{figure}

\begin{figure}
\includegraphics[scale=0.475]{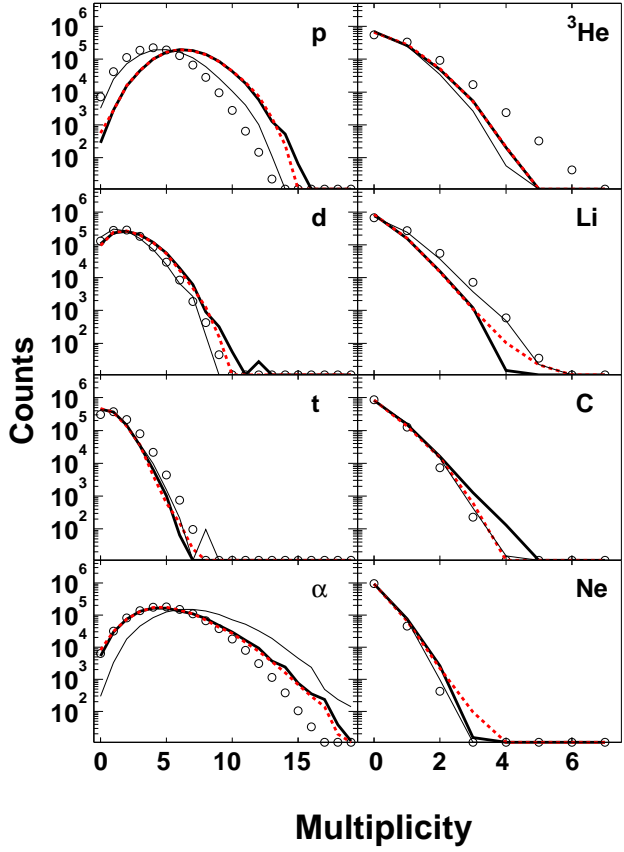}
\caption{\footnotesize Multiplicity distributions of selected particles 
for the ``Violent" collisions for $^{92}$Mo at 47A MeV. 
Particles are indicated in each figure. See also the caption of 
Fig.~\ref{CPMul}. 
}
\label{CPMul_individual}
\end{figure}

\begin{figure}
\includegraphics[scale=0.475]{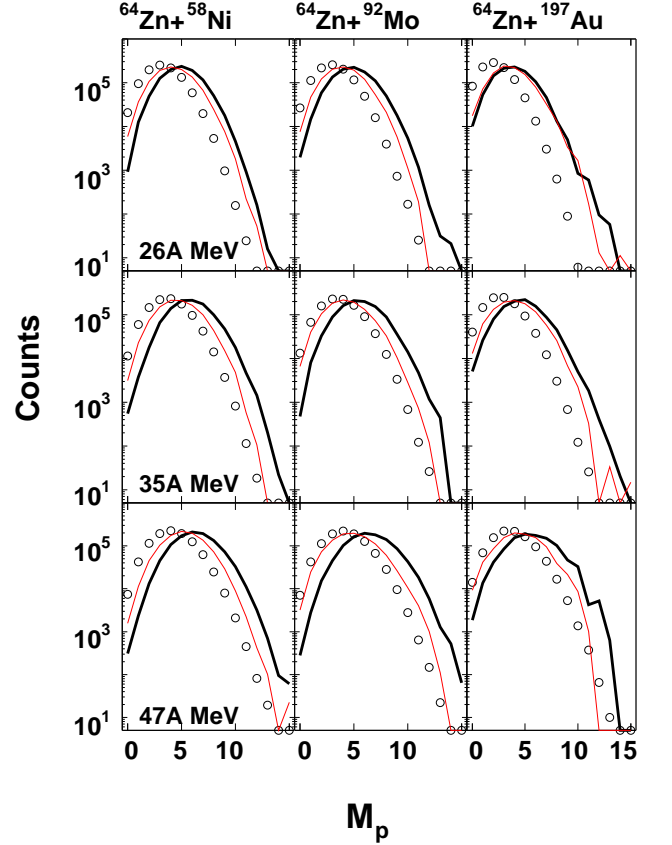}
\caption{\footnotesize Proton multiplicity distributions for the ``Violent" 
collisions of all reactions. Figures 
in the same column show the results from the same reaction system and 
figures in the same row show those at the same incident energy. 
The reaction system is indicated at the top of the figure and the incident 
energy is indicated in each row. Experimentally observed 
multiplicities are shown by the circles and calculations 
with soft EOS + NN$_{emp}$ and stiff EOS + NN$_{LM}$ are shown by 
thin and thick solid lines, respectively. 
}
\label{pMul}
\end{figure}

\begin{figure}
\includegraphics[scale=0.475]{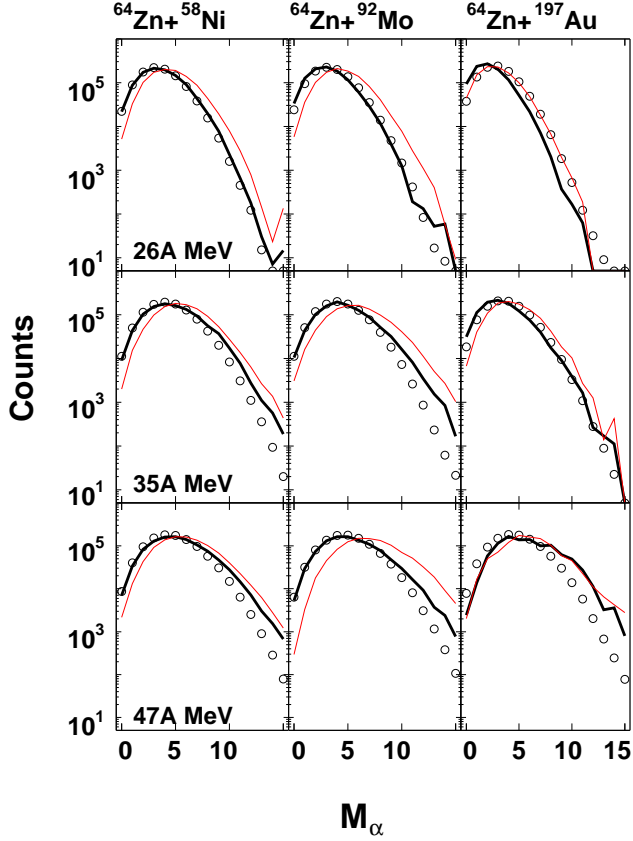}
\caption{\footnotesize Similar plot to Fig.~\ref{pMul} but for $\alpha$ 
particles.
}  
\label{aMul}
\end{figure}

Neutron and charged particle multiplicity distributions are presented in 
Figs.~\ref{NeutronMul} to \ref{aMul}.
In Fig.~\ref{NeutronMul} experimental neutron multiplicity distributions 
are compared with the calculated results for the ``Violent" collisions of 
the reactions at 47A MeV. 
In each figure the experimental and the calculated 
results for the three different sets of parameters are shown. 
Overall the experimental results show broader distributions than those 
of the calculations for all reactions. For $^{197}$Au
a shoulder is observed above the neutron multiplicity M$_{n} > 40 $. It 
originates from pile-up in which two reactions occur 
during the 10$\mu$sec beam period before the shut-off of the beam.
The pile-up is related to the reaction rate and is different for different
reaction systems and incident energies (it can also fluctuate with time). 
Therefore in this work the pile-up effect is not taken into account in the 
experimental filter.
The mean values of the calculated distributions  
for $^{58}$Ni and $^{92}$Mo agree with those of the 
experiments within 1-2 neutrons, whereas that for $^{197}$Au 
produces $\sim$3 additional neutrons for the soft EOS and 
$\sim$5 additional neutrons for the stiff EOS. While the mean value for
$^{58}$Ni and $^{92}$Mo is slightly better reproduced by the calculations 
with the stiff EOS, that for $^{197}$Au is better fit 
with the soft EOS. For all reactions the 
calculated multiplicities for the stiff EOS, but different NN cross sections, 
show almost identical distributions.

\begin{figure}
\includegraphics[scale=0.45]{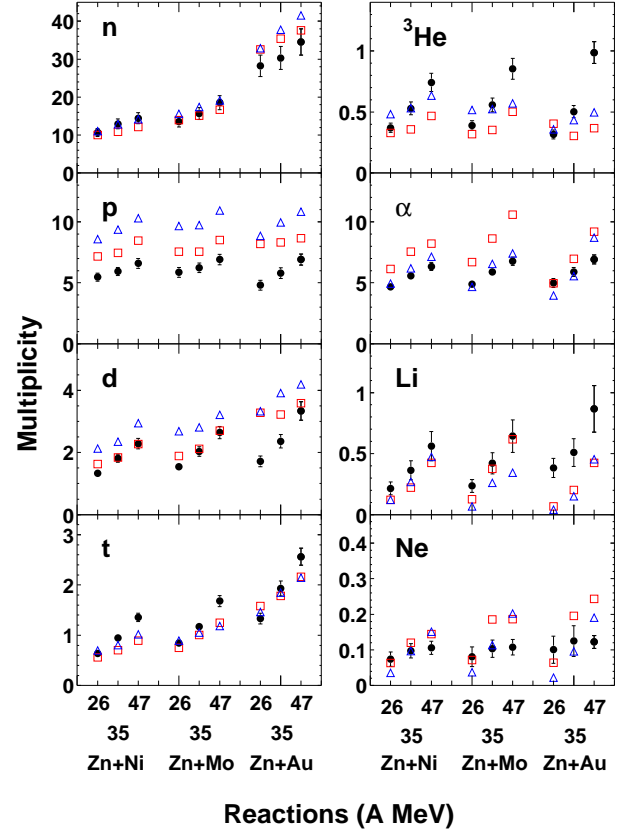}
\caption{\footnotesize Summary of mean multiplicities 
of selected particles for the ``Violent" class of events for all 
reactions. Experimental values are efficiency corrected but calculated 
ones are non-filtered values.
Reaction systems and incident energies are indicated on the x axis 
and particle is indicated in each panel. Experimental results are shown 
by solid dots and calculated results with soft EOS + NN$_{emp}$ and 
stiff EOS + NN$_{LM}$ are shown by open squares and triangles, respectively. 
}
\label{AvMul}
\end{figure}

Similar trends are also observed for other reactions. In the top-left of 
Fig.~\ref{AvMul}, the mean multiplicities of neutrons, corrected for the  
efficiency, are summarized for ``Violent" collisions for all 
reactions studied here. The calculated results for Soft+NN$_{emp}$ and 
Stiff+NN$_{LM}$ are compared to the experiment. The experimental 
mean values for $^{58}$Ni and $^{92}$Mo 
are reasonably well reproduced by the calculations but 
for $^{197}$Au the experimental mean values are exceeded 
by about 10-20\% at all incident energies. 

Typical charged particle multiplicity distributions for light charged 
particles 
(Z $\leq$ 2) and heavier fragments are shown separately in Fig.~\ref{CPMul}.
For collisions in ``Violent", ``Semi-Violent" and ``Semi-peripheral" classes 
for $^{92}$Mo at 47A MeV, both experimental and 
calculated results are shown. Calculations with three different parameter 
sets lead to very similar distributions for all cases. 
For the ``Violent" and ``Semi-Violent" collisions the light charged particle 
multiplicity is overpredicted by 20\% in the calculations. The 
heavier charged particle multiplicity is also overpredicted by about one unit
in the mean value. The widths of the distributions are well reproduced for 
all cases. For the ``Semi-Peripheral" collisions, the light charged particle 
multiplicities are well reproduced both in the mean value and width, 
but for the heavier charged particle multiplicities 
all calculations show a shoulder at the multiplicities around 4-5. 

Typical individual charged particle multiplicities for the ``Violent" 
collisions are shown in Fig.~\ref{CPMul_individual} for the same reaction. 
There is an interesting connection between the multiplicity of
protons and $\alpha$ particles. For the soft EOS, the proton 
multiplicity distribution is reproduced quite well, both in the mean value 
and width. On the other hand 
the mean value of the $\alpha$ distribution is overpredicted by more than 2 
units. On the contrary, for the stiff EOS the $\alpha$ multiplicity 
distribution is well reproduced, but the proton multiplicity 
distribution is overpredicted by more than two units. 
Since the differences are small for the other light charged particles 
and their experimental distributions
are reasonably reproduced, the overprediction of the calculated multiplicity 
distributions for light charged particles seen in Fig.~\ref{CPMul} is 
mainly caused by alpha particles for the soft EOS and by protons for the 
stiff EOS. For Li fragments the calculated distributions show slightly smaller
values than the experimental values for all three calculations. Those for 
Be fragments, not shown, also show a similar trend. For heavier IMF with
5 $\le$ Z $\le$ 15, all calculations overpredict the experimental 
multiplicities by 1-2 units, as seen in Carbon and Neon cases in 
Fig.~\ref{CPMul_individual}. Here also one can see that the calculated 
results for the stiff EOS with different NN cross sections show almost 
identical distributions for all cases. 

Proton and $\alpha$ multiplicities for all reactions are shown in 
Figs.~\ref{pMul} and \ref{aMul} for ``Violent" collisions, respectively. 
For most reactions,
calculations with the soft EOS are favored for the proton multiplicity 
distributions and those with the stiff EOS are favored for the $\alpha$ 
multiplicity distributions. For $^{197}$Au at 26A and 35A MeV, 
however, the calculated proton multiplicities are overpredicted for both
EOS's while the experimental $\alpha$ multiplicity distributions are rather 
well reproduced by the calculations with the soft EOS.  

For the ``Violent" class the results are also shown in Fig.~\ref{AvMul}.  
In the figure the efficiency corrected experimental multiplicities and
non-filtered calculated multiplicities are shown. The experimental detection 
efficiency for each particle is evaluated from the calculated events. The
differences between calculations for a given particle agree within a few \%. 
The error bars in the figure include the systematic errors.
The increase of the experimental multiplicity with energy and target mass is 
reasonably reproduced for all particles by both calculations, 
whereas the calculated absolute mean values deviate from the experimental 
multiplicities by about 20-30\% for some cases. 
For Li, the experimental mean values are overpredicted for most 
cases and the deviation becomes significant for $^{197}$Au. 
For the heavier fragments the calculated mean values become
larger than the experimental ones in general. For Ne, 
the experimental mean values are well reproduced at 26A MeV, but 
overpredicted at 47A MeV by both calculations. 

\subsection*{B. Charge distributions}

\begin{figure}
\includegraphics[scale=0.48]{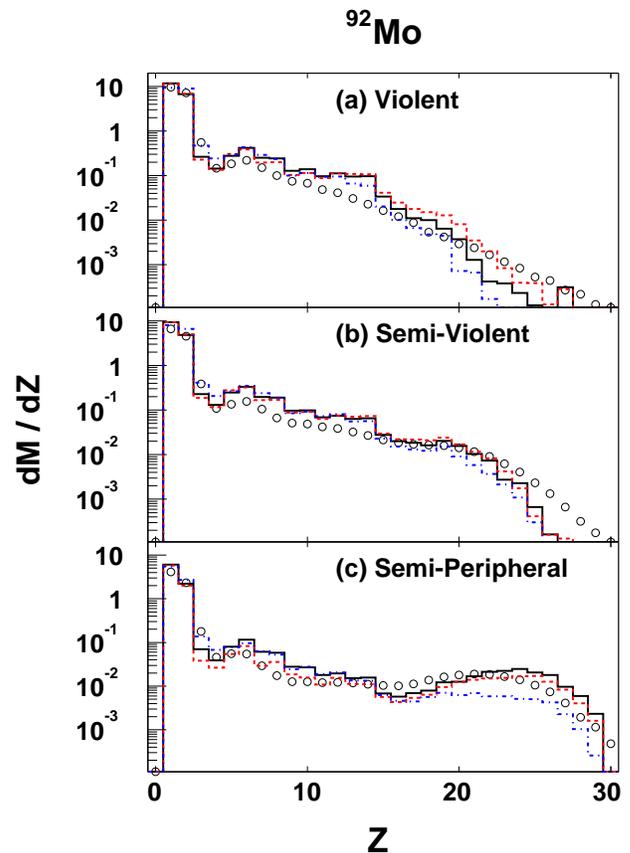} 
\caption{\footnotesize Angle integrated charge distributions of events 
in (a) ``Violent", (b) ``Semi-Violent" and (c) ``Semi-Peripheral" classes 
are shown for $^{92}$Mo at 47A MeV. Experimental results are shown by circles. 
Dot-dashed, dashed and solid lines correspond to soft EOS + NN$_{emp}$, 
stiff EOS + NN$_{LM}$ and stiff EOS + NN$_{emp}$, respectively.      
}
\label{Z_znmo47}
\end{figure}

\begin{figure}
\includegraphics[scale=0.45]{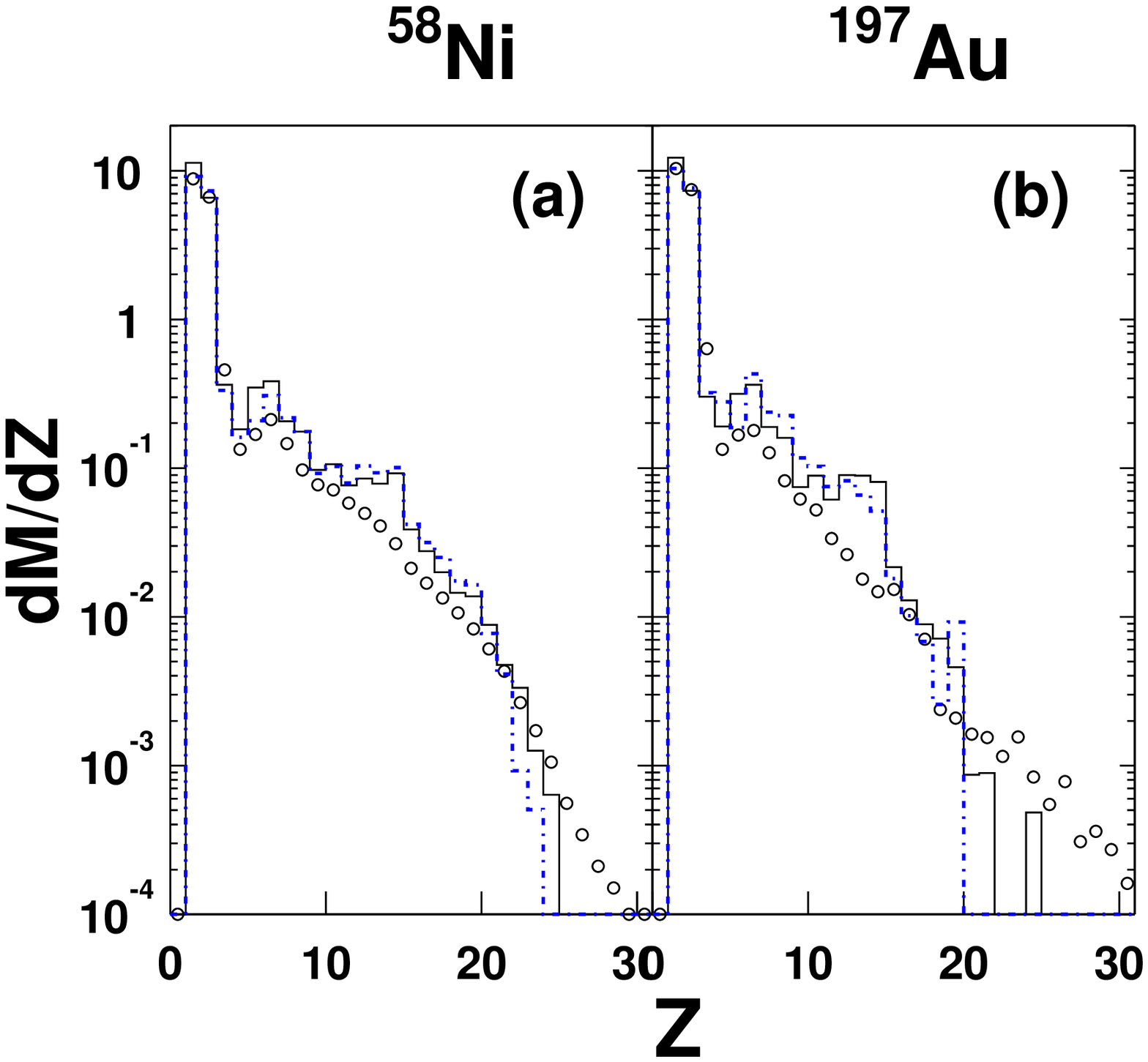}
\caption{\footnotesize Angle integrated charge distributions of events
in violent collisions for (a) $^{58}$Ni and (b) $^{197}$Au at 47A MeV.
Dot-dashed and solid lines correspond to soft EOS + NN$_{emp}$ and 
stiff EOS + NN$_{LM}$, respectively. 
}
\label{Z_znni47_znau47}
%\end{figure}

%\begin{figure}
\includegraphics[scale=0.45]{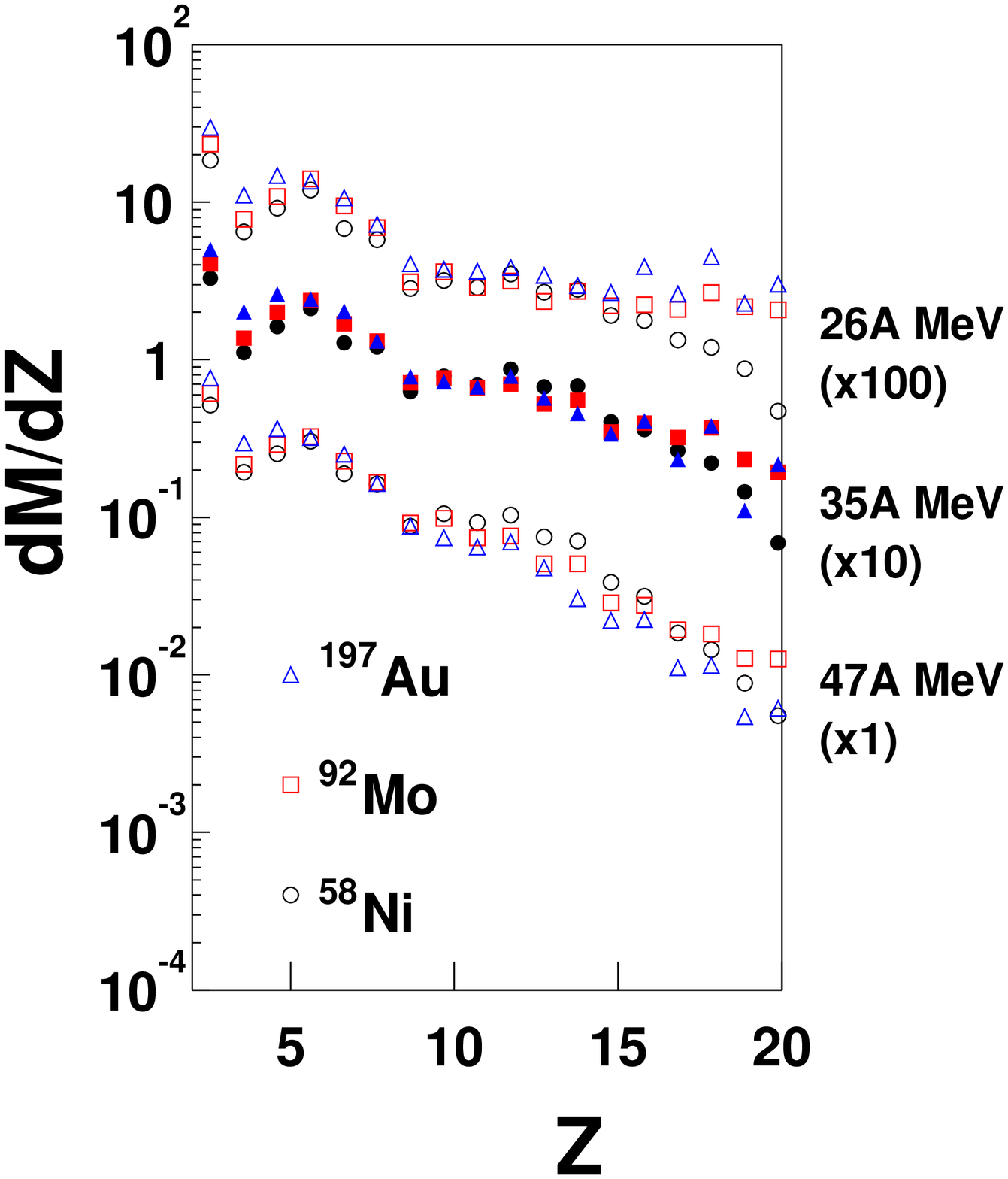}
\caption{\footnotesize Efficiency corrected experimental charge distributions 
for fragments with 3 $\leq$ Z $\leq$20. Each group shows results with three 
different targets at a given incident energy. The incident energy is shown 
on the right. The differential multiplicity is given in an absolute scale at
47A MeV and premultiplied by factors of 100 and 10, respectively, for results at 26A and 35A MeV.
}
\label{Z_Exp}
\end{figure}

Charge distributions for different classes of centrality for $^{92}$Mo 
at 47A MeV are shown in Fig.~\ref{Z_znmo47}. The experimental charge 
distributions evolve to the larger Z side when collisions become less 
violent. This trend is well reproduced by all calculated results, as one can 
see in the figure.  No significant difference is observed between the  
calculated results with different parameter sets, except for fragments 
with 20 $\le$ Z $\le$ 30 
in the ``Semi-Peripheral" class. For most of the cases the calculated 
fragment multiplicities with $5 \le Z \le 15$ are overpredicted by a factor 
of 1.5-2. A similar discrepancy is observed at lower incident energies. 
In Fig.~\ref{Z_znni47_znau47}, the charge distributions for $^{58}$Ni and 
$^{197}$Au are compared with the calculations. For $^{58}$Ni the discrepancy 
is similar to $^{92}$Mo and, for $^{197}$Au, the discrepancy becomes larger. 
For $^{197}$Au all calculated results are overpredicted by a factor of 2-3. 
As one can recognize in the figure, the charge distributions for the two
targets are quite similar both in the experiments and calculations. This is 
also true, in some extent, for all different reactions studied in the present 
work. In Fig.~\ref{Z_Exp}, experimental efficiency corrected charge 
distributions are compared with each other for light fragments with 
3 $\leq$ Z $\leq$ 20. The detection efficiency correction is made by comparing 
the calculated results with and without the experimental filter. 
Each group shows results for the three different targets at a 
given incident energy. For $^{197}$Au fragment multiplicity tends to be 
slightly smaller at 35 and 47A MeV, but the shapes of the global distributions 
are rather similar to each other, even for different incident energies. 
A similar trend is also seen in the calculated events.  
The resemblance of the relative charge distributions is quite interesting 
because the reaction mechanisms are quite different as discussed later. 

\subsection*{C. Energy spectra}

\begin{figure*}
\includegraphics[scale=0.7]{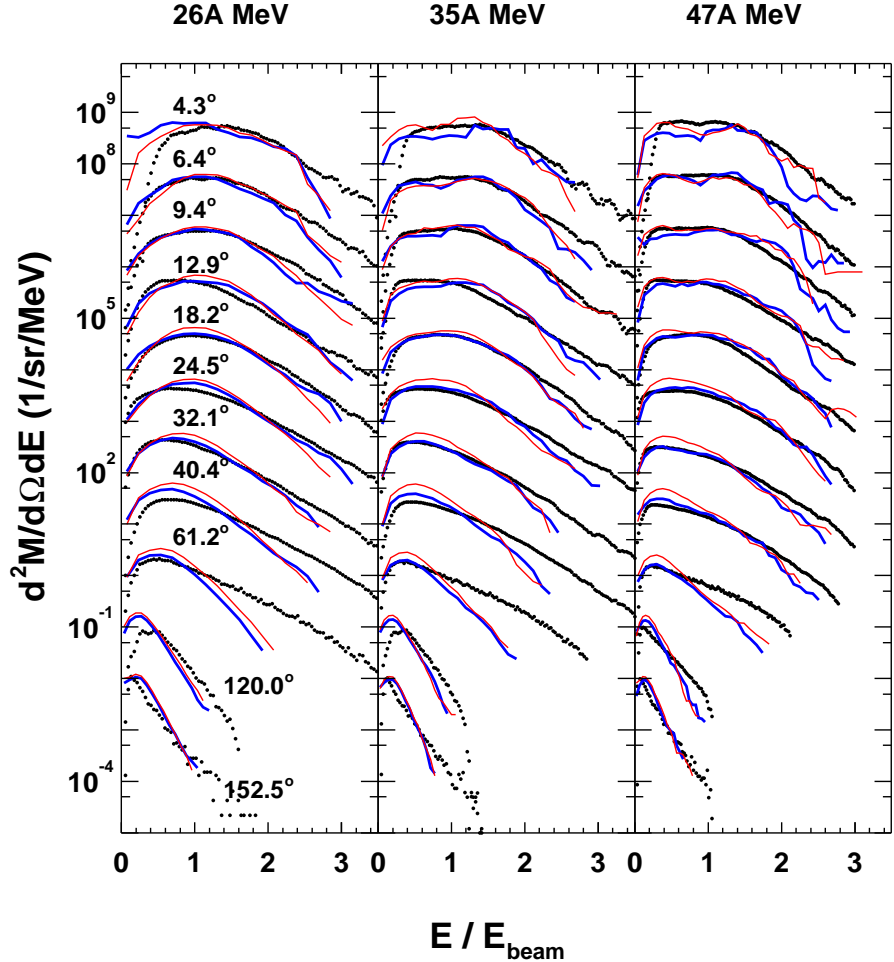} 
\caption{\footnotesize Proton energy spectra for the violent collisions for
$^{92}$Mo at three incident energies. The incident 
energy is indicated at the top of each figure. The spectra in each column
correspond to the energy spectra at different angles, indicated 
in the left column.  
Experimental results are shown by dots. Thick 
and thin solid lines correspond to calculated results for Soft+NN$_{emp}$ and 
Stiff+NN$_{LM}$, respectively. For all cases the differential 
multiplicity is given in absolute units. In order to avoid the overlap of the
data spectra are multiplied by factors of 10$^{n}$ (n=0,1,2,3,4,5,6,7,8,9,10)
from bottom to top. No experimental filter is used for the calculated events.
}
\label{Ep_znmo}
\end{figure*} 

\begin{figure*}
\includegraphics[scale=0.70]{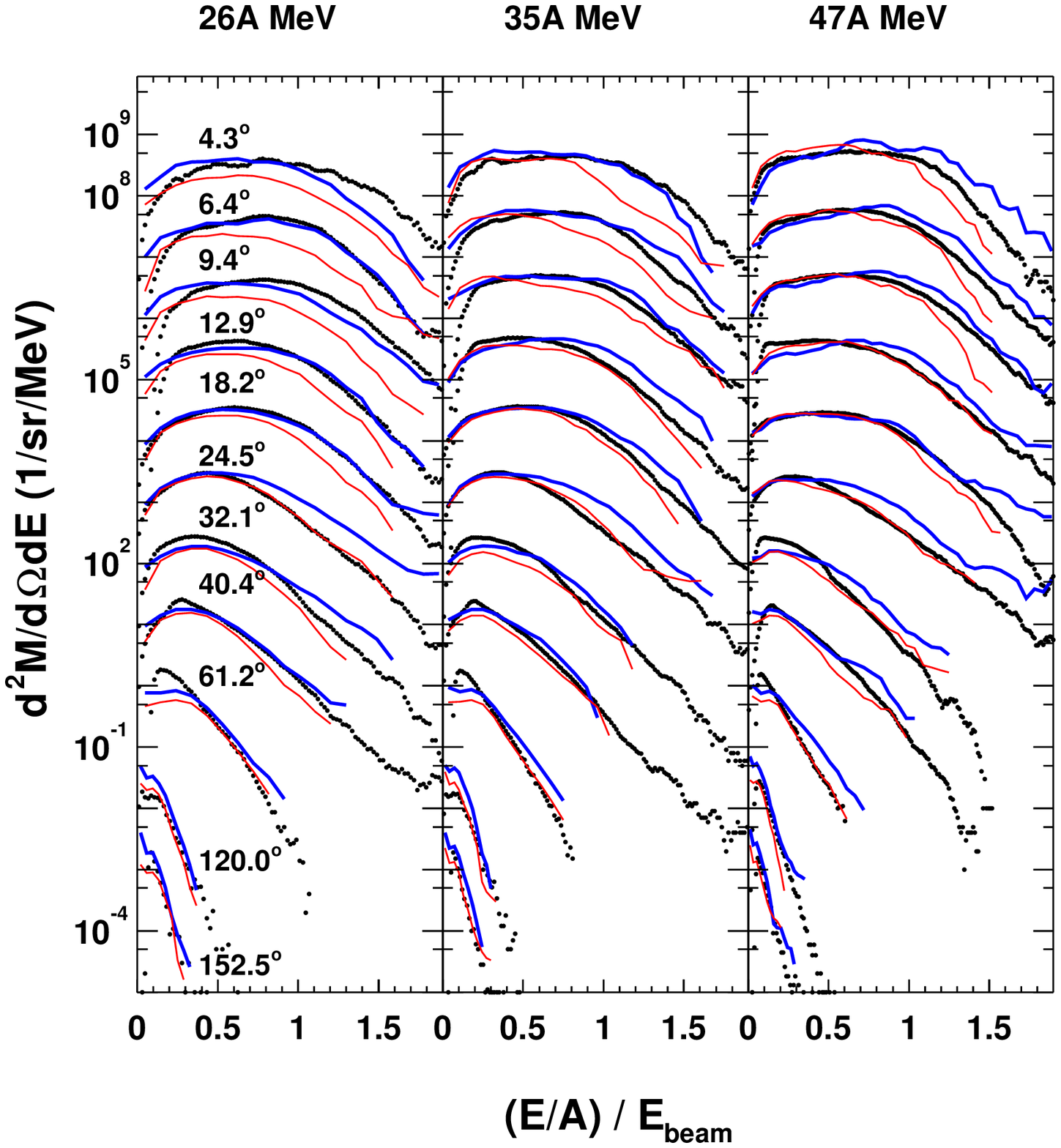} 
\caption{\footnotesize Similar plots to Fig.~\ref{Ep_znmo}, but for $\alpha$
particles.
}
\label{Ea_znmo}
\end{figure*} 

\begin{figure*}
\includegraphics[scale=0.70]{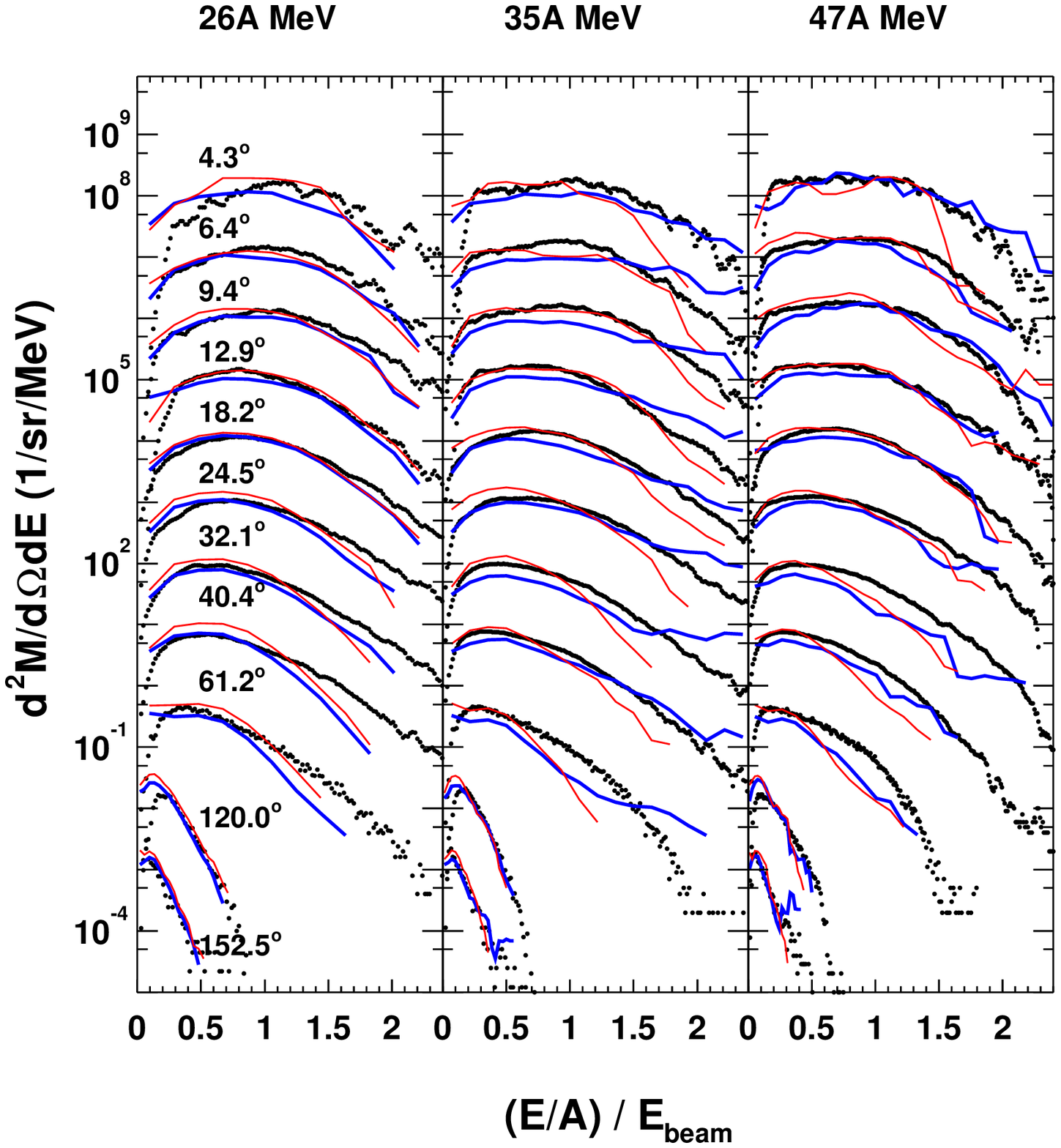} 
\caption{\footnotesize Similar plots to Fig.~\ref{Ep_znmo}, but for deuterons. 
}
\label{Ed_znmo}
\end{figure*}

\begin{figure*}
\includegraphics[scale=0.70]{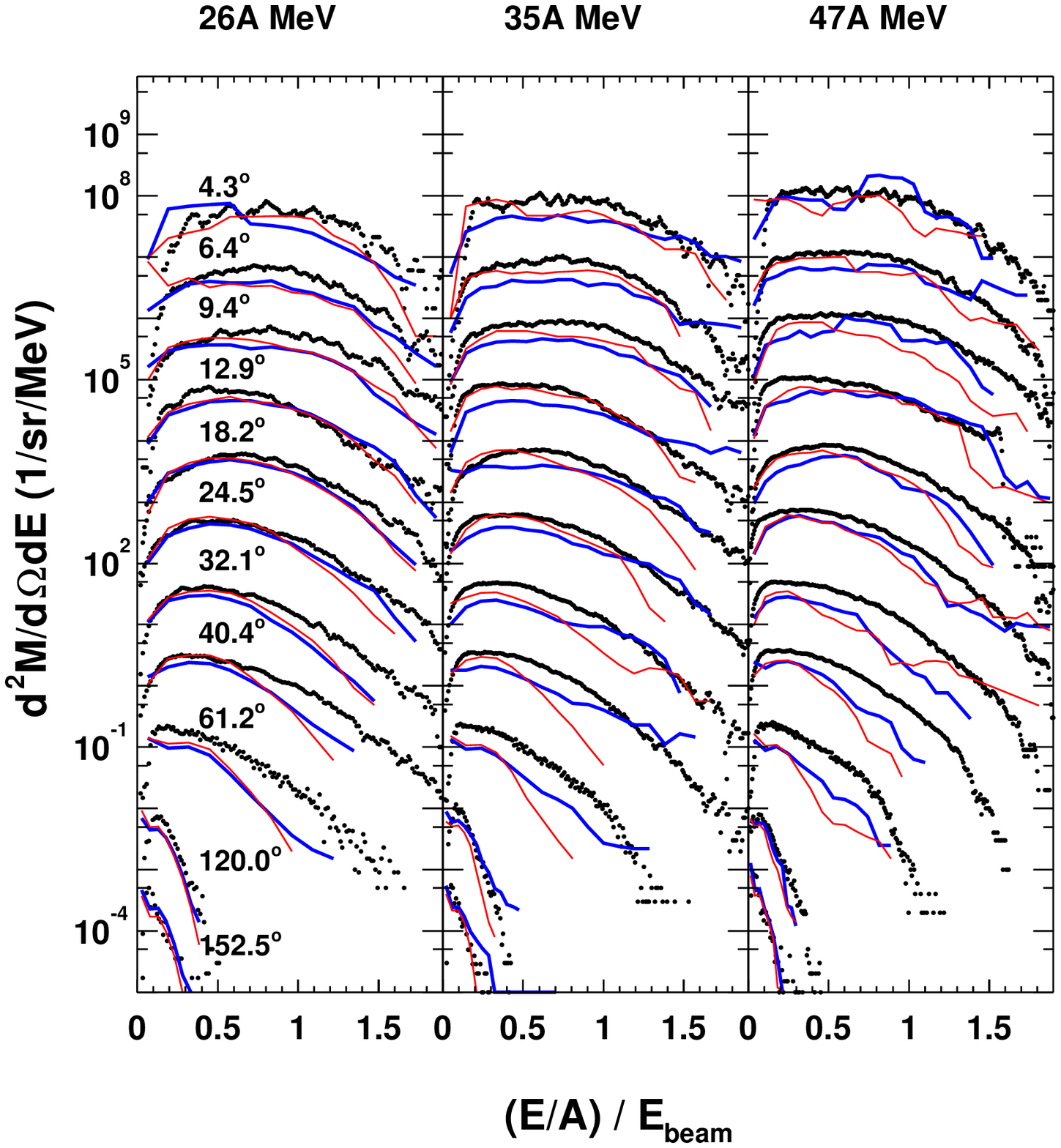} 
\caption{\footnotesize Similar plots to Fig.~\ref{Ep_znmo}, but for tritons. 
}
\label{Et_znmo}
\end{figure*}

For the violent collisions for $^{92}$Mo at three 
different incident energies, typical energy spectra of light charged particles 
are shown in Figs.~\ref{Ep_znmo} to ~\ref{Et_znmo}. 
Experimental spectra and calculated spectra for the soft EOS + NN$_{emp}$ 
and the stiff EOS + NN$_{LM}$ are shown. Energy is scaled by 
the beam energy. The vertical axis is the differential
multiplicity and all results are given in an absolute scale. As one can see for
all cases, the experimental spectra, at three incident energies, are very 
similar to each other in shape and angular dependence. This indicates that  
these energy spectra can be parametrized as emission from moving sources 
with similar source velocities and apparent temperatures, but scaled by the 
incident energy.  
In the present work, however, no such analysis was performed. Instead the  
experimental energy spectra and angular distributions are compared with
calculated AMD-V events. As seen in Figs.~\ref{Ep_znmo} to ~\ref{Et_znmo}, 
a reasonable agreement is obtained for all cases.
For protons, however, the experimental slopes of the high energy 
tails in Fig.~\ref{Ep_znmo} tend to be harder than those in calculations, 
especially at 26A and 35A MeV. The deviation becomes significant at angles 
of $\theta=40.4^o$ and 61.2$^{o}$ which correspond to emission near $\theta=90^o$ 
in the center of mass system. A similar discrepancy is also observed 
at 47A MeV. 
For the calculations with the stiff EOS, an excess of 
low energy protons is clearly observed at angles between 
24.5$^o \le \theta \le 60.2^o$ at both 35 and 47A MeV. This is slightly less 
prominent at 26A MeV. This is the main cause of the excess of the 
proton multiplicity in the calculations with the stiff EOS, observed 
in Fig.~\ref{pMul}.

The experimental $\alpha$ energy spectra are compared with calculated spectra 
in Fig.~\ref{Ea_znmo}. At 35 and 47A MeV the experimental spectra are well 
reproduced by the calculations with the stiff EOS, except at the most 
forward angle, whereas the calculations with the soft EOS overpredict 
the yields on higher energy side at most of the forward angles. This  
overproduction causes the overprediction of the alpha multiplicity observed 
for the soft EOS in Fig.~\ref{aMul}. On the other hand, at 26A MeV, the 
calculation with the soft EOS reproduces the experimental spectra quite well 
at most angles. The stiff EOS significantly 
underestimates the yields in higher energy side.

Deuteron and triton energy spectra are shown in Fig.~\ref{Ed_znmo} and 
Fig.~\ref{Et_znmo}. At 26A MeV both calculations underpredict the yields 
for both deuterons and tritons of higher energies. This is 
similar to observations for the proton spectra. For the deuteron spectra 
at higher incident energies, the calculations with the stiff EOS reproduce 
the experimental spectra better than those with the soft EOS at the four most 
forward angles. The overprediction of low energy deuterons,    
that is seen in proton spectra, is also observed at angles 
between 24.5$^o \le \theta \le 60.2^o$. This excess is not observed for the 
triton spectra, but the corresponding statistical fluctuations are 
significantly large in that case.

\begin{figure}
\includegraphics[scale=0.5]{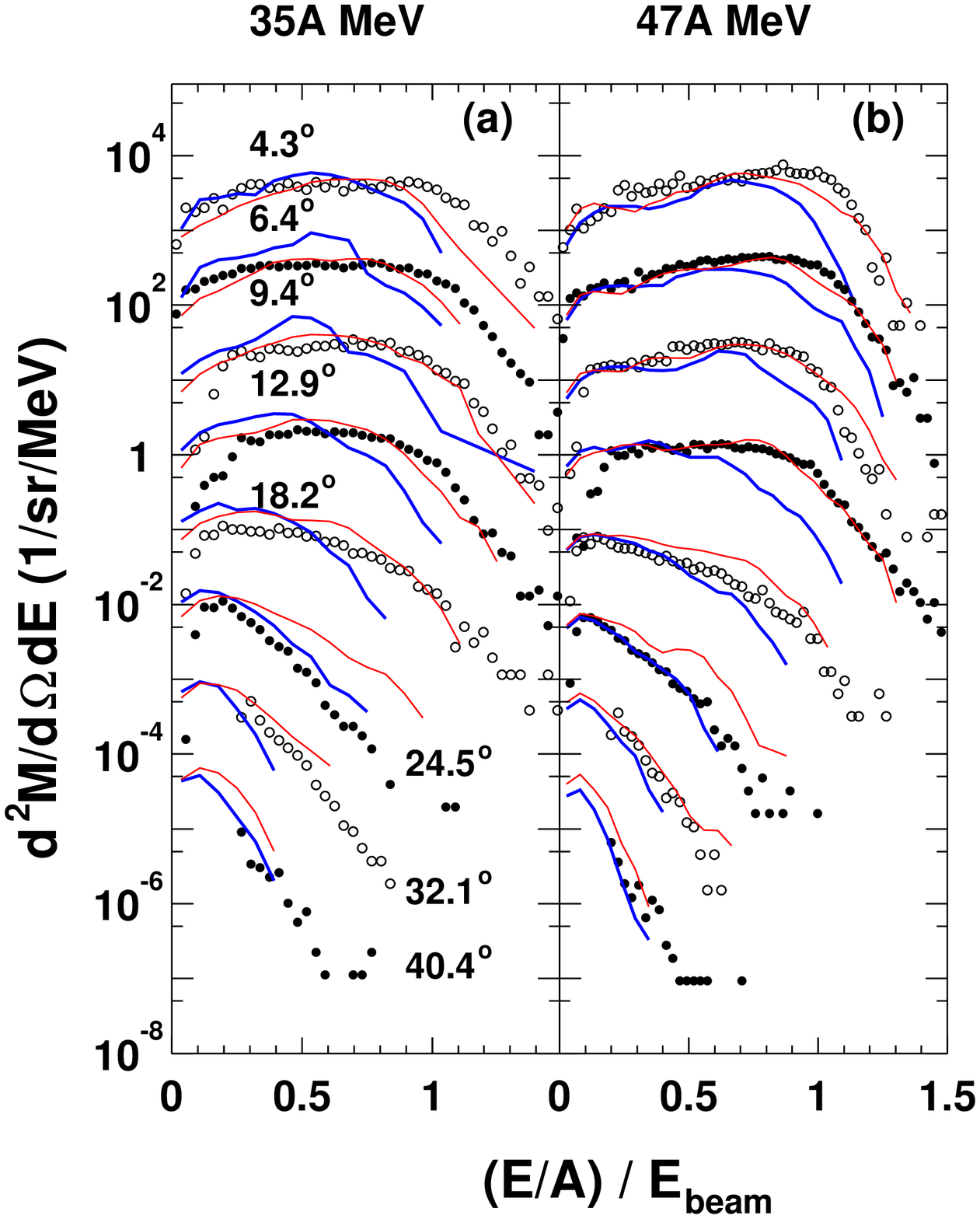} 
\caption{\footnotesize Inclusive energy spectra of IMF with Z=8 at different 
angles for violent collisions with $^{92}$Mo (a) at 
35A and (b) at 47A MeV. Experimental spectra are shown 
by symbols and calculated results with soft EOS + NN$_{emp}$ and 
stiff EOS + NN$_{LM}$ are shown by thick and thin solid lines, 
respectively. The experimental spectra are plotted in an absolute scale, 
whereas each calculated result is normalized to the experimental spectra 
at $\theta=4.3^o$.}
\label{Ez8_znmo}
\end{figure}

Typical inclusive energy spectra of heavier fragments are shown for Oxygen 
fragments in Fig.~\ref{Ez8_znmo} for $^{92}$Mo at 35A and 47A MeV. 
As discussed in ref~\cite{wada00}, 
the energy spectra of IMF depend little on the multiplicity class selection.
The experimental spectra, scaled by the beam energy, show very similar shapes 
and angular distributions for both incident energies, except at the most 
forward angle. This again indicates that these spectra can be 
described by moving source (or sources) parametrizations 
with similar source velocities 
and apparent temperatures, scaled by the incident energy. 
In AMD-V most of experimental spectra are better 
reproduced by the calculations with the stiff EOS, especially in the higher 
energy side. 
   
\subsection*{D. Velocity distributions}

\begin{figure}
\includegraphics[scale=0.5]{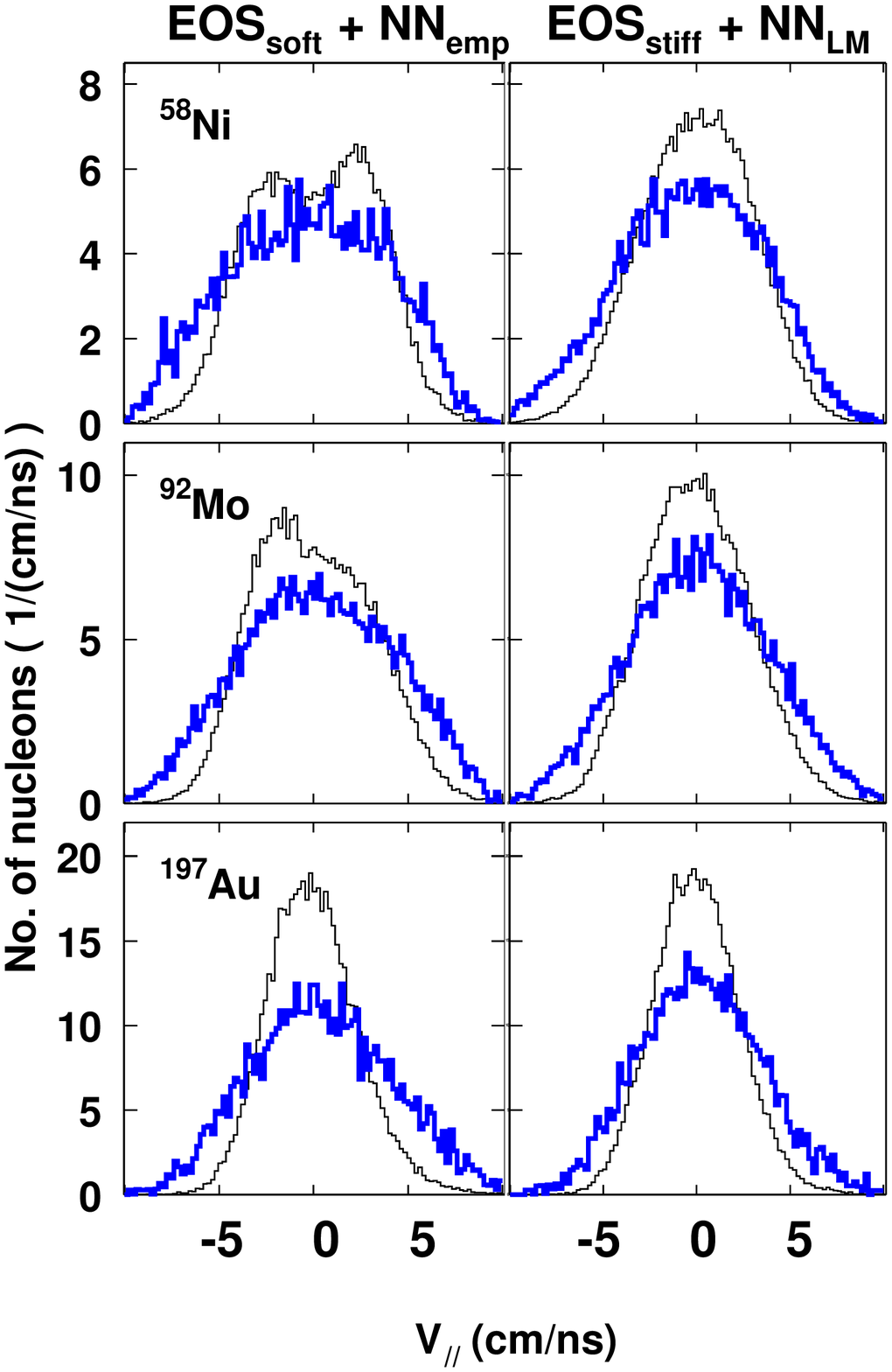} 
\caption{\footnotesize Calculated parallel velocity distributions of protons 
in the center of mass system for different reactions at 47A MeV. 
Each row corresponds to the same reaction which is specified 
by target as indicated in each panel on the left. The calculated results
with soft EOS + NN$_{emp}$ are plotted on the left and those with
stiff EOS + NN$_{LM}$ are on the right.
The distributions of all free and bound protons before the afterburner 
at 280 fm/c are displayed by thin lines. The distributionis of free 
protons after the afterburner are shown by thick 
lines. The normalization for the latter is arbitrary.
}
\label{PVpara_simu}
\end{figure}

As shown in Fig.~\ref{Vpara}, for the parallel velocity distributions of all 
nucleons at early stages of reactions, AMD-V predicts a distinct difference 
between the soft EOS and stiff EOS calculations for reactions with the 
lighter targets at 47A MeV. The calculated distributions of all free and 
bound protons also show similar differences as seen in Fig.~\ref{PVpara_simu}. 
However, in order to make the distribution for all protons 
experimentally, all fragments have to be identified at all angles
with very low energy threshold. This is a very difficult task in experiments.
Instead we compare the experimental and calculated distributions of free 
protons. 

We first compare the calculated parallel velocity 
distribution of all free and bound protons before the afterburner and that of 
only free protons after the afterburner. The results are shown in 
Fig.~\ref{PVpara_simu} for the reactions at 47A MeV. The results with the soft
EOS are shown on the left and those with the stiff EOS are shown on the right.
For free protons after the afterburner the parallel velocity 
distributions for the soft EOS show slightly broader distributions than those 
for the stiff EOS, but the shape becomes very similar. Especially
for $^{58}$Ni, on the left, the two-peak structure seen for free and bound 
protons before the afterburner becomes a single broad peak. As a result, 
no distinct difference is observed between calculated results with the 
soft EOS and the stiff EOS for the distributions of free protons after the 
afterburner . This indicates that the statistical decay process 
in the afterburner changes the distributions quite drastically. 

\begin{figure}
\includegraphics[scale=0.45]{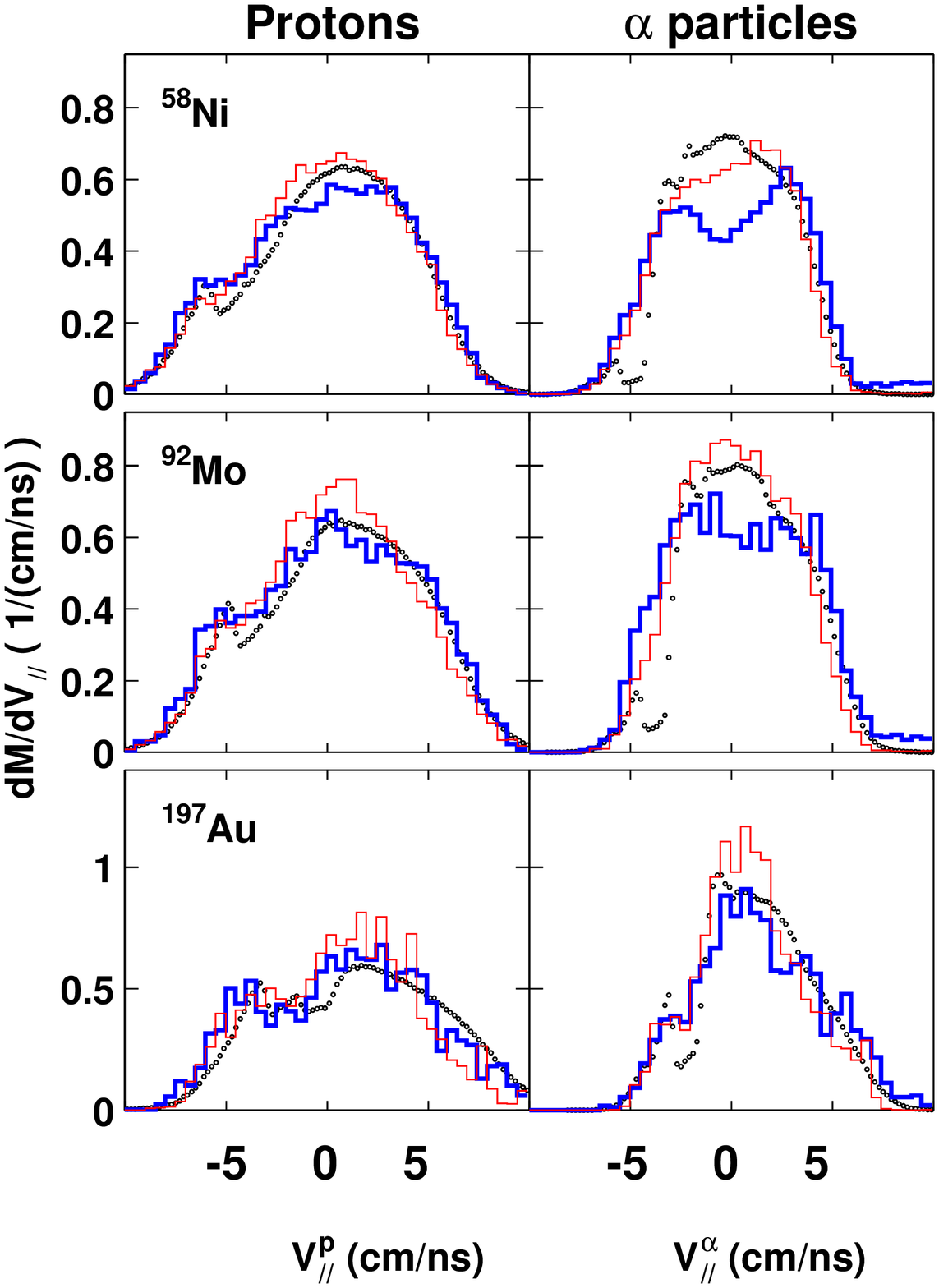} 
\caption{\footnotesize Parallel velocity distributions of protons (left) and
$\alpha$ particles (right) in the center of mass system for different 
reactions at 47A MeV. Reaction is specified by target as indicated 
in each panel on the left. Experimental results are shown by dots and 
the calculated results with the soft EOS +NN$_{emp}$ are plotted by 
histograms of thick lines and those for the stiff EOS + NN$_{LM}$
by histograms of thin lines.
The calculated results are normalized to the experimental distributions to 
obtain the same number of protons at V$_{//}$ $\geq$ 0. in each cases. 
}
\label{PAVpara}
\end{figure}

In the left 
column of Fig.~\ref{PAVpara} the experimental parallel velocity distributions 
of free protons are compared with those of the calculations. The dips 
in the negative velocity side in the experimental distributions are caused by 
the target shadow to the detectors at 90$^{o}$. As one can see, both 
calculations with the soft EOS and the stiff EOS reproduce the experimental 
proton distributions quite well and no distinct difference is observed 
between the two calculations. On the other hand the parallel velocity 
distributions of $\alpha$ particles do show some differences between the 
two calculations seen in the right column of Fig.~\ref{PAVpara} where the 
experimental results for $\alpha$ particles are compared with those 
of the calculations. In the calculated results for $^{58}$Ni, 
the two-peak structure is clearly observed in the simulation with
the soft EOS, whereas a single broad peak is seen in that with the stiff EOS.  
This result is consistent with the calculations reported by 
Ono~\cite{ono00} 
for $^{40}$Ca + $^{40}$Ca at 35A MeV. This trend is also observed for 
$^{64}$Zn + $^{92}$Mo in Fig.~\ref{PAVpara}, but is slightly less prominent. 
For $^{64}$Zn + $^{197}$Au no difference is observed between the 
two calculations. The overall experimental trend  
for $\alpha$ particles favors the stiff EOS. 

\begin{figure}
\includegraphics[scale=0.45]{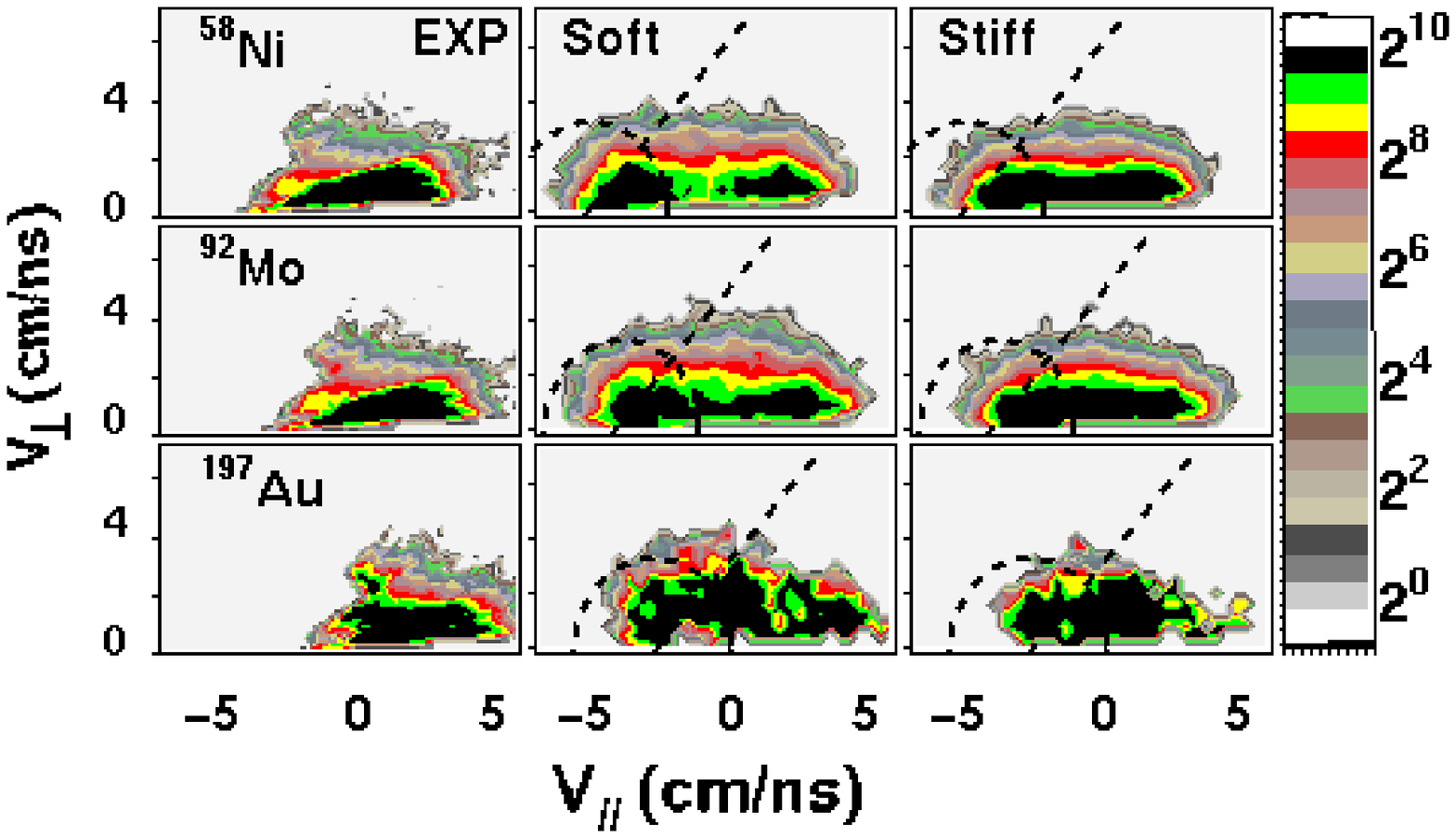} 
\caption{\footnotesize Two dimensional Galilean invariant plots of parallel 
velocity {\it vs} perpendicular velocity of all fragments with Z=7 and 8 
for the reactions at 47A MeV. Plots are in the center 
of mass system. Each row shows the results for the same reaction system 
and reaction is specified by the target, indicated in the left
figure of each row. The experimental distributions are shown on the left 
column. The calculated distributions for soft EOS + NN$_{emp}$ are shown 
in the middle column and those for stiff EOS + NN$_{LM}$ are plotted on the 
right column. No experimental filter is applied for the calculations, except 
the forward angle detection limit (3.5$^{o}$). Dashed lines in the middle 
and right columns indicate the detection limit of the polar angle. Dashed 
circles indicate approximate energy thresholds for these fragment.
}
\label{HFVpVt}
\end{figure}

Since the differences observed in the calculations with the soft EOS 
and stiff EOS in Fig.~\ref{Vpara} are not reflected in the free 
proton distributions but appear in the $\alpha$ distributions,
one can expect that the difference is related to the velocity 
distribution of heavier fragments. In Fig.~\ref{HFVpVt}, bi dimensional
plots of parallel velocity versus perpendicular velocity are shown for
fragments with Z=7,8 for the reactions at 47A MeV. The velocity distributions
for calculations with the soft EOS show a clear two-peak structure for both  
$^{58}$Ni and $^{92}$Mo. 
On the other hand those of the calculations with the stiff EOS show an 
elongated distribution. Although in the experiment about 50\% of 
fragments are not detected because of the energy threshold and the detection 
angles, as shown by dashed lines in the calculated results, one can still 
clearly see that the experimental IMF distributions favor the calculations 
with the stiff EOS.   

\subsection*{E. Collective flow analysis}

\begin{figure}
\includegraphics[scale=0.45]{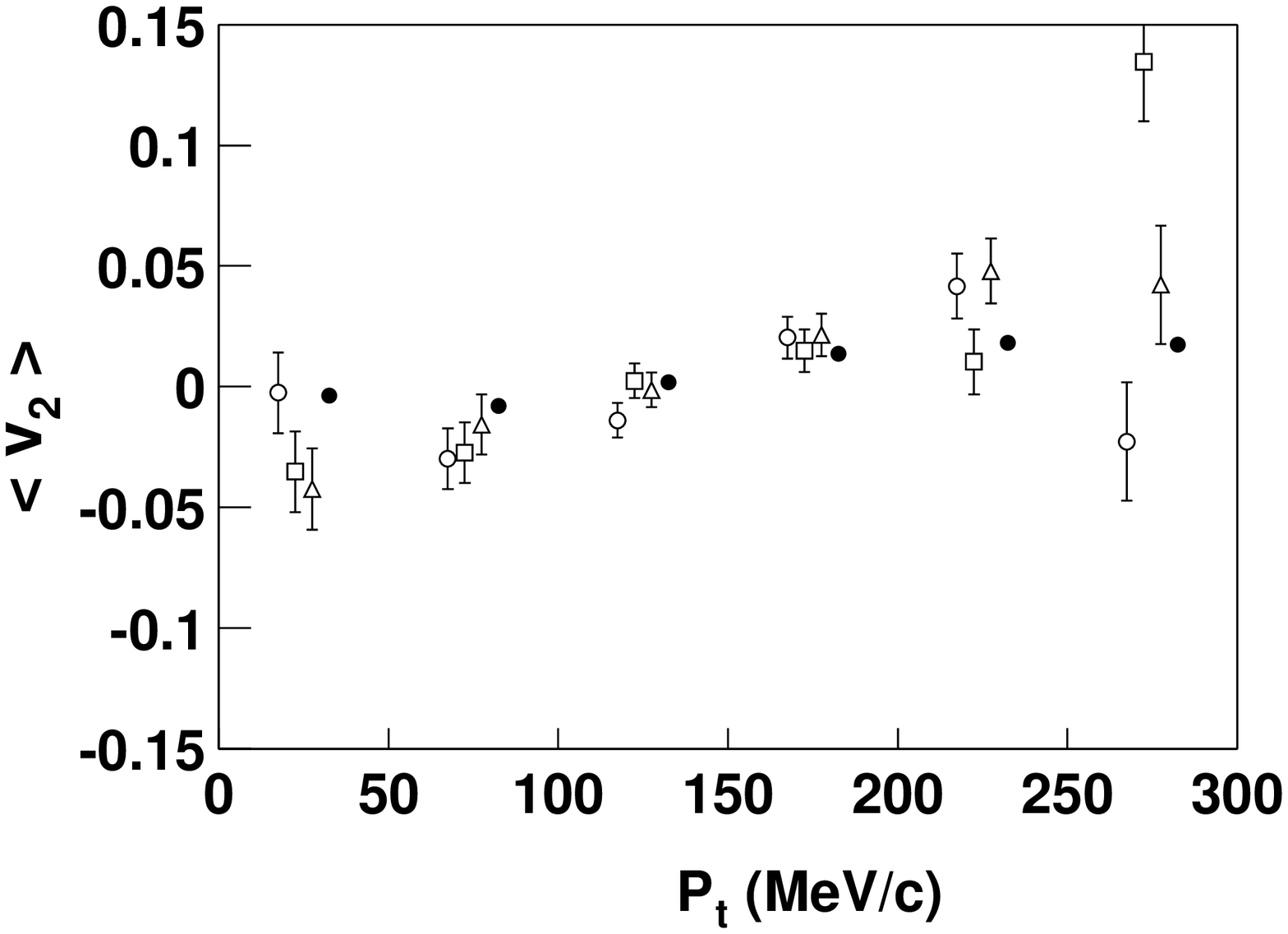} 
\caption{\footnotesize Averaged differential elliptic flow for protons for
$^{58}$Ni at 47A MeV as a function of the transverse momentum $p_t$. Events 
in the ``Violent" and Semi-Violent classes are used. Experimental results
are shown by dots and calculated results for stiff EOS + NN$_{LM}$, 
stiff EOS + NN$_{emp}$ and soft EOS + NN$_{emp}$ are shown by open circles, 
squares and triangles, respectively.  
$<v_{2}>$ values are calculated in 50 MeV/c steps in $p_t$. 
Calculated results are plotted around the center value of each p$_{t}$
with a small shift to avoid overlap of results. 
}
\label{EFlow}
\end{figure} 

In collective flow analyses, the directed flow, azimuthal angle 
correlation and elliptic flow have been 
studied~\cite{Shen93,Shen98,Ollitrault98}. These flows are essentially 
governed by particle emission at early stages. Therefore, the event 
multiplication in the afterburner does not reduce the statistical errors on the 
results. Flow is also a rather small effect and a large number of events
has to be generated in AMD-V to get reasonable comparisons 
to the experimental results. Of the three different flow analyses, the
differential elliptic flow is an averaged observable and requires a smaller
number of events for reasonable comparisons. Therefore, we studied 
the differential elliptic flow $v_2$ of protons~\cite{BALi01}, given by  
the averaged value of $v_2$, and described by
\begin{eqnarray}
<v_{2}(p_{t})> = {1\over{N(p_{t})}}{\sum_{j=1}^{N(p_{t})} { {p_{jx}^2 - p_{jy}^2}\over{p_{ 
jt}^2}}},
\end{eqnarray}
where N(p$_{t}$) is the number of protons in a given p$_{t}$ range and 
p$_{jt}$ is the transverse momentum of the j-th proton. 
p$_{jx}$ and p$_{jy}$ are
its perpendicular momenta in and out of the reaction plane, 
respectively. The reaction plane is determined, using transverse momentum of 
emitted particles as proposed by Danielewicz and Odyniec~\cite{Danielewicz85}. 
In Fig.~\ref{EFlow} $<v_{2}>$ values of the experimental and calculated 
results are shown for protons for $^{58}$Ni at 47A MeV. Events in the 
``Violent" and ``Semi-Violent" classes are summed in order to increase 
statistics. Only protons in the mid-rapidity range 
(0.25 $\leq$ y/y$_{beam}$ $\leq$ 0.75) are used. Errors are evaluated from 
the deviations from zero for $<v_{2}>$ values when the azimuthal angle of the 
reaction plane is randomized. As one can observe, the experimental $<v_{2}>$ 
value shows a small deviation from zero in the energy 
range studied in this paper. No significant difference is observed between 
the values extracted from the calculated events for different parameter sets.
The calculated values are consistent with the experimental ones within
error bars, except the most highest p$_t$, where the statistics becomes poor. 

\subsection*{F. Discussion}

As seen in Fig.~\ref{Vpara}, a prominent difference is observed between 
the calculated results with the soft EOS and with the stiff EOS 
in the parallel velocity distribution of all free and bound nucleons before 
the afterburner (t=280 fm/c) for $^{58}$Ni and $^{92}$Mo at 47A MeV. 
The distribution of all protons also shows a two-peak (or a shoulder) 
structure as seen in Fig.~\ref{PVpara_simu}. However this two-peak  
structure becomes a broad single-peak distribution for the free protons 
after the afterburner. The experimental distributions are well reproduced 
by both calculations with the soft EOS and with the stiff EOS. These good 
agreements suggest that the prominent difference, seen in the velocity 
distribution at the early stages of the reaction, may be reflected in the bound 
nucleons in clusters. 
This is interesting because, in this case, the difference which appears in 
the velocity distribution of the clusters may not depend on the details 
of the multifragmentation process. In other words, the signal of the stiffness 
of the EOS remains as a fossil signal in the velocity distribution 
of fragments. 

As seen in section V.B, our experimental multiplicity of fragments with 
Z $\ge$ 4 is overpredicted in all calculations. A similar result has also been
observed in the $^{129}$Xe+Sn reaction at 50A MeV~\cite{ono02}. In order to 
resolve this discrepancy between the experimental and calculated results, 
an extension was made in the reference, so that the shrinking as 
well as the diffusion of the wave packets would be taken into account during 
their propagation in the mean field. This treatment with the
new parameter suppresses the dynamical decay of the excited fragments at early 
stages and reproduces the experimental charge distribution quite well. 
However the above observation for the parallel velocity distribution indicates
that the signature of the stiffness of EOS remains as a footprint in the 
fragment velocity distributions with or without taking into account this 
kind of fine tuning of the model.

In the present work, the Gogny interaction with the stiff EOS is favored by
the experimental velocity distributions of the clusters. On the other hand 
the Gogny force with the soft EOS gave a better description for the giant 
monopole resonance studies~\cite{Youngblood01}. However one should note the 
following: 
(1) In the giant resonance studies, normal nuclear matter is studied, 
whereas in this work the nature of compressed nuclear matter at high 
temperature is treated;
(2) Our result strongly depends not only on the stiffness 
of the EOS but also on the reaction dynamics, especially the
Pauli blocking during the reaction. In AMD-V Pauli blocking is 
taken into account exactly for the propagation of the wave packets 
at all times. For stochastic nucleon-nucleon collisions, the 
Pauli-blocking is automatically taken into account in the transformation 
between the physical coordinate space and AMD space. However this 
transformation, which is given in Eq.\ (\ref{eq:AMDphysicalK}), is an 
approximation. One of the consequences of this approximation is seen in the 
charged particle multiplicity distributions for collisions near the surface, 
discussed in detail in ref~\cite{wada00}. When the Pauli-blocking becomes 
very significant as in the reactions studied in this work, this procedure may 
cause some errors. In fact at 26A MeV the calculated results for both soft 
and stiff EOS significantly underpredict the slopes of energy spectra  
for protons, deuterons and tritons, as seen in 
Figs.~\ref{Ep_znmo},\ref{Ed_znmo},\ref{Et_znmo}. 
This suggests that the AMD-V calculations lead to too much transparency at 
lower energies, resulting in less compression and/or excitation energy 
at the early stage of the collisions. Therefore it is appropriate to only 
conclude that, in the framework of the present AMD-V calculations, 
the stiff EOS is preferred to reproduce the present experimental results.

Another interesting observation in this study is that the different 
NN collision cross sections do not alter the nuclear semi-transparency.
Two formulations have been employed in the present calculations. 
Although the cross sections are different by a factor of 2-3 on average 
in the range of proton energies relevant to this study, the apparent effect 
on the nuclear stopping is rather small, as seen in Fig.~\ref{Vpara}. 
This result is quite surprising because
in QMD studies of the nuclear collective flow, the effect of 
different NN cross sections on the balance energy is of a similar order to 
the effect of different stiffnesses of the EOS~\cite{Magestro}. 
Ono et al. studied collective flow in the $^{40}$Ar + $^{27}$Al 
reaction, using AMD without the diffusion process~\cite{ono95}. In that study 
the strength of flow was reduced by about 30\% at 45A MeV when $\sigma$, 
the empirical NN cross section given by Eq.\ (\ref{eq:NNempirical}), 
is increased by 50\%. 

The results of the differential elliptic flow analysis are consistent with the 
observation of the parallel velocity distributions, although the deviations 
of the experimental and calculated results from the isotropic distribution are
rather small. These results suggest that the diffusion process in AMD-V 
plays a significant role, not only in the multifragmentation process, 
but also in the wave packet propagation. This diffusion process tends to 
randomize the trajectory of wave packets and to smear out collective flow 
in this energy region.  

\section*{VI. MULTIFRAGMENTATION MECHANISM}

In the previous work on the $^{64}$Zn + $^{58}$Ni reactions at 35A-79A MeV, 
we reported that the semi-transparency plays an important role for the 
multifragmentation process~\cite{wada00}. However for the 
$^{64}$Zn + $^{197}$Au reactions the calculations indicate that all projectile
nucleons essentially stop in the target nuclei at all incident energies, 
as seen in Fig.~\ref{Vpara}. Therefore it is expected
that transparency plays little role in the multifragmentation process
for these reactions and that other mechanisms, such as expansion and 
statistical multifragmentation processes, may play a dominant role in the 
disintegration of the system. On the other hand the similarity of the 
charge distributions for all reactions studied here, 
as seen in Fig.~\ref{Z_Exp}, suggests that there is a common feature 
for the multifragment production in these reactions in which quite different 
dynamics are involved. 

Since the overall experimental results are reproduced reasonably well 
by AMD-V with the stiff EOS, in the following we will analyze central 
collision events ( b $\leq$ 3 fm ) for the different systems, calculated with 
the stiff EOS, in order to elucidate the multifragmentation mechanism in great 
detail. No afterburner is applied in the analysis in this section.
  
\subsection*{A. Global character of reactions}

\begin{figure}
\includegraphics[scale=0.5]{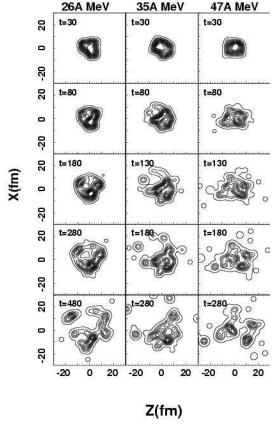} 
\caption{\footnotesize Time evolution of nuclear density distributions, 
projected on the reaction plane (x-z plane, z being the beam direction), 
for a calculated event with b $\sim$ 2 fm at different reaction 
times for $^{197}$Au. Incident energy is indicated at the top of each column. 
Reaction time is indicated in the unit of fm/c in each panel. 
The time zero is set at the time 
when the projectile and the target touch each other. The z axis is taken as 
the beam direction and the contour scale is in linear. The smallest circle 
indicates a nucleon. 
}
\label{Density-plot}
\end{figure}

In Fig.~\ref{Density-plot}, density distributions for $^{197}$Au are plotted 
as a function of time.  
The projectile and the target are fully overlapped around t $\sim 30$ fm/c
and the system starts to expand and undergoes multifragmentation 
for all incident energies. One should note that prefragments are 
already recognized at early stages of the reaction 
( t $\sim$ 80$\sim$180 fm/c ). This is quite different from a statistical
multifragmentation picture, in which a hot system expands and clusterizes at
a low freezeout density~\cite{Bondorf95}. It should be also noted that 
the fragment sizes are very similar with each other for three incident energies.
Nucleon emissions are identified as early as t $\sim$ 100 fm/c at 35 A and 47A 
MeV and t $\sim$ 200 fm/c at 26A MeV.

\begin{figure}
\includegraphics[scale=0.475]{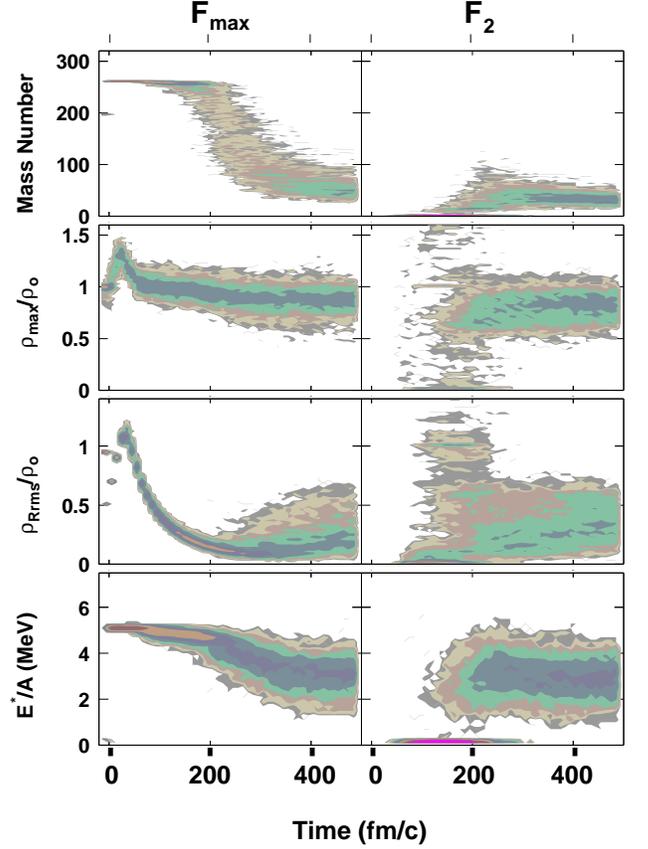} 
\caption{\footnotesize Time evolution of mass number, density and excitation 
energy of the largest fragment (left) and the second largest fragment with 
Z > 2 (right) from top to bottom, respectively for $^{197}$Au at 35A MeV. 
See details in the text.  Contours are in logarithmic scale.
}
\label{AmaxAmax2}
\end{figure} 
 
In Fig.~\ref{AmaxAmax2}, the mass number, density and excitation energy of the 
largest fragment (F$_{max}$) and of the second largest fragment 
(F$_{2}$, Z $> 2$) are 
plotted as a function of the reaction time for $^{197}$Au at 35A MeV. In 
order to evaluate these quantities, each calculated event is clusterized in 
coordinate space with a coalescence radius of 5 fm at a given time. 
In the first row, the mass numbers of F$_{max}$ and F$_{2}$ are shown.
At t $\sim$ 200-300 fm/c, F$_{max}$ undergoes multifragmentation. The generated
F$_{2}$ has a similar size to the F$_{max}$ after this multifragmentation. 

In the second row the maximum density $\rho_{max}$ is plotted. In the time of 
overlap, $\rho_{max}$ reaches around 1.4$\rho_{o}$ and quickly returns 
to the normal density at t $\sim$ 100 fm/c. After that $\rho_{max}$ stays 
around 0.85$\rho_{o}$, $\rho_{max}$ of F$_{2}$ also shows a similar value to 
that of F$_{max}$. This indicates that prefragments, as seen in 
Fig.~\ref{Density-plot}, have density close to the normal density when they 
are formed at early stages and keep this density during the expansion of the 
system.

In the third row, the density calculated from the root-mean-square radius 
(Rrms) is plotted. 
$\rho_{Rrms}$ is calculated by assuming that the fragment has a spherical 
shape, i.e.,
\begin{equation} 
\rho_{Rrms} = \rho_{o} \frac{A}{A_{o}} \{ \frac{R_{o}}{R_{Rrms}} \} ^{3}. 
\end{equation}
Here $A_{o}$, $\rho_{o}$ and R$_{o}$ are the mass number, density and 
r.m.s. radius of the initial projectile nucleus, respectively.
$\rho_{Rrms}$ reaches 1.3$\rho_{o}$ at t $\sim$ 30 fm/c , a similar value to 
$\rho_{max}$, indicating that the shape of the composite system at the time of 
overlap is nearly spherical. $\rho_{Rrms}$ decreases rapidly after that and
reaches 0.2$\rho_{o}$ at t $\sim$ 200 fm/c, although $\rho_{max}$ is close to
the normal density. This indicates that the system at this time 
has a very deformed shape and non-uniform density. $\rho_{Rrms}$ of 
F$_{2}$ is also small, indicating that emitted fragments have also an odd 
shape. 

In the fourth row the excitation energies are shown.   
The excitation energy of a fragmenti, E$^{*}$, is calculated by subtracting the 
calculated binding energy from the internal energy, i.e., 
\begin{equation}
E^{*}= E_{int} - E_{B.E.} 
= \langle T\rangle + \langle V\rangle + \langle V_{C}\rangle - E_{B.E.}
\label{eq:Ex}
\end{equation}
$\langle T\rangle$ is the expectation value of kinetic energy in the rest frame 
of the fragment, $\langle V\rangle$ is that of the effective interaction 
given in Eq.~\ref{eq:E_internal} for the case of the stiff
EOS, for example, $\langle V_{C}\rangle$ is that of the Coulomb 
interaction and E$_{B.E.}$ is the calculated binding energy.
The excitation energy of F$_{max}$ is E$^{*}$/A $\sim$ 5 MeV 
at the time of overlap and gradually decreases with time. 
An interesting observation 
is that the excitation energy of F$_{2}$ distributes around 3 MeV/nucleon 
independently on the emission time and shows significantly lower than that of
F$_{max}$ at t $\leq$ 300 fm/c. At t $\geq$ 400 fm/c, 
the excitation energy of F$_{2}$ becomes similar to that of the F$_{max}$. 
Some of F$_{2}$ at t 
$\leq$ 300 fm/c have excitation energies of less than 1 MeV/nucleon. These 
are Li isotopes, which have $|E_{B.E.}| <$ 5 MeV/nucleon. A more detailed 
discussion about cold fragment emission will be given later.

\subsection*{B. Light particle emission}

\begin{figure}
\includegraphics[scale=0.45]{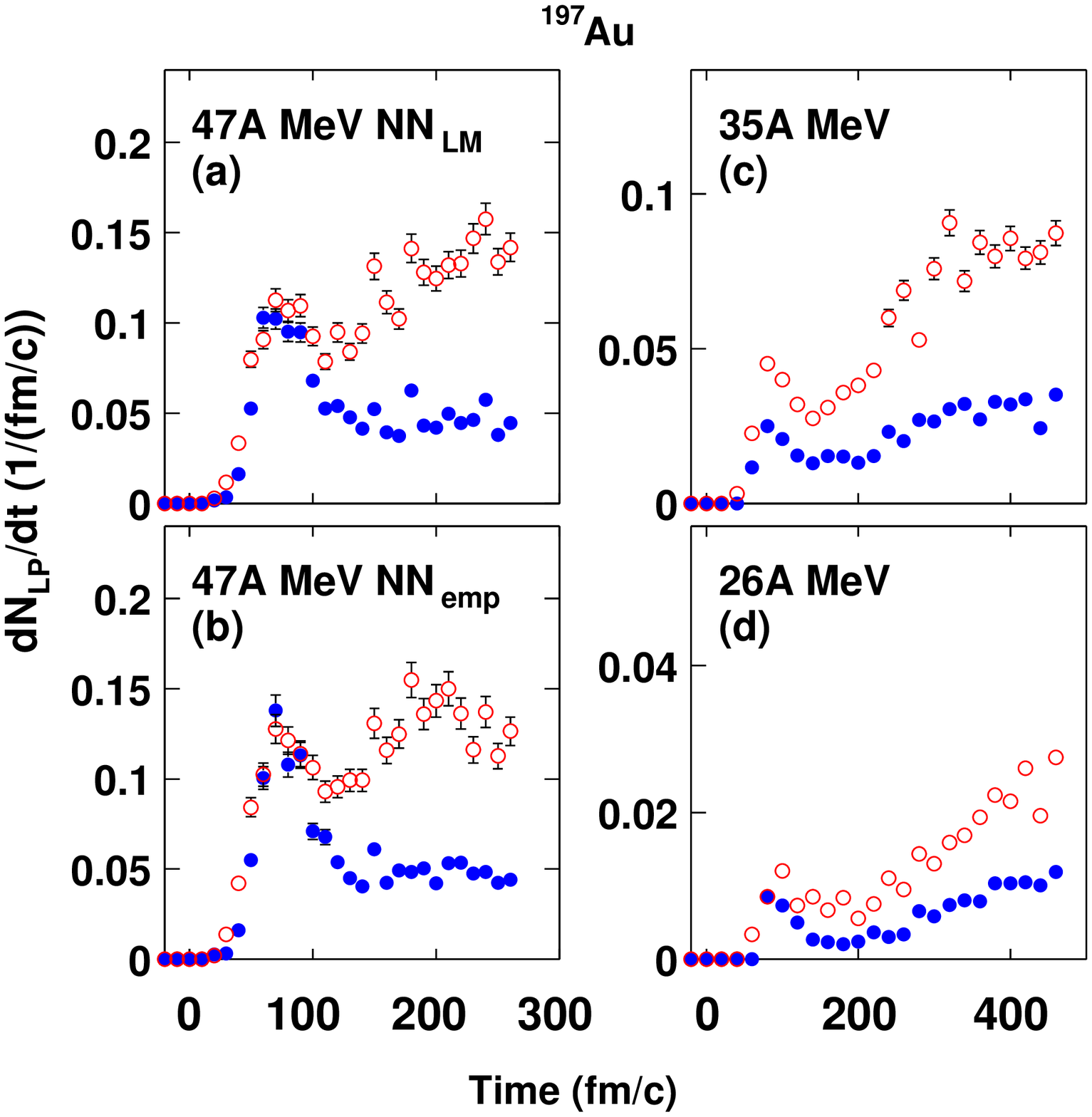} 
\caption{\footnotesize Nucleon emission rate for $^{197}$Au is 
plotted as a function of time. (a) 47A MeV with NN$_{LM}$ (b) 47A MeV with
NN$_{emp}$ (c) 35A MeV with NN$_{LM}$ (d) 26A MeV with NN$_{LM}$. 
Nucleons emitted as light particles (Z $\leq$ 2) are selected. 
Open circles indicate nucleons from the target and solid dots indicate  
those from the projectile. }
\label{NEmissionT}
%\end{figure} 

%\begin{figure}
\includegraphics[scale=0.475]{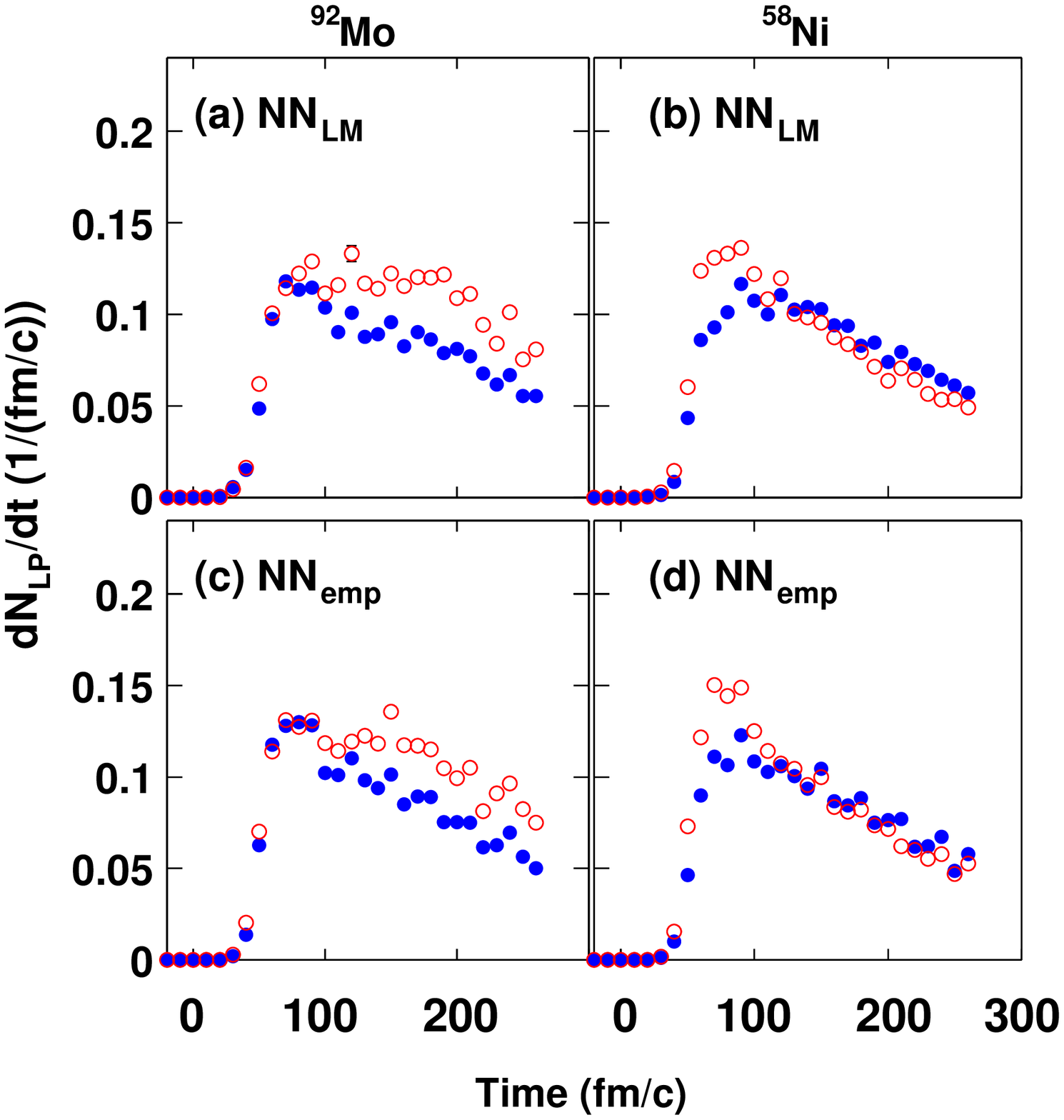} 
\caption{\footnotesize Similar plots to Fig.~\ref{NEmissionT}, but for 
$^{92}$Mo (left) and $^{58}$Ni (right) at 47A MeV. 
} 
\label{NEmissionT2}
 \end{figure} 
 
In order to elucidate reaction mechanisms, the emission of light particles 
(Z $\leq$ 2) is studied. The reactions with $^{197}$Au are examined first.
In Fig.~\ref{NEmissionT} the emission rates of nucleons emitted as a light 
particle are plotted as a function of time 
separately for nucleons from the projectile and for those from the target.
At t $\sim$30 fm/c, which corresponds to the full overlap 
time of the projectile and the target, light particles start to emerge  
and the emission rate shows a bump at t $\sim$ 80 fm/c 
for all incident energies. At 47A MeV, on the left side of 
Fig.~\ref{NEmissionT}, the emission rate at 
this bump remains almost equal for nucleons from the projectile and those 
from the target. This indicates that these nucleons are emitted from a source 
which consists of equal numbers of the projectile nucleons and the target 
nucleons, that is, the overlap region of the projectile and target. 
As seen in Fig.~\ref{NEmissionT} (a) and (b), the emission 
rates stay more or less the same when the in-medium NN cross section is changed 
from NN$_{LM}$ to NN$_{emp}$. At the relative NN energy of 50 MeV, the 
Li-Machaliedt formulae gives about 50\% larger average NN cross section than 
the empirical formula. This increase of the NN cross section enhances the NN 
collisions but also reduces the mean free path of the scattered nucleons. The 
above observation, therefore, suggests that these two effects are more or less 
balanced and the emission rate stays more or less same at 47A MeV. 
As discussed later, 
phenomenologically these nucleons can be described by emission from a
moving source with half the beam velocity, which is experimentally identified 
as the intermediate velocity source. 
 
At lower energies a similar trend is observed, though the emission rates 
of the target nucleons in the bump are about 50\% larger at 35A MeV and 20\% 
larger at 26A MeV than that of the projectile nucleons. These 
differences are still smaller than the 3:1 ratio of the target mass to the 
projectile mass. The differences may result from the fact that the absolute 
emission rate at these energies becomes much smaller than that at 47A MeV and 
it significantly depends on the details of the emission 
process, such as the location of the overlap region in the composite system.
      
A similar trend is also observed in the other reaction systems. 
In Fig.~\ref{NEmissionT2} results for $^{58}$Ni and $^{92}$Mo at 47A MeV 
are shown both for NN$_{LM}$ and NN$_{emp}$. All emission rates show a bump  
at t $\sim$80 fm/c for nucleons from the projectile and those from the target. 
The strength of the emission rate at the bump is $\sim$1.2 both 
for NN$_{LM}$ and NN$_{emp}$. This value is very similar to that for 
$^{197}$Au. This is consistent with the above conclusion that these 
preequilibrium particles originate from the overlap zone.
 
After the emission of these preequilibrium particles, as seen in 
Fig.~\ref{NEmissionT}, the emission rate of nucleons 
from the target increases much faster than that from the projectile. 
For the reaction at 47A MeV, the ratio between the two reaches a ratio of 
3 to 1, essentially equal to the ratio of the target mass to the projectile 
mass, at t $\sim$ 150$\sim$200 fm/c.
This time becomes t $\sim$250$\sim$300 fm/c for the reaction at 35A MeV 
and t $\geq$ 500 fm/c for the reaction at 26A MeV.

\begin{figure}
\includegraphics[scale=0.425]{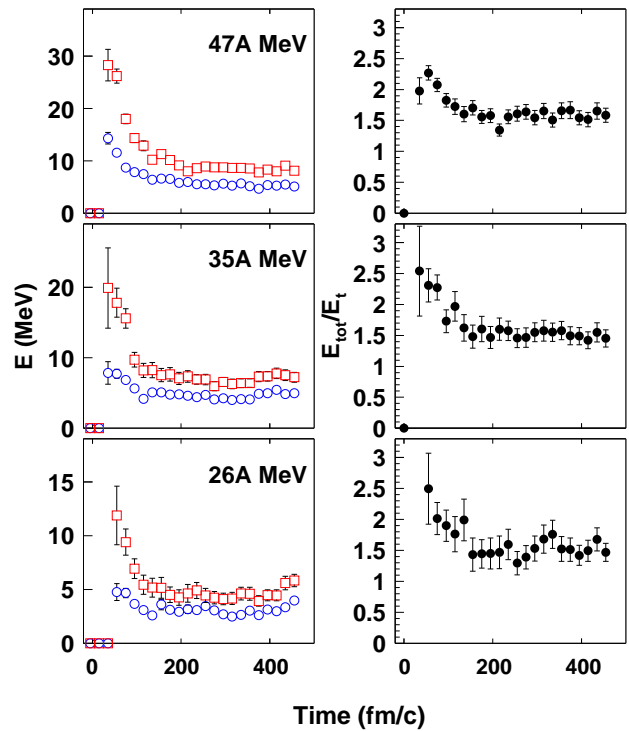} 
\caption{\footnotesize (Left)Average kinetic energy of neutrons 
in the center
of mass system as a function of time for $^{197}$Au at three incident 
energies. The incident energies are indicated in each panel. Total
and transverse energy are shown by open squares and circles, 
respectively. Energies and times are those when neutrons are identified 
for the first time. (Right) Ratios of the total to transverse energies of
the left figure.}
\label{NEkEt}
\end{figure}

The energy spectra of the light particles also provide valuable 
information on the reaction mechanism. In order to avoid the Coulomb energy 
complications for particle emissions, neutron energy spectra are studied. 
In Fig.~\ref{NEkEt} kinetic 
energies of neutrons are plotted as a function of time. Total and transverse 
kinetic energies are plotted on the left. For all incident energies, 
both energies decrease rapidly as time increases up to t $\sim$ 120 
$\sim$150 fm/c and decrease very slowly after that. As seen in 
Fig.~\ref{AmaxAmax2} at 35A MeV, the composite system undergoes 
multifragmentation at t $\sim200\sim300$ fm/c, but
the energy spectra change very smoothly around this time range. 
No effect is observed in the energy spectra. This indicates 
that the fragments are already formed before this time. On the right, the 
ratio of the two energies are plotted. The ratio is above 2 at t $\leq$ 100 
fm/c and rapidly decreases to a value of 3/2 at around 120-150fm/c, 
slightly depending on the incident energy. Then the ratio becomes more or less
constant after that. The value of 3/2 indicates that the neutron emission 
source is fully thermalized. 

\begin{figure}
\includegraphics[scale=0.44]{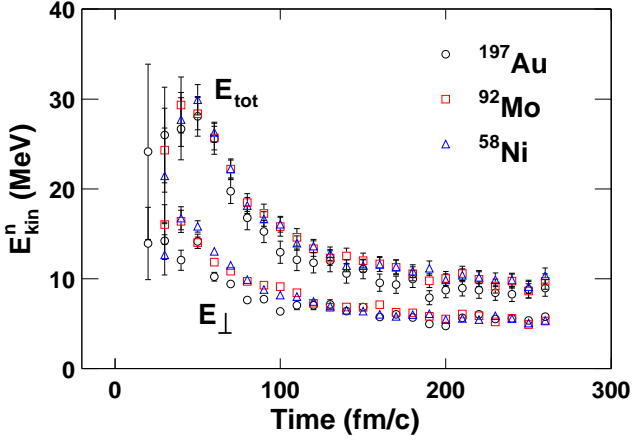} 
\caption{\footnotesize Average kinetic energy of neutrons as a function of time 
at 47A MeV.
Different symbols correspond to different targets, as indicated in the 
figure. Upper set of results shows the total kinetic energy E$_{tot}$ and 
lower set for the transverse energy (E$_{\perp}$). 
}
\label{NEkall}
\end{figure}

In Fig.\ref{NEkall} total and transverse kinetic energies of neutrons are 
compared for reactions with different targets at 47A MeV. 
Shapes of the spectra are very similar to each other  
for all three targets. The energies for the lighter targets are slightly 
higher, reflecting the semi-transparency for these reactions. 
The similarities in energies indicate that the preequilibrium emission 
mechanism for different reactions is similar, 
even though the semi-transparency is 
prominent for the reactions with the lighter targets.
  
\subsection*{C. Fragment emission}

\begin{figure}
\includegraphics[scale=0.425]{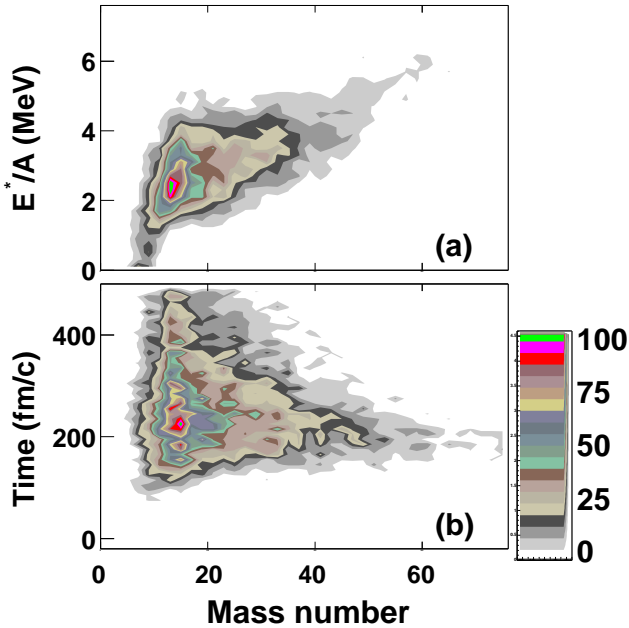} 
\caption{\footnotesize (a) Excitation energy and (b) 
emission time of fragments with A $\leq$ 100 are plotted as 
a function of fragment mass number for $^{197}$Au at 47A MeV. The largest 
fragment is excluded in these plots.
}
\label{FragExTimeA}
\end{figure} 

\begin{figure}
\includegraphics[scale=0.575]{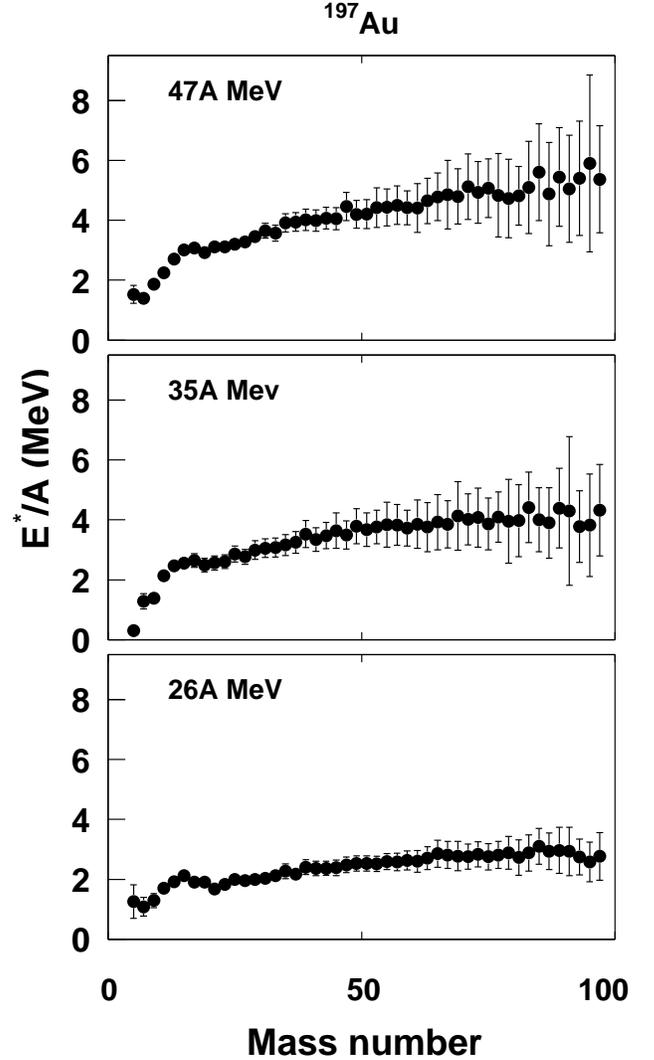} 
\caption{\footnotesize Average excitation energy of fragments with Z $\geq$ 3
is plotted as a function of the fragment mass number for $^{197}$Au  
at three incident energies. The incident energy is indicated in each panel. 
}
\label{FragAvEx}
\end{figure}

\begin{figure}
\includegraphics[scale=0.475]{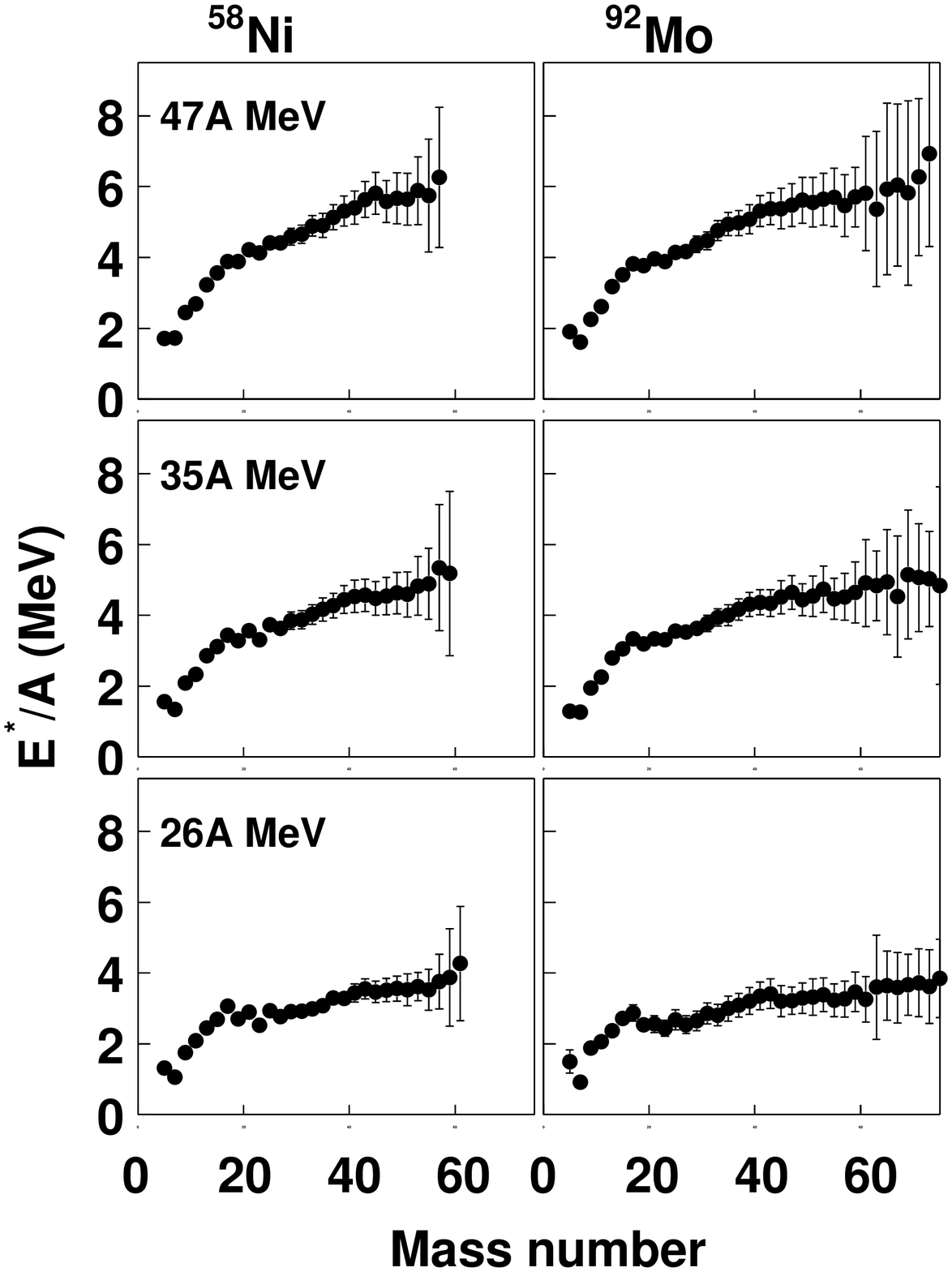} 
\caption{\footnotesize Similar plots to Fig.~\ref{FragAvEx}, but for 
$^{58}$Ni on left and $^{92}$Mo on right sides, respectively. 
}
\label{FragAvEx2}
\end{figure}

One of the interesting observations in the calculated results 
in Fig.~\ref{AmaxAmax2} is that the excitation energy of the second largest 
fragment is much lower than that of the largest fragment. In 
order to study this in detail, the excitation energies and emission times of 
fragments are examined.  Results for $^{197}$Au at 47A MeV are plotted as a 
function of fragment mass number in Fig.~\ref{FragExTimeA}. These values are 
evaluated at the time when each fragment is identified for the first time. 
A fragment, identified at a certain time step, often breaks up in the later 
time steps. A fragment is considered as a newly identified fragment 
when the fragment 
appears with a mass number difference of more than 4 mass units between the 
time span of 10fm/c. In that case, the end of the time span is taken as 
the emission time of the new fragment. 
As seen in the upper panel, the distribution shows a peak around
E$^{*}$/A $\sim$ 2.5 MeV at A $\sim$ 15. For the lighter fragments their
excitation energy rapidly decreases with decreasing mass number.
For the heavier fragments, on the other hand, the excitation energy
slowly increases with increasing mass number and reaches
E$^{*}$/A $\sim$5$\sim$6 MeV at A $\sim$ 60, which is similar to that of the
largest fragments at t $\sim$ 200 fm/c. The lower panel indicates that
lighter fragments are identified in a broad range of time from 100 fm/c to
500 fm/c, peaking around 250 fm/c, whereas the heavier fragments of F$_{2}$
are identified mainly around t $\sim$ 200 fm/c.
  
This trend is essentially the same for the reactions at lower incident
energies. In Fig.~\ref{FragAvEx}, the average
excitation energy of fragments is compared for $^{197}$Au at all incident
energies. For all cases the excitation energy shows a similar trend.
At 47A MeV the average excitation energy of the lightest fragment starts
less than 2 MeV/nucleon and increases rapidly to about 3 MeV/nucleon
Then it increases slowly with increasing
mass number. At 26A MeV, the average excitation energies are about
2 MeV/nucleon lower than those at 47A MeV.

Similar observations are also made in the other reaction systems. 
In Fig.~\ref{FragAvEx2} the average excitation energies of fragments are shown 
for $^{58}$Ni and $^{92}$Mo at all incident energies. The excitation energies 
of fragments at a given incident energy are almost identical and very similar 
to that for $^{197}$Au. This indicates that, in the AMD-V calculations,  
cold fragment emission is a common feature of intermediate heavy ion 
reactions. 
  
\begin{figure}
\includegraphics[scale=0.45]{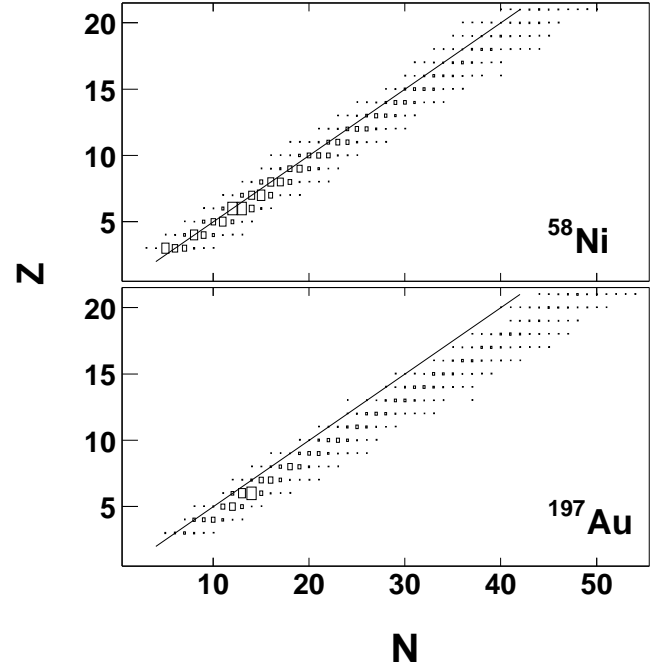} 
\caption{\footnotesize Isotope distributions of fragments for $^{58}$Ni 
(upper) and $^{197}$Au (lower). Lines indicate N=2Z. 
}
\label{FragNZ}
\end{figure}

\begin{figure}
\includegraphics[scale=0.475]{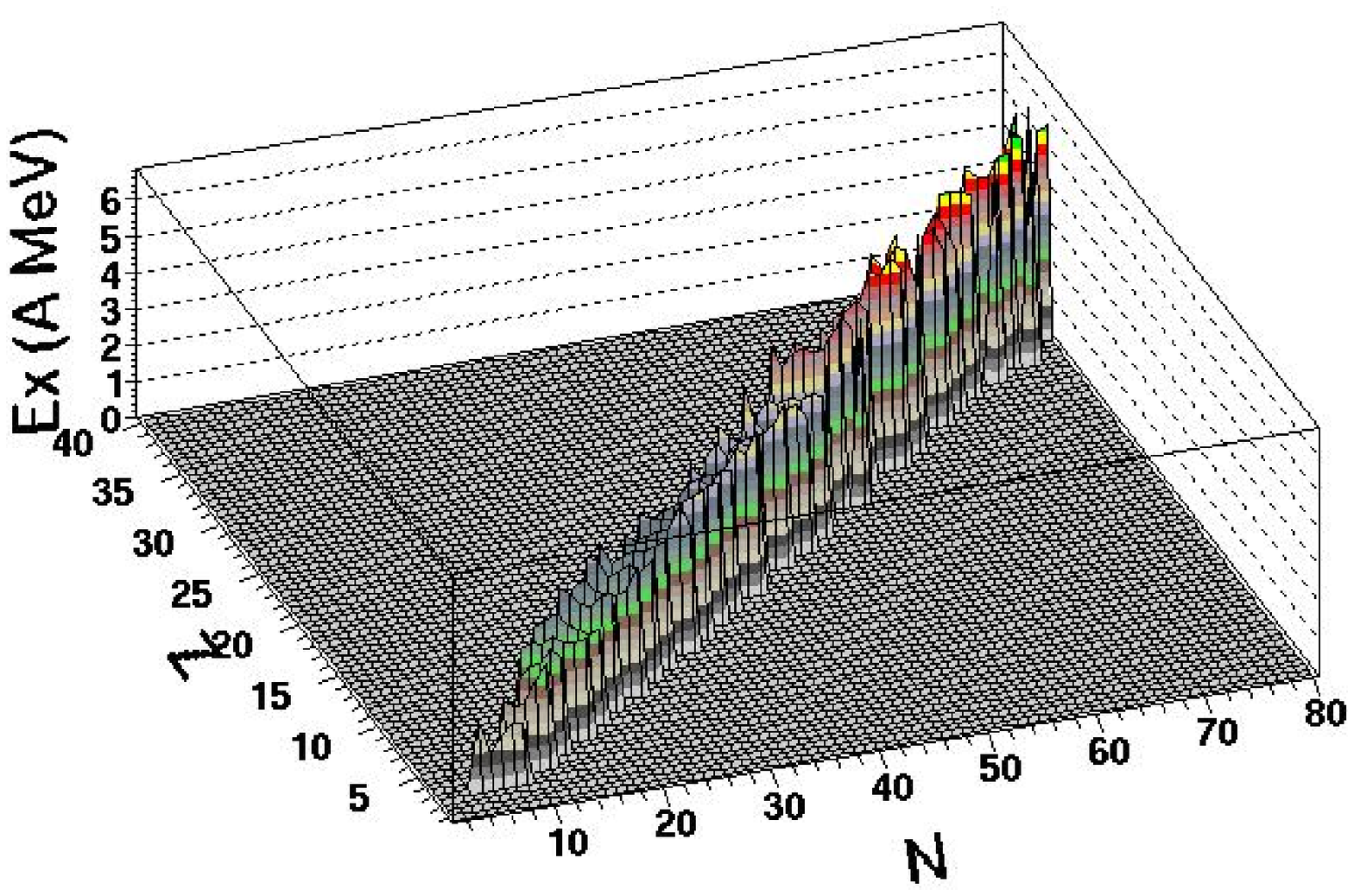}
\includegraphics[scale=0.475]{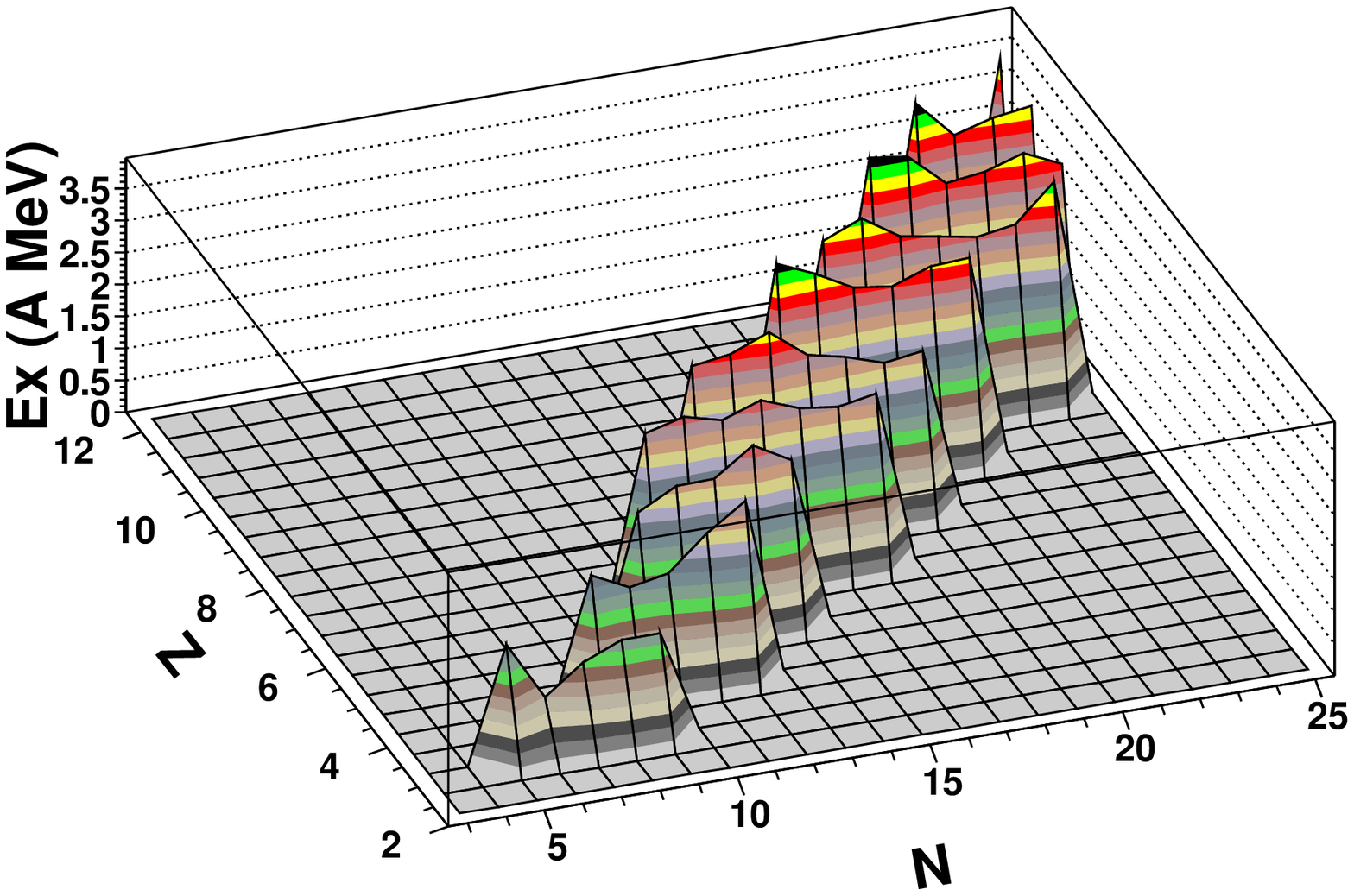}
\caption{\footnotesize Average excitation energy distributions of isotopes 
for $^{197}$Au. The lower plot is the same figure as the upper, but expanded 
for the lighter fragments. 
}
\label{FragExNZ}
\end{figure}

It is also interesting to see the excitation energy distribution for different 
isotopes with a given atomic charge Z. The distribution of produced 
isotopes significantly depends on the neutron-proton ratio of the system.
In Fig.~\ref{FragNZ} isotope distributions are compared between $^{197}$Au 
and $^{58}$Ni at 47A MeV. As one can see, fragments produced in the reaction 
with $^{197}$Au are distributed much more on the neutron rich side than those 
with $^{58}$Ni. 
In Fig.~\ref{FragExNZ}, the average excitation energy of these isotopes is 
plotted for $^{197}$Au. As seen in the figure, the average excitation energy 
of isotopes for a given Z shows only a small variation. This variation is 
typically within less than 1 MeV/nucleon. 
Very neutron rich isotopes, such as $^{21}$O and $^{24}$F, have 
excitation energies similar to those near the $\beta$ stability line.

\begin{figure}
\includegraphics[scale=0.45]{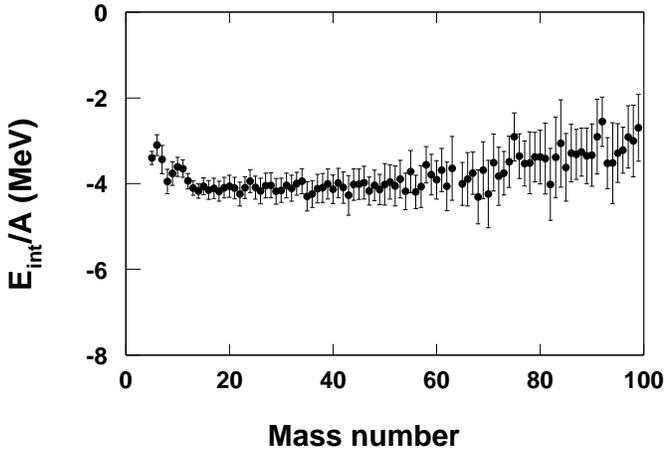}
\caption{\footnotesize Average internal energy distribution of fragments with 
A $\geq$ 5 for $^{197}$Au at 47A MeV. The energy is evaluated when the 
fragments are identified at the first time. 
}
\label{FragEintA}
\end{figure}

The excitation energies studied above are the calculated internal energies 
corrected by the binding energy, as given by Eq.~\ref{eq:Ex}. The 
binding energy varies, depending on the mass and charge of fragments. This 
variation becomes significant for lighter fragments. It is, therefore, 
interesting to see the 
distribution of the internal energy without the correction of the binding 
energy, in order to study the energy partition to the fragments. 
In Fig.~\ref{FragEintA}, the calculated internal energies are plotted as a 
function of mass number for $^{197}$Au at 47A MeV. The distribution shows a 
quite different trend, comparing to those seen in the excitation energies in 
Figs.~\ref{FragAvEx} and \ref{FragAvEx2}, especially for fragments with 
A $\leq$ 15. The internal energies of fragments are almost constant 
and even slightly increase for the lighter fragments, whereas their excitation 
energies rapidly decrease. This is because these fragments are less and less 
bound with decreasing mass number. 
For heavier fragments with A $\geq$ 60, the internal 
energy increases slowly with increasing mass number, which is also seen 
in the excitation energy distribution. This observation indicates that each 
fragment carries away more or less the same amount of internal energy per 
nucleon, which is about -4 MeV/nucleon at 47A MeV. This energy becomes about 
-5 MeV/nucleon at 26A MeV.

\begin{figure}
\includegraphics[scale=0.45]{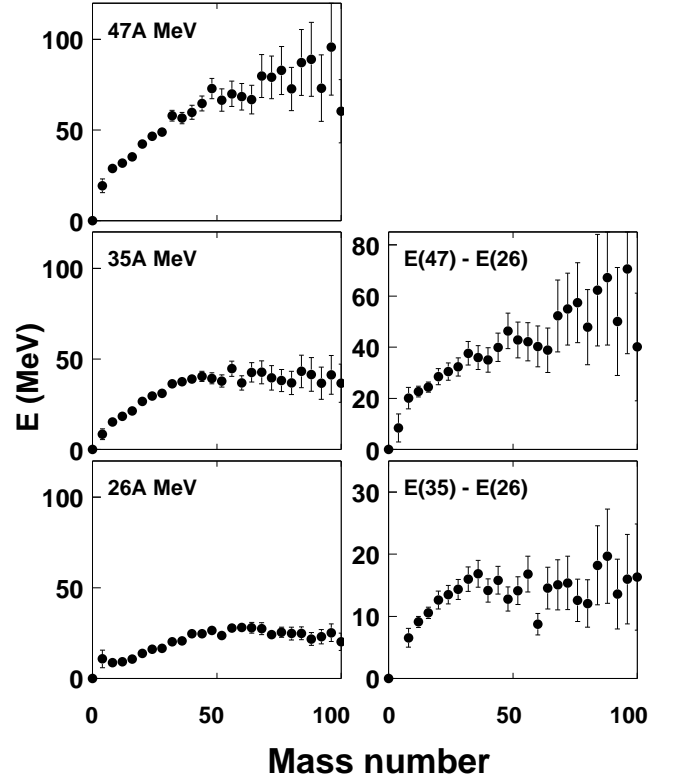} 
\caption{\footnotesize (Left) Average kinetic energy of fragments 
in the center of mass system as a function of fragment mass number
for $^{197}$Au. 
Incident energy is indicated in each panel. Energy
is evaluated at the time when the fragment is identified at the first time.
(Right) Subtracted kinetic energy of fragments for E$_k$ at 47A MeV - E$_k$ 
at 26A MeV in the upper panel and for E$_k$ at 35A MeV - E$_k$ at 26A MeV 
in the lower one. 
} 
\label{FragEkA}
\end{figure}

The energy spectra of fragments also provide valuable information about 
reaction mechanisms. 
In Fig.~\ref{FragEkA} the average kinetic energies of fragments are shown 
on the left column for $^{197}$Au. The average kinetic energy 
increases almost linearly as mass increases. This linear increase 
results partially from the increase of the Coulomb energy 
for larger fragments and partially from a possible expansion energy. 
The Coulomb energy contribution should be similar for the three incident 
energies, because the fragment charge distributions are similar as seen in 
Fig.~\ref{Density-plot}. 
In order to eliminate the Coulomb energy contribution from the kinetic energy,
the kinetic energy at 26A MeV is subtracted from those at 47A MeV and 
35A MeV. The subtracted results are shown on the right column in the figure. 
At 47A MeV they still show a significant linear 
increase as mass increases, whereas at 35A MeV, the linear increase becomes 
less prominent. As a thermal energy contribution to the fragment kinetic 
energy should be constant for different mass fragments, this remaining linear 
increase indicates that these fragments have gone through a significant 
expansion process. The difference of the expansion energy between 47A MeV and 
26A MeV is about 0.5A MeV and 0.1A MeV between 35A MeV and 26A MeV. The sharp 
drop of the energy difference between these two cases suggests that the 
expansion energy at 26A MeV is very small ($\ll$ 0.1A MeV). Therefore we can 
conclude that the approximate expansion energy is $\sim$0.5A MeV 
at 47A MeV and $\sim$0.1A MeV at 35A MeV. 

\begin{figure}
\includegraphics[scale=0.45]{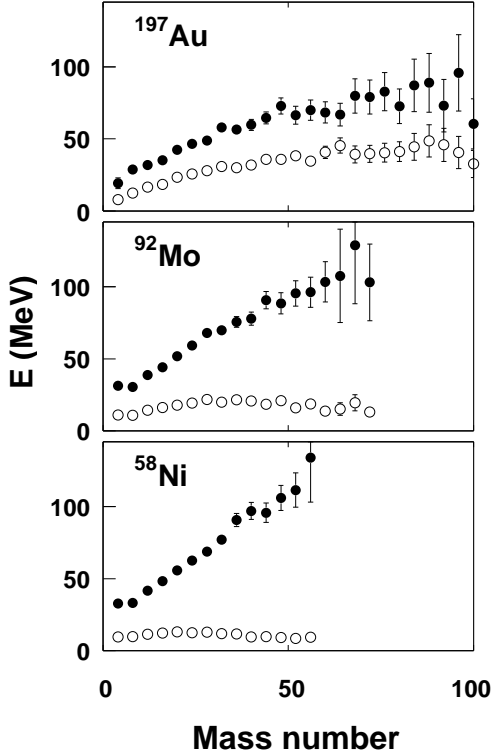} 
\caption{\footnotesize Average total and transverse fragment energies as a 
function of fragment mass number in the center of mass system for reactions 
at 47A MeV with the different targets. The target is indicated in each figure. 
Dots indicate the total energy and open circles indicate the transverse energy.
} 
\label{FragEkEt}
\end{figure}

The distributions of fragment kinetic energies are quite different for 
$^{58}$Ni and $^{92}$Mo at 47A MeV, reflecting different reaction mechanisms. 
In Fig.~\ref{FragEkEt} the average total kinetic and transverse energies of 
fragments are compared for the different targets at 47A MeV. For $^{197}$Au, 
the ratio of the total to the transverse energy is between 1/2 $\sim$ 3/2, 
which indicates the existence of a significant radial expansion. 
For $^{58}$Ni and $^{92}$Mo at 47A MeV, the total kinetic energy 
increases more rapidly as fragment mass increases. At A $\sim$50, the 
average kinetic energy for the $^{58}$Ni target becomes almost double of that 
for the $^{197}$Au target. On the other hand the transverse energy does not 
increase as the total kinetic energy increases and stays around $\sim$10 MeV. 
This indicates that the fragment distribution is stretched along the beam axis. This results from the semi-transparency, as reported in the previous 
work~\cite{wada00}.

\subsection*{D. Thermal equilibration time}

In Fig.~\ref{NEmissionT} for $^{197}$Au, the ratio of the emission rate
between the target nucleons and the projectile nucleons reaches a factor
of 3 at t $\sim$200 fm/c at 47A MeV, t $\sim$300 fm/c at 35A MeV 
and t $\geq$ 500 fm/c at 26A MeV. This factor of 3 is roughly the ratio of 
the target mass to the projectile mass and expected for nucleon emissions from 
a thermalized system. 
However this observation does not 
mean that the system reaches thermal equilibrium until 
these late stages, because a significant amount of the
target nucleons are carried away by cold light fragments before these times.
These fragments decay very slowly and do not contribute to the emission rate 
at the early stages. 

\begin{figure}
\includegraphics[scale=0.5]{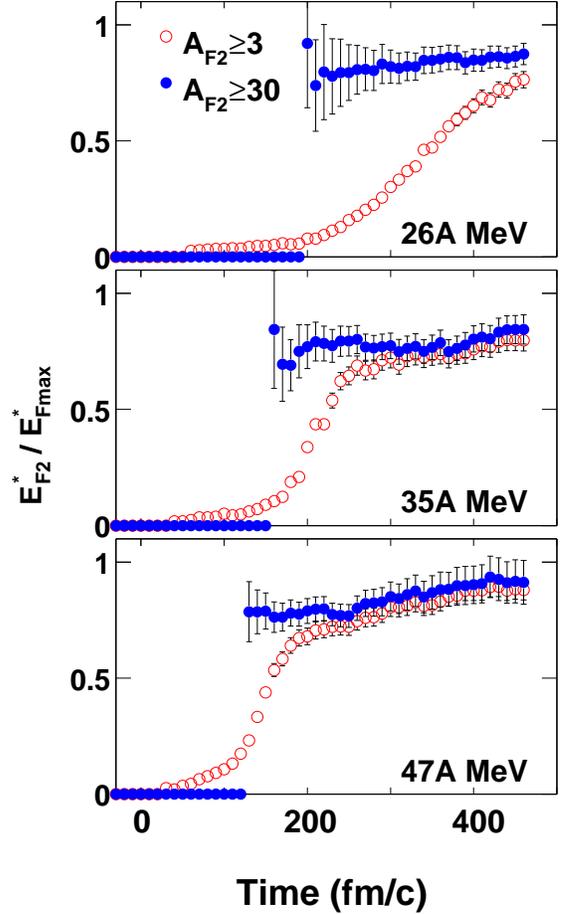} 
\caption{\footnotesize Ratio of the excitation energy of the second largest 
fragment ( F$_{2}$ ) and that of the largest fragment ( F$_{max}$ ) as a 
function of time for $^{197}$Au. Open circles represent results when all 
fragments with Z $\geq$ 3 are taken into account as F$_{2}$ and solid dots 
indicate results when only fragments 
with A $\geq$ 30 are taken into account as F$_{2}$. Incident energy is 
indicated in each figure.
}
\label{F2ExFmaxEx}
\end{figure}

In order to elucidate the thermal equilibration time, 
the ratio of the excitation energies of the largest 
fragment (F$_{max}$) and the second largest fragment (F$_{2}$) is plotted in 
Fig.~\ref{F2ExFmaxEx}. When all fragments with Z $\geq$ 3 are taken into 
account for F$_{2}$ (circles), the ratio gradually increases and reaches a 
plateau at t $\sim$200 fm/c at 47A MeV, t $\sim$250 fm/c at 35A MeV 
and t $\geq$ 500 fm/c at 26A MeV. These times are very similar to those for 
which the nucleon emission rates become equal to the 3:1 ratio. 
On the other hand when only fragments 
with A $\geq$ 30 are taken into account for F$_{2}$, the ratio behaves in a  
drastically different way and stays more or less constant from the 
earliest time when these heavier fragments are identified. This
time is t = 130 fm/c at 47A MeV, 160 fm/c at 35A MeV and 180 fm/c at 
26A MeV. The values of the ratios are slightly lower than 1.0, 
because the excitation energies of A $\sim$ 30 $\sim$ 40 are still slightly 
lower than those of the heavier fragments, as is seen in Fig.~\ref{FragAvEx}. 
This observation indicates that the system is already thermally equilibrated 
at the earliest times that these heavier fragments can be identified.

From the nucleon emission rates in Figs.~\ref{NEmissionT} and \ref{NEmissionT2},
one can see that the preequilibrium emission ceases at t $\sim$120 fm/c 
at 47A MeV and slightly later at lower incident energies.  
The ratios of the neutron total to transverse energies on the right side of 
Fig.~\ref{NEkEt} 
reach a value of 3/2 at these times and remain constant after that.
The value 3/2 suggests that the neutron emission source is fully thermalized.
These times are consistent with those observed from 
the earliest emission times of F$_{2}$ 
with A $\geq$ 30. Therefore from these observations we conclude that the 
excited nucleus, remaining after the preequilibrium emission, reaches thermal 
equilibrium at t $\sim$120 fm/c at 47A MeV, t $\sim$140 at 35A MeV and 
t $\sim$160 fm/c at 26A MeV. 
   
\subsection*{E. Multifragmentation and cold fragment emission}

In this subsection we briefly summarize the observations for the 
calculated central collision events and draw a scenario 
for multifragmentation in Fermi energy heavy ion reactions. 

For light particle emission we summarize as follows:
\begin{itemize}{}
\item {The emission rate of nucleons emitted as light particles shows a  
maximum at t $\sim$ 80 fm/c for all reactions studied here. At a given 
incident energy all reactions show similar rates for 
these preequilibrium particles emitted from the projectile and 
from the target.
The emission rates are insensitive to the change of the in-medium NN cross 
section. The absolute emission rate decreases significantly as the incident 
energy decreases, but even at 26A MeV a similar rate is observed for the 
projectile and target nucleons.  
These facts suggest that the preequilibrium light particles are emitted from 
the overlap zone of the projectile and the target. 
}
\end{itemize}

For fragment emission we summarize as follows:
\begin{itemize}{}
\item {Charge distributions of fragments with Z $\leq$ 20 are very similar 
to each other in 
the experiments and in the calculations, regardless of the target. 
}
\item {Cold fragment emission is generally observed  
for all reactions studied here. The excitation energy for A $\leq$ 10 is 
E$^{*}$/A = 1$\sim$2 MeV. It increases linearly as mass increases and 
reaches to that of the largest fragment at A $\sim$ 60, remaining essentially 
constant for heavier fragments. The rapid increase of the excitation energy 
for fragments with A $\leq$ 15 originates from the variation of the binding
energy. The internal energies of fragments show a flat distribution 
for A $\leq$ 60. 
}
\item {For $^{197}$Au a significant radial expansion is 
observed at 47A MeV. The expansion energy decreases quickly as the incident 
energy decreases. The approximate expansion energy is $\sim$0.5A MeV at 
47A MeV and $\sim$0.1A MeV at 35A MeV. 
}
\item{For the reactions with $^{58}$Ni and $^{92}$Mo at 47A MeV,
a significant semi-transparency is observed.
} 
\end{itemize}
 
For the thermal equilibration time we summarize as follows: 
\begin{itemize}{}
\item {
The thermal equilibrium of the system is established around t $\sim$ 120
fm/c at 47A MeV, t $\sim$ 140 fm/c at 35A MeV and t $\sim$ 160 
fm/c at 26A MeV. 
} 
\end{itemize}
       
From these observations we can draw the following scenario for the 
multifragmentation in the Fermi energy domain: 
 
\begin{list}{}
\item {(1) The projectile enters into the target nucleus and creates a 
hot overlap zone.
}  
\item {(2) The overlap zone decays quickly by emitting fast nucleons before 
thermal equilibrium with the surrounding target nucleons is established.
}
\item {(3) During the preequilibrium emission, the remaining system starts to 
equilibrate. Nucleons close to each 
other in the phase space start to form cold fragments which 
coexist with a hotter nucleon gas. 
In this stage, cold 
fragments share almost an equal internal energy per nucleon with each other.  
}
\item {(4) The system continues to  
expand and undergoes multifragmentation with cold fragment emission. This 
multifragmentation process is accelerated by additional dynamical 
processes, such as the semi-transparency for $^{58}$Ni and $^{92}$Mo 
at 47 A MeV and 
the radial expansion for $^{197}$Au at 47A MeV. These additional dynamical 
processes drastically change the kinetic properties of the emitted 
fragments.
} 
\end{list}
%\end{itemize}

\section*{VII. DISCUSSION}

For the above multifragmentation scenario, the existence of a hot
overlap zone and cold fragment formation play important roles. For the
existence of the hot overlap zone, many experimental evidences have been 
accumulated. Since the pioneering work of Awes {\it et al.}, moving source 
analyses have been applied to many reaction systems at intermediate energies
~\cite{Awes82,Westfall84,wada89,Prindle98,hagel00,Santonocito02}. The existence 
of the intermediate velocity source has been commonly observed in these 
analyses. The source velocity extracted from these analyses is about 
half the beam velocity, regardless of the target mass. This has been 
interpreted as indicating that 
the source consists of equal amounts of nucleons from the projectile and the 
target. The apparent temperature of the source increases linearly as the 
incident energy increases, and is more or less independent of the target mass 
and the centrality of the reaction~\cite{wada89,Prindle98,Santonocito02}. These 
observations are consistent with preequilibrium emission of light
particles from the overlap zone, discussed in the previous section.  

A more direct indication for the preequilibrium particle emissions from the 
overlap zone was reported recently by Verde {\it et al.}~\cite{Verde02}.  
In that work, two-proton correlation functions were studied using 
a newly developed imaging source technique.
The source image of two protons was derived numerically 
from the two proton correlation function, without assuming a source shape, 
such as a Gaussian distribution. This method is capable of determining 
the source size in the environment in which fast (preequilibrium) and slow 
(evaporated) components coexist in spectra. The model was
applied to $^{14}$N + $^{197}$Au at 35A MeV and the source size of the 
preequilibrium protons was extracted. The extracted source size is 
R$_\frac{1}{2} \sim$ 3 fm for all proton energy ranges.  
(R$_{\frac{1}{2}}$ is the radius at a half density). 
This source size is comparable to the size of the projectile and much smaller 
than the size of the target.

Cold IMF emission has been reported in many heavy ion collisions. 
The temperature of IMFs has been determined from the population of the excited 
states~\cite{Pochodzalla87,Xu89,Nayak92,Schwartz93}.
In these studies temperatures of T $\sim$3$\sim$5.5 MeV have been typically 
obtained in a variety of intermediate heavy ion reactions.
If one uses the relation E$^{*}$=aT$^{2}$ with a=A/10, these temperatures 
lead to fragment excitation energy of  E$^{*}$/A = 1$\sim$3 MeV. 
Marie {\it et al.} also reported cold fragment emissions observed in 
LCP(light charged particle)-IMF correlation studies~\cite{marie98}. 
The average charged particle multiplicities for each emitted IMF 
were evaluated in the $^{129}$Xe + Sn reaction at 50A MeV. 
Using a statistical cascade code, the average excitation energy 
was determined for each IMF. When the neutron to proton 
ratio of IMFs is assumed to that of the system, the excitation energy of 
E$^{*}$/A $\sim$ 3 MeV is obtained for fragments 
with 3 $\leq Z \leq 20$. This energy 
is significantly lower than the excitation energy ( E$^{*}$/A $\sim$ 12 MeV) 
of the system, but consistent with the value calculated in this study.
The correlation experiment has recently been extended from 25A MeV to 150A 
MeV in the same reaction system~\cite{Hudan00}. The extracted excitation 
energy of fragments is E$^{*}$/A $\sim$ 3 MeV, independent of the incident 
energy. 

In each of the studies in Refs.~\cite{marie98} and \cite{Hudan00}, however, the 
extracted excitation energies are slightly decreasing as Z increases,
whereas the results of the calculations in Figs.~\ref{FragAvEx} 
and \ref{FragAvEx2} 
show a clear increase of the excitation energy as fragment mass increases, 
especially for the lighter fragments. This difference may result 
from the fact that, in both of the experimental studies, the fragment isotope 
distribution was neglected and neutron emission from the fragment was not 
measured. Thus the average primary fragment mass for a given average primary 
fragment Z was calculated by (1+f)Z, where f is N/Z of the  
fragment and it was assumed in Refs.~\cite{marie98} and \cite{Hudan00} that  
N/Z is that of the composite system. If N/Z is assumed to be that of the valley
of the stability rather than that of the composite system, the extracted 
excitation energies become less than 1.5 MeV/nucleon for most of the fragments. 
As seen in Fig.~\ref{FragNZ}, for a given Z the calculated distributions of 
fragment mass are broad   
and the N/Z value at the peak of the mass distribution is slightly less than 
that of the system. For example, the distribution of Oxygen isotopes has
a peak at $^{18}$O, or N/Z = 1.25, for $^{64}$Zn + $^{197}$Au, which has a
composite system of N/Z = 1.39. Since the experimental values are
sensitive to the determination of the primary fragment mass and hence the 
number of emitted neutrons, one needs to take into 
account the isotope distribution of the fragments for the evaluation of the 
average excitation energy in order to make 
more detailed comparisons between the experiments and the calculations. 

Recently Cussol suggested that the limitation of the excitation energy of 
fragments may be related to the neutron (or proton) separation energies of the 
fragments, using a Classical Molecular Dynamics model~\cite{Cussol03}. 
In most heavy ion reactions, 
the composite system is neutron rich. When light fragments are formed in 
an equilibrated system, the neutron/proton ratio of light fragments 
tends to be far away from the $\beta$ stability line and these fragments have
small neutron separation energies. These fragments can not hold 
excitation energy higher than the neutron separation energy and therefore 
they limit
the average excitation energy of the fragments. In AMD-V simulations, however, 
no systematic correlation is observed between the neutron/proton ratio and the
excitation energy, as shown in Fig.~\ref{FragExNZ}. The calculated results 
also show no significant difference between the neutron rich system 
($^{64}$Zn + $^{197}$Au ) and more or less symmetric system 
($^{64}$Zn + $^{58}$Ni).

The coexistence of cold fragments and gaseous nucleons
has recently been 
suggested by Campi {\it et al.}, using a Classical Molecular Dynamics (CMD) 
model~\cite{Campi02}. In their CMD model, particles interact through a 
Lennard-Jones potential, which is given by 
\begin{equation} 
V(r) = 4\epsilon \{ (\frac{\sigma}{r})^{12} - (\frac{\sigma}{r})^{6}\}, 
\end{equation}
where $\epsilon$ and $\sigma$ are energy and length scales, respectively.
In the reference a thermalized system is provided with a given number of 
particles and a fixed excitation energy within a small container in high 
density.
At time=0 the container is removed and the system is allowed to expand 
freely. It undergoes multifragmentation. Fragment size and 
temperature are examined during the process. Since the particles interact 
through the above interaction, they tend to move at a distance r$_{o}$ from 
each other ( V(r) becomes minimum at r=r$_{o}$ ). 
Using proximity in the phase space and separating gaseous
nucleons by the Hill criterium (potential energy + kinetic energy $\geq$ 0.), 
clusters are identified in the gaseous particle environment. By examining the 
temperature of gaseous particles and the internal temperatures of the clusters
during the expansion, it is revealed that the temperatures
are much lower than the temperature of the gaseous particles.
The distribution of cluster size is almost independent of 
the time of the expansion, indicating that the cluster size is independently 
determined of the particle density. 

\begin{figure}
\includegraphics[scale=0.45]{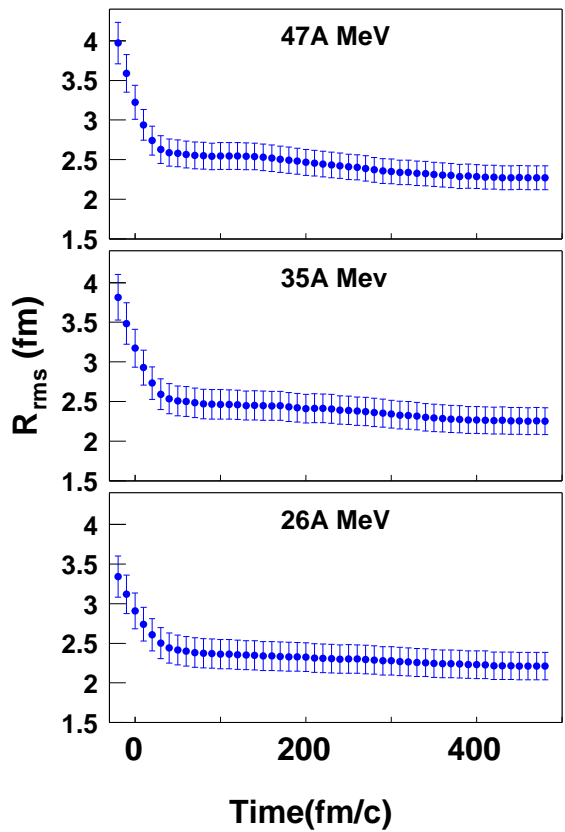}
\caption{\footnotesize Time evolution of the average root-mean-square radius 
of 14 nucleons, which end up $^{14}$C at t=480 fm/c for $^{197}$Au. 
See details in the text.
} 
\label{C14Rrms}
\end{figure}

In AMD-V, nucleons are moving freely in a mean field, but 
anti-symmetrization tends to maintain clusters with nucleons which are close 
each other in the phase space. This is illustrated in Fig.~\ref{C14Rrms}. 
In the figure the root-mean-square radius ( R$_{rms}$ ) of $^{14}$C, which 
is the most abundant light fragment for $^{197}$Au, is examined. 14 nucleons 
in $^{14}$C are identified at t=480 fm/c. For these identified 14 nucleons, 
R$_{rms}$ is calculated as a function of time. At t $\leq$ 0, 
the time before the two 
nuclei touch each other, R$_{rms}$ shows large 
values because about 3$\sim$4 nucleons on average originate from the 
projectile and the rest from the target. At 47A MeV, R$_{rms}$ reaches 2.6 fm 
at t = 30 fm/c, the time of the full overlap, and after that, R$_{rms}$ 
decreases very slowly. The change in R$_{rms}$ between t = 30 fm/c and 
t = 480 fm/c is $\sim$10\%. This trend is essentially the same for lower 
incident energies. At 26A MeV, R$_{rms}$ becomes 2.4 fm at t = 50 fm/c and 
2.2 fm at 480 fm/c. The slightly smaller R$_{rms}$, compared to that 
at 47A MeV, reflects the fragment lower excitation 
energy at 26 A MeV. This indicates that nucleons in light fragments are 
already close together at the time of the overlap and move together until 
they are identified as an isolated fragment. The excitation energies of light 
fragments are small, independently of the emission time, as seen in 
Fig.~\ref{FragExTimeA}(b). This indicates that the fragments stay 
cold and coexist with the hot nucleon gas. 
      
Different dynamics involved in the reactions, such as radial expansion and 
semi-transparency, drastically change the kinematic
characteristics of the emitted fragments. A drastic change in kinematic 
distributions has been reported in $^{40}$Ar 
induced reactions~\cite{Colin98}. In that experiment
$^{40}$Ar beam was bombarded on Cu, Ag and Au targets over a wide range of 
incident energy (8A -115A MeV). For the central collisions below 44A MeV, 
the observed nature of the reactions indicates a fusion-like process, 
whereas a multi-body spray of IMFs is observed at higher incident energies. 
This observation is consistent with the present results for $^{58}$Ni. 
At 26A MeV the projectile fully stops in the target and fragment emissions
become more or less isotropic. On the other hand at 47A MeV, a significant
semi-transparency spreads the fragments at forward and backward directions.

The existence of a significant expansion has been suggested based upon the 
determination of the average
kinetic energy of IMFs by the INDRA collaboration~\cite{Marie97,Frankland01}.  
The extracted energy of the possible expansion is about 0.5A MeV for Xe+Sn 
at 50A MeV, consistent with the observation in Fig.~\ref{FragEkA} at 
47A MeV. However
the value in the above references has been evaluated based on the 
statistical multifragmentation model, which is not supported in the present 
work. 
 
\section*{VIII. SUMMARY}

A detailed study has been presented for three 
reaction systems, $^{64}$Zn + $^{58}$Ni, $^{92}$Mo and $^{197}$Au, at three 
incident energies, 26A, 35A, 47A MeV. Multiplicity distributions, charge 
distributions, energy spectra and velocity distributions of the reaction 
products have been measured. Detailed comparisons have been 
made between the experimental results and those of the calculations with two 
different effective interactions, corresponding to a soft EOS (K=228 MeV) 
and a stiff EOS (K=360 MeV). 
For light particles the main characteristic features of the above observables 
are generally well described by all three calculations, although some 
significant discrepancies are observed in all cases. 
Experimental proton multiplicity distributions, 
for example, favor the calculations with the soft EOS, whereas those of the 
$\alpha$ particles favor the calculations with the stiff EOS. 
For the fragments with Z $\ge$ 3 the experimental energy spectra clearly
favor the calculations with the stiff EOS, although the multiplicity of 
IMF's is overpredicted by a factor of 1.5-2 for 
$^{58}$Ni and $^{92}$Mo and 2-3 for $^{197}$Au.
The velocity distributions of $\alpha$ particles and IMF's also show a 
clear difference between calculations with the soft EOS and the stiff EOS, 
relating to the velocity distribution of nucleons in an early stage of the 
reaction. The experimental velocity distributions of these particles clearly 
favor the calculations with the stiff EOS. Two different formulations for the 
in-medium NN cross sections were used in this study, but the effect of 
changing the NN cross section is rather small on the observables and 
no conclusive evidence favoring one of these formulations has been found.

The mechanism of the multifragmentation process was explored, 
using central collision events of
AMD-V, calculated with the stiff EOS. Preequilibrium light particle emission 
from the overlap zone and multifragmentation with cold fragment emission are
commonly observed in all reactions. The thermal equilibration times 
observed in the calculated results are t $\sim$ 120 fm/c at 47A MeV, 
t $\sim$ 140 fm/c 
at 35A MeV and t $\sim$ 160 fm/c at 26A MeV. Cold fragments are formed  
by nucleons close to each other in phase space at early stages and they stay 
cold in a hotter nucleon gas, exhibiting an almost equal internal energy per 
nucleon. Then the 
system expands and undergoes multifragmentation with cold fragment emission.  
For $^{197}Au$ at 47A MeV a significant radial expansion 
takes place and for $^{58}$Ni and $^{92}$Mo at 47A MeV, semi-transparency 
becomes significant. The kinematic characteristics of emitted 
fragments change drastically, depending on the additional dynamics involved. 
Many existing experimental results are consistent with the 
mechanisms suggested by the calculated AMD-V events.  
   
\section{ACKNOWLEDGMENTS}

We thank the staff of the Texas A\&M Cyclotron facility for their support 
during the experiment. We also acknowledge the staff of the RIKEN VPP700E 
supercomputer facility and Dr. I. Tanihata and Dr. Y. Yano for allowing the
use of the facility. We further thank Dr. R. J. Charity for 
providing us the GEMINI code.  This work was
supported by the U.S. Department of Energy under Grant No. 
DE-FG03-93ER40773 and the Robert Welch Foundation.

\end{document}